\documentclass[%
 aip,
 amsmath,amssymb,
 preprint,% 单栏
]{revtex4-1}

\usepackage{graphicx}% Include figure files
\usepackage{dcolumn}% Align table columns on decimal point
\usepackage{bm}% bold math
\usepackage[utf8]{inputenc} % support chinese and greek
\usepackage[T1]{fontenc}
\usepackage{CJKutf8}
\usepackage{mathrsfs}

\usepackage{etoolbox}

\usepackage[colorlinks=true, citecolor=red]{hyperref}
\begin{document}
	\preprint{AIP/123-QED}
	\title{\Large A transformer-based neural operator for large-eddy simulation of turbulence }
	% Force line breaks with	
	\author{Zhijie Li(\begin{CJK}{UTF8}{gbsn}李志杰\end{CJK})} % 
	\affiliation{Department of Mechanics and Aerospace Engineering, Southern University of Science and Technology, Shenzhen 518055, China}
	
	\author{Tianyuan Liu(\begin{CJK}{UTF8}{gbsn}刘天源\end{CJK})} %  
	\homepage{Authors to whom correspondence should be addressed: wangjc@sustech.edu.cn and tianyuan@pku.edu.cn}
	\affiliation{College of Engineering, Peking University, Beijing, 100091, China}

	\author{Wenhui Peng(\begin{CJK}{UTF8}{gbsn}彭文辉\end{CJK})} %  
	\affiliation{The Hong Kong Polytechnic University, Department of Applied Mathematics, 999077, Hong Kong Special Administrative Region of China}
	
	\author{Zelong Yuan(\begin{CJK}{UTF8}{gbsn}袁泽龙\end{CJK})} %  
	\affiliation{Harbin Engineering University Qingdao Innovation and Development Base, Qingdao, 266000, China}
	
	\author{Jianchun Wang(\begin{CJK}{UTF8}{gbsn}王建春\end{CJK})}%  
	\homepage{Authors to whom correspondence should be addressed: wangjc@sustech.edu.cn and tianyuan@pku.edu.cn}
	\affiliation{Department of Mechanics and Aerospace Engineering, Southern University of Science and Technology, Shenzhen 518055, China}

%	\date{\today}
%------------------------------------------------------------------------------------------------------
\begin{abstract}
Predicting the large-scale dynamics of three-dimensional (3D) turbulence is challenging for machine learning approaches. This paper introduces a transformer-based neural operator (TNO) to achieve precise and efficient predictions in the large-eddy simulation (LES) of 3D turbulence. The performance of the proposed TNO model is systematically tested and compared with LES using classical sub-grid scale (SGS) models, including the dynamic Smagorinsky model (DSM) and the dynamic mixed model (DMM), as well as the original Fourier neural operator (FNO) model, in homogeneous isotropic turbulence (HIT) and free-shear turbulent mixing layer. The numerical simulations comprehensively evaluate the performance of these models on a variety of flow statistics, including the velocity spectrum, the probability density functions (PDFs) of vorticity, the PDFs of velocity increments, the evolution of turbulent kinetic energy, and the iso-surface of the Q-criterion. The results indicate that the accuracy of the TNO model is comparable to the LES with DSM model, and outperforms the FNO model and LES using DMM in HIT. In the free-shear turbulence, the TNO model exhibits superior accuracy compared to other models. Moreover, the TNO model has fewer parameters than the FNO model and enables long-term stable predictions, which the FNO model cannot achieve. The well-trained TNO model is significantly faster than traditional LES with DSM and DMM models, and can be generalized to higher Taylor-Reynolds number cases, indicating its strong potential for 3D nonlinear engineering applications.

\end{abstract}

\maketitle
%-------------------------------------------------------------------------------------------------
\section{Introduction}
% LES + SGS models 
Large-eddy simulation (LES) has gained increasing attention as a valuable tool for turbulent flow prediction thanks to its balance between computational accuracy and efficiency. LES is renowned for its capability to simulate the evolution of large-scale structures and model sub-grid scale (SGS) motions\cite{smagorinsky1963,lilly1967,germano1992,garnier2009large,deardorff1970numerical}. LES can model complex flows where Reynolds-averaged Navier-Stokes (RANS) is not successful, showing superior accuracy in such cases. Additionally, it demonstrates superior computational efficiency compared to direct numerical simulation (DNS)\cite{durbin2018some,pope1975more,choi2012grid,yang2021grid,bin2023constrained}. Properly addressing the unclosed subgrid-scale (SGS) terms is crucial for obtaining accurate results in LES. Since the pioneering work, various SGS models have been proposed\cite{meneveau2000,fan2023eddy,chen2012reynolds}, including the dynamic Smagorinsky model (DSM)\cite{moin1991dynamic,germano1991,germano1992}, the dynamic mixed model (DMM)\cite{shi2008constrained,erlebacher1992toward}, the velocity gradient model (VGM)\cite{clark1979evaluation}, and the approximate deconvolution model (ADM)\cite{maulik2018data,yuan2021dynamic}.

%NN model
Recently, neural networks (NNs) have become increasingly popular in computational fluid dynamics (CFD)\cite{brunton2020machine,duraisamy2019turbulence,lienen2023generative}. Deep neural networks have shown remarkable capabilities in approximating highly non-linear functions, making them particularly suitable for modeling complex phenomena\cite{lecun2015deep}. Well-trained ``black-box'' neural network  models can quickly make inferences on modern computers, surpassing the efficiency of traditional CFD methods. Various techniques based on NNs have been developed to improve the simulation of turbulence in RANS and LES methods\cite{maulik2018data,yuan2021dynamic,wang2021artificial,beck2019deep,xie2019artificial,wang2018investigations,gamahara2017searching,zhou2019subgrid,li2021data,ling2016reynolds}. Zhou et al. employed an artificial neural network to create a novel SGS model for LES of isotropic turbulent flows\cite{zhou2019subgrid}. Beck et al. introduced convolutional neural networks (CNNs) and residual neural networks (RNNs) to develop precise SGS models for LES\cite{beck2019deep}. Guan et al. employed transfer learning and a deep convolutional neural network to enhance the precision and stability of turbulence prediction\cite{guan2022stable}. Han et al. introduced a hybrid deep neural network model to capture the spatial-temporal features of the unsteady flow fields\cite{han2019novel}. Furthermore, several researchers have investigated the integration of supplementary physical knowledge into deep learning\cite{cai2021physics,lanthaler2022error,karniadakis2021physics,raissi2017physics,yang2019predictive}. Raissi et al. proposed a physics-informed neural network (PINN) to solve general nonlinear partial differential equations (PDEs)\cite{raissi2019physics}. Chen et al. introduced a method that combines theory-guided principles with hard constraint projection. This technique transforms governing equations into a more manageable form through discretization and then applies hard constraint optimization via projection within a localized patch\cite{chen2021theory}. Jin et al. developed Navier-Stokes flow nets (NSFnets) by integrating governing equations, initial conditions, and boundary conditions into the loss function\cite{jin2021nsfnets}.  

% NO model，but
Although previous neural network architectures have shown proficiency in learning mappings within finite-dimensional Euclidean spaces, their ability to generalize across various initial or boundary conditions is constrained\cite{raissi2019physics,wu2020data,xu2021deep,lu2022comprehensive,kovachki2023neural,goswami2022deep}. Li et al. recently proposed a novel framework known as the Fourier neural operator (FNO), which enables efficient learning of mappings between infinite-dimensional spaces using input-output pairs\cite{li2020fourier}. The FNO model has demonstrated superior performance in predicting two-dimensional (2D) turbulence compared to current state-of-the-art models including U-Net\cite{chen2019u}, TF-Net\cite{wang2020towards}, and ResNet\cite{he2016deep}. Wen et al. introduced an enhanced version of the FNO model called U-FNO, which combines the U-Net architecture with FNO for improved accuracy and efficiency in solving multiphase flow problems\cite{wen2022u}. Choubineh et al. evaluated the FNO model to predict pressure distribution over a small, specified shape-data problem\cite{choubineh2023fourier}. Peng et al. presented an improved FNO model coupled with attention, which showed highly effective in accurately reconstructing the statistical properties and instantaneous flow structures of 2D turbulence, especially at high Reynolds numbers\cite{peng2022attention}. While there has been a considerable amount of research on FNO-based models\cite{li2022fouriergeo,jiang2023fourier,tran2021factorized,renn2023forecasting,li2021physics,guibas2021adaptive,hao2023gnot,benitez2023fine,meng2023fast,deng2023temporal}, most of these works have concentrated on 2D problems. Modeling three-dimensional (3D) turbulence using deep neural networks poses a greater challenge due to the substantial increase in data size and parameters compared to 2D problems\cite{momenifar2022dimension}. Training models with a large number of parameters can be computationally expensive and memory-demanding, presenting difficulties given hardware limitations.

% 然后找3D transformer和它的一些发展，最后在提出一种3D TNO
Mohan et al. proposed two reduced models for 3D homogeneous isotropic turbulence (HIT) and scalar turbulence. These models were developed using deep learning techniques, specifically the convolutional generative adversarial network (C-GAN) and compressed convolutional long-short-term-memory (CC-LSTM) network\cite{mohan2020spatio}. Nakamura et al. integrated a 3D convolutional neural network autoencoder (CNN-AE) with a long short-term memory (LSTM) network to predict the 3D channel flow\cite{nakamura2021convolutional}. Li et al. applied the FNO approach for the prediction of large-scale dynamics of 3D homogeneous isotropic turbulence\cite{li2022fourier}. Peng et al. introduced a linear attention coupled Fourier neural operator (LAFNO) for simulating 3D isotropic turbulence and free-shear turbulence\cite{peng2023linear}. Li et al. proposed an implicit U-Net enhanced Fourier neural operator (IU-FNO) for the long-term prediction of large-scale dynamics of turbulence\cite{li2023long}. Li et al. proposed a geometry-informed neural operator (GINO) to efficiently learn the solution operator of large-scale PDEs with varying geometries\cite{li2023geometry}.

Due to the outstanding performance, the transformer model has received significant attention and application in natural language processing\cite{vaswani2017attention,keskar2019ctrl,dai2019transformer,hao2023gnot,li2022transformer}. In recent years, an increasing number of researchers have started to adopt this powerful model in the domain of CFD to explore its potential\cite{drikakis2024generalizability,xu2023super,momenifar2021emulating,bi2022pangu,wang2024towards,janny2023eagle}. Patil et al. proposed a convolutional encoder-decoder-based transformer model for autoregressive training on spatio-temporal data of turbulent flows\cite{patil2022autoregressive}. Dang et al. introduced a learned simulator based on the transformer architecture, namely turbulence neural transformer (TNT), to predict turbulent dynamics on coarsened grids\cite{dang2022tnt}. Li et al. proposed a factorized transformer to accurately and efficiently simulate 2D Kolmogorov flow and 3D smoke buoyancy\cite{li2023scalable}. Jiang et al. proposed a transformer-based decoding architecture for flow field prediction\cite{jiang2023transcfd}. However, general transformer methods often employ grid downsampling techniques to reduce computational costs. Unfortunately, this approach tends to sacrifice high-frequency information of turbulence, leading to a decrease in simulation accuracy. In this work, we develop a transformer-based neural operator (TNO) as a surrogate model for LES of 3D turbulence. The TNO model enables direct inference and prediction of flow dynamics at a million-grid scale, ensuring the preservation of multiscale features of turbulence. It achieves stable, efficient, and accurate predictions on large-scale dynamics of turbulence. 

% section 
The rest of the paper is structured as follows. Section \ref{sec:2} introduces the governing equations for LES and the conventional SGS models. Section \ref{sec:3} briefly introduces the Fourier neural operator and some modifications. In Section \ref{sec:4}, we introduce a transformer-based neural operator model, namely TNO model. Section \ref{sec:5} focuses on conducting numerical simulations for forced homogeneous isotropic turbulence (HIT) and free-shear mixing layer turbulence. This section includes a description of the data generation process, an evaluation of the proposed model's performance through a \textit{posteriori} analysis, and an assessment of the computational efficiency achieved. Finally, Section \ref{sec:6} gives some further discussions on the proposed method and concludes the paper.

%======================================================================================================
%--------------------------------------------------------------------------------------------------
\section{\label{sec:2}The large-eddy simulation and sub-grid scale models}
In this section, the large-eddy simulation (LES) of the incompressible turbulence and classical models for the unclosed subgrid-scale (SGS) stress are briefly introduced.

\subsection{\label{sec:2.1}The governing equations for LES of the turbulence}
For three-dimensional incompressible turbulence, the mass and momentum equations are given by\cite{pope2000,Ishihara2009}
	\begin{equation}
	\frac{\partial u_i}{\partial x_i}=0,
	\label{eq1}
	\end{equation}
	\begin{equation}
	\frac{\partial u_i}{\partial t}+\frac{\partial\left(u_i u_j\right)}{\partial x_j}=-\frac{\partial p}{\partial x_i}+\nu \frac{\partial^2 u_i}{\partial x_j \partial x_j}+\mathcal{F}_i.
	\label{eq2}
	\end{equation}
Here, $u_i$ is the $i$th velocity component, $p$ denotes the pressure divided by the constant density, $\nu$ represents the kinematic viscosity, and $\mathcal{F}_i$ stands for the $i$th large-scale force component. Besides, the root mean square (rms) velocity is defined by $u_{\text{rms}}=\sqrt{\left\langle u_i u_i\right\rangle}$, and $\langle\rangle$ denotes a spatial average of the computational domain. $\varepsilon=2 v\left\langle S_{i j} S_{i j}\right\rangle$ is the average dissipation rate of kinetic energy, where $S_{i j}=\frac{1}{2}\left(\partial u_i / \partial x_j+\partial u_j / \partial x_i\right)$ is the strain rate tensor. With these variables, the Kolmogorov length scale $\eta$, the Taylor length scale $\lambda$, and the Taylor-scale Reynolds number $Re_\lambda$ can be respectively defined as\cite{wyp2022,pope2000}
	\begin{equation}
	\eta=\left(\frac{\nu^3}{\varepsilon}\right)^{1 / 4}, \quad \lambda=\sqrt{\frac{5 \nu}{\varepsilon}} u_{\text{rms}}, \quad Re_\lambda=\frac{u_{\text{rms}} \lambda}{\sqrt{3} \nu}.
	\label{eq3}
	\end{equation}

Furthermore, the integral length scale $L_I$ and the large-eddy turnover time $\tau$ are given, respectively, by\cite{pope2000}
\begin{equation}
	L_I = \frac{3 \pi}{2\left(u_{\text{rms}}\right)^2} \int_0^{\infty} \frac{E(k)}{k} d k,\quad \tau=\frac{L_I}{u_{\text{rms}}},
	\label{eq4}
\end{equation}
where $E(k)$ is the energy spectrum and the kinetic energy $E_k$ is given by $E_k=\int_0^{\infty}E(k)dk=\frac{1}{2}\left(u_{\text{rms}}\right)^2$. Here, $k$ represents the wave number in Fourier space.

In order to derive the governing equations for the large-scale flow field, a filtering operation $\bar{f}(\mathbf{x})=\int_D f(\mathbf{x}-\mathbf{r}) G(\mathbf{r}, \mathbf{x} ; \Delta) d \mathbf{r}$ is applied to the Navier-Stokes equations. Here, $G$ and $\Delta$ are the filter kernel and filter width, respectively\cite{pope2000,sagaut2006}. $f$ denotes a variable in physical space, and $D$ is the entire domain. For any variable $f$ in Fourier space, a filtered variable is given by $\bar{f}(\boldsymbol{k})=\hat{G}(\boldsymbol{k}) f(\boldsymbol{k})$\cite{pope2000}. Here, $\hat{G}(\boldsymbol{k})$ is the filter transfer function. Therefore, the filtered incompressible Navier-Stokes equations can be derived as follows\cite{pope2000,sagaut2006}
	\begin{equation}
	\frac{\partial \bar{u}_i}{\partial x_i}=0,	
	\label{eq5}
	\end{equation}
	\begin{equation}
	\frac{\partial \bar{u}_i}{\partial t}+\frac{\partial\left(\bar{u}_i \bar{u}_j\right)}{\partial x_j}=-\frac{\partial \bar{p}}{\partial x_i}-\frac{\partial \tau_{i j}}{\partial x_j}+\nu \frac{\partial^2 \bar{u}_i}{\partial x_j \partial x_j}+\overline{\mathcal{F}}_i .
	\label{eq6}
	\end{equation}
Here, $\tau_{ij}$ is the unclosed SGS stress defined by $\tau_{i j}=\overline{u_i u_j}-\bar{u}_i \bar{u}_j$. In order to ensure the solvability of the LES equations, it is crucial to model the SGS stress as a function of the filtered variables.

%Here, an overbar denotes the filtering at scale $\Delta$, and a tilde represents the test filtering operation at the double-filtering scale $\tilde{\Delta}=2 \Delta$. The spectral filter is employed to double-filtering in HIT, and a Gaussian filter is utilized in free-shear turbluent mixing layer.
%--------------------------------------------------------------------------------------------------
\subsection{The sub-grid models for LES}
The subgrid-scale (SGS) models aim to construct an approximate constitutive equation for the unclosed SGS terms based on the resolved variables, and accurately capture the intricate nonlinear interactions between the resolved large scales and unresolved small scales\cite{moser2021,johnson2022}. One of the most commonly used SGS models is the Smagorinsky model, given by \cite{smagorinsky1963,lilly1967,germano1992}
	\begin{equation}
	\tau_{i j}^A=\tau_{i j}-\frac{\delta_{i j}}{3} \tau_{k k}=-2 C_s^2 \Delta^2|\bar{S}| \bar{S}_{i j},	
	\label{eq7}
	\end{equation}
where $\delta_{i j}$ is the Kronecker delta operator, $\Delta$ denotes the filter width, $\bar{S}_{ij}=\frac{1}{2}\left(\partial \bar{u}_i / \partial x_j+\partial \bar{u}_j / \partial x_i\right)$ stands for the strain rate of the filtered velocity, and $|\bar{S}|=\left(2 \bar{S}_{i j} \bar{S}_{i j}\right)^{1 / 2}$ represents the characteristic filtered strain rate. The coefficient $C_s^2$ can be determined empirically or through theoretical analysis\cite{lilly1967}. This coefficient can also be dynamically established using the Germano identity and a least-squares algorithm, yielding the dynamic Smagorinsky model (DSM) with the coefficient given by\cite{germano1991,lilly1992}
	\begin{equation}
	C_s^2=\frac{\left\langle \mathcal{L}_{i j} \mathcal{M}_{i j}\right\rangle}{\left\langle\mathcal{M}_{k l} \mathcal{M}_{k l}\right\rangle}.	
	\label{eq8}
	\end{equation}
Here, $\mathcal{L}_{i j}={\widetilde{\bar{u}_i \bar{u}}}_j-\tilde{\bar{u}}_i \tilde{\bar{u}}_j$, $\mathcal{M}_{i j}=\tilde{\alpha}_{i j}-\beta_{i j}$, $\alpha_{i j}=2 \Delta^2|\bar{S}| \bar{S}_{i j}$, and $\beta_{i j}=2 \tilde{\Delta}^2|\tilde{\bar{S}}| \tilde{\bar{S}}_{i j}$, where an overbar denotes the filtering at scale $\Delta$, and a tilde represents coarser filtering $\tilde{\Delta}=2 \Delta$.

Another commonly used SGS model, namely the dynamic mixed model (DMM), combines the scale-similarity model with the dissipative Smagorinsky term, and is given by\cite{liu1994properties,shi2008constrained}
\begin{equation}
	\tau_{i j}=C_1 {\Delta}^2|\bar{S}| \bar{S}_{i j}+C_2\left(\widetilde{\bar{u}_i \bar{u}_j}-\tilde{\bar{u}}_i \tilde{\bar{u}}_j\right) .
	\label{eq9}
\end{equation}
Similar to the DSM model, the coefficients $C_1$ and $C_2$ of the DMM model are dynamically estimated using the Germano identity through the least-squares algorithm. $C_1$ and $C_2$ are expressed as\cite{yuan2020deconvolutional}
\begin{equation}
	C_1=\frac{\left\langle N_{i j}^2\right\rangle\left\langle L_{i j} M_{i j}\right\rangle-\left\langle M_{i j} N_{i j}\right\rangle\left\langle L_{i j} N_{i j}\right\rangle}{\left\langle N_{i j}^2\right\rangle\left\langle M_{i j}^2\right\rangle-\left\langle M_{i j} N_{i j}\right\rangle^2} ,
	\label{eq10}
\end{equation}
\begin{equation}
	C_2=\frac{\left\langle M_{i j}^2\right\rangle\left\langle L_{i j} N_{i j}\right\rangle-\left\langle M_{i j} N_{i j}\right\rangle\left\langle L_{i j} M_{i j}\right\rangle}{\left\langle N_{i j}^2\right\rangle\left\langle M_{i j}^2\right\rangle-\left\langle M_{i j} N_{i j}\right\rangle^2}.
	\label{eq11}
\end{equation}
Here $M_{ij}=H_{1, i j}-\tilde{h}_{1, i j}$, $H_{1, i j}=-2 \tilde{\Delta}^2|\tilde{\bar{S}}| \tilde{\bar{S}}_{i j}$, $h_{1, i j}=-2 {\Delta}^2|\bar{S}| \bar{S}_{ij}$, and $N_{i j}=H_{2, i j}-\tilde{h}_{2, i j}$, $H_{2, i j}=\widehat{\tilde{\bar{u}}_{i} \tilde{\bar{u}}_{j}} -\hat{\tilde{\bar{u}}}_i \hat{\tilde{\bar{u}}}_j$, $h_{2,ij}=\widetilde{\bar{u}_i \bar{u}_j}-\tilde{\bar{u}}_i \tilde{\bar{u}}_j$. The hat denotes the filter at scale $\hat{\Delta}=4\Delta$.

%#=================================================================================================
%------------------------------------------------------------------------------------------------
\section{\label{sec:3}The Fourier neural operator}
\begin{figure*}
	\includegraphics[width=1\linewidth]{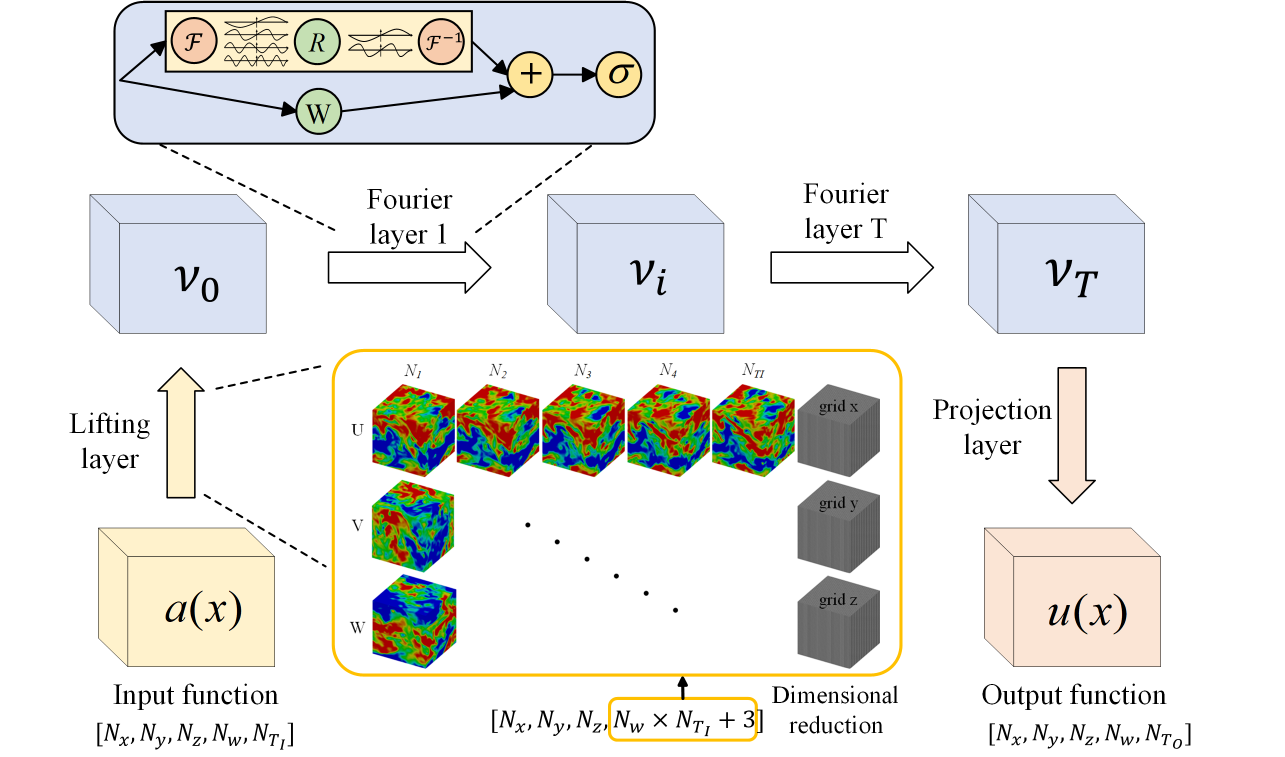}
	\caption{The Fourier neural operator (FNO) architecture.}
	\label{NNFNO}
	\end{figure*}

The Fourier neural operators (FNO) train the model on a finite set of input-output pairs to establish a mapping between two infinite-dimensional spaces. Denote the non-linear mapping as $G^{\dagger}: \mathcal{A} \rightarrow \mathcal{U}$, where $\mathcal{A}=\mathcal{A}\left(D ; \mathbb{R}^{d_a}\right)$ and $\mathcal{U}=\mathcal{U}\left(D ; \mathbb{R}^{d_u}\right)$ are separable Banach spaces of function taking values in $\mathbb{R}^{d_a}$ and $\mathbb{R}^{d_u}$, respectively\cite{beauzamy2011}. Here, $D \subset \mathbb{R}^d$ is a bounded open set, $\mathbb{R}$ denotes real number space, $\mathbb{R}^{d_a}$ is the value sets of input $a(x)$, and $\mathbb{R}^{d_u}$ represents the value sets of output $u(x)$. The Fourier neural operators learn an approximation of $G^{\dagger}$ by constructing a mapping parameterized by $\theta \in \Theta$. The optimal parameters $\theta^{\dagger} \in \Theta$ are determined through data-driven methods\cite{vapnik1999}. The architecture of FNO is shown in Fig.~\ref{NNFNO} which consists of three main steps.

Firstly, the input $a(x)$ is transformed to a higher-dimensional representation $v_0(x)=P(a(x))$ through a local transformation $P$ which is commonly parameterized by a shallow fully connected neural network.

Then, the higher-dimensional representation $v_0(x)$ is updated iteratively by
	\begin{equation}
	v_{t+1}(x)=\sigma\left(W v_t(x)+\left(\mathcal{K}(a ; \phi) v_t\right)(x)\right), \quad \forall x \in D.	
	\label{eq12}
	\end{equation}
Here, $\mathcal{K}$ is a non-local integral operator, $\mathcal{K}: \mathcal{A} \times \Theta_{\mathcal{K}} \rightarrow \mathcal{L}\left(\mathcal{U}\left(D ; \mathbb{R}^{d_v}\right), \mathcal{U}\left(D ; \mathbb{R}^{d_v}\right)\right)$ maps to bounded linear operators on $\mathcal{U}\left(D ; \mathbb{R}^{d_v}\right)$ and is parameterized by $\phi \in \Theta_{\mathcal{K}}$, $\sigma: \mathbb{R} \rightarrow \mathbb{R}$ is non-linear activation function, and $W: \mathbb{R}^{d_v} \rightarrow \mathbb{R}^{d_v}$ is a linear transformation. The neural operators employ iterative architectures $v_0 \mapsto v_1 \mapsto \ldots \mapsto v_T$ where $v_i$ for $i=0,1, \ldots, T-1$ is a sequence of function\cite{li2020b}. 

Finally, the output $u(x)=Q\left(v_T(x)\right)$ is generated by applying a projection $Q$ to the higher-dimensional representation $v_T(x)$, where $Q$ is a fully connected layer that maps from $\mathbb{R}^{d_v}$ to $\mathbb{R}^{d_u}$\cite{li2020fourier}. 

Let $\mathcal{F}$ and $\mathcal{F}^{-1}$ denote the Fourier transform and its inverse transform of a function $f: D \rightarrow \mathbb{R}^{d_v}$, respectively. By replacing the kernel integral operator in Eq.~\ref{eq12} with a convolution operator defined in Fourier space, the Fourier integral operator can be rewritten as
\begin{equation}
	\left(\mathcal{K}(\phi) v_t\right)(x)=\mathcal{F}^{-1}\left(R_\phi \cdot\left(\mathcal{F} v_t\right)\right)(x), \quad \forall x \in D.	
	\label{eq13}
\end{equation}
Here, $R_\phi$ is the Fourier transform of a periodic function $\mathcal{K}: \bar{D} \rightarrow \mathbb{R}^{d_v \times d_v}$ parameterized by $\phi \in \Theta_{\mathcal{K}}$. Assuming that the frequency mode $k \in D$ is periodic, it can be represented using a Fourier series expansion that consists of discrete modes denoted as $k \in \mathbb{Z}^d$. Finite-dimensional parameterization can be obtained by truncating the Fourier series at a maximal mode $k_{\max }=\left|Z_{k_{\max }}\right|=\mid\left\{k \in \mathbb{Z}^d:\left|k_j\right| \leq k_{\max , j}\right.$, for $\left.j=1, \ldots, d\right\} \mid$. $\mathcal{F}\left(v_t\right) \in \mathbb{C}^{n \times d_v}$ can be obtained by discretizing domain $D$ with $n \in \mathbb{N}$ points, where $v_t \in \mathbb{R}^{n \times d_v}$\cite{li2020fourier}.  $\mathcal{F}\left(v_t\right) \in \mathbb{C}^{k_{\max } \times d_v}$ can be obtained by truncating the higher modes, where $\mathbb{C}$ is the complex space. $R_\phi$ is parameterized as complex-valued weight tensor containing a collection of truncated Fourier modes $R_\phi \in \mathbb{C}^{k_{\max } \times d_v \times d_v}$. Therefore, the following equation can be derived as:
\begin{equation}
	%	\begin{multline} % use for double columns, + \\ 
	\left(R_\phi \cdot\left(\mathcal{F} v_t\right)\right)_{k, l}=\sum_{j=1}^{d_v} R_{\phi k, l, j}\left(\mathcal{F} v_t\right)_{k, j},\quad k=1, \ldots, k_{\max }, \quad j=1, \ldots, d_v.
	\label{eq14}
	%	\end{multline}
\end{equation}

For a flow that is uniformly discretized with resolution $s_1 \times \cdots \times$ $s_d=n$, the fast Fourier transform (FFT) can be applied to $\mathcal{F}$. For $f \in \mathbb{R}^{n \times d_v}, k=$ $\left(k_1, \ldots, k_d\right) \in \mathbb{Z}_{s_1} \times \cdots \times \mathbb{Z}_{s_d}$, and $x=\left(x_1, \ldots, x_d\right) \in D$, the FFT $\hat{\mathcal{F}}$ and its inverse $\hat{\mathcal{F}}^{-1}$ are given by
	\begin{equation}
	\begin{aligned}
	& (\hat{\mathcal{F}} f)_l(k)=\sum_{x_1=0}^{s_1-1} \cdots \sum_{x_d=0}^{s_d-1} f_l\left(x_1, \ldots, x_d\right) e^{-2 i \pi \sum_{j=1}^d \frac{x_j k_j}{s_j}}, \\
	& \left(\hat{\mathcal{F}}^{-1} f\right)_l(x)=\sum_{k_1=0}^{s_1-1} \cdots \sum_{k_d=0}^{s_d-1} f_l\left(k_1, \ldots, k_d\right) e^{2 i \pi \sum_{j=1}^d \frac{x_j k_j}{s_j}}. 
	\end{aligned}
	\label{eq15}
	\end{equation}

The FNO model stands out from traditional numerical methods and other neural operator approaches by showcasing remarkable adaptability and generalization in effectively handling high-dimensional and large-scale data\cite{li2022fourier,guibas2021adaptive,rashid2022learning,pathak2022fourcastnet,li2022fouriergeo}. The main idea of FNO is to utilizes Fourier transform to map high-dimensional data to the frequency domain and employs neural network to learn the relationships between Fourier coefficients, enabling the approximation of nonlinear operators. This approach makes FNO to efficiently learn the governing rules of entire families of partial differential equations (PDEs)\cite{li2022fourier}.

Unlike the approach employed in our previous study\cite{li2022fourier}, where direct FFT was applied to the four-dimensional data, we can improve accuracy and reduce computational cost by first reducing the dimensionality of the input and then applying FFT to the three-dimensional data. More specifically, Fig.~\ref{NNFNO} depicts the tensor input with the following dimensions: $N_x$, $N_y$, and $N_z$ represent the number of grid points along the three respective directions, $N_w$ indicates the velocity components in the three directions, and $N_{T_I}$ is a time dimension of five in the input. We combine the three velocity components $N_w$ with the five input time steps $N_{T_I}$, incorporating the spatial grid position information along the three directions. This integration allows us to map the data using lifting layer $P$ into a higher-dimensional space for learning purposes. By employing this dimensionality reduction method through dimension merging, we effectively reduce the number of network parameters and address the issue of limited channel width due to memory constraints. This further enhances accuracy and computational efficiency. Therefore, all subsequent discussions of the FNO model refer to the modified version of the FNO model.

%=================================================================================================
%------------------------------------------------------------------------------------------------
\section{\label{sec:4}A Transformer-based neural operator}
We develop a transformer-based neural operator (TNO) as a surrogate model for LES to achieve stable, efficient, and accurate predictions on large-scale dynamics of turbulence. The TNO model aims to learn the mapping relationship between input $a(x)$ and output $\hat{u}(x)$ through a data-driven approach. This mapping relationship can be expressed as
\begin{equation}
	\hat{u}(x) = \mathcal{G} ( \textbf{x}, a(x) ).
	\label{eq-tno-a}
\end{equation}
Here, $\textbf{x}$ represents the spatial coordinate encoding, $a(x)$ denotes the input function, which is typically the acquired velocity field at multiple time steps. $\mathcal{G}$ refers to the TNO network, and $\hat{u}(x)$ represents the predicted velocity field for the subsequent time step.

\begin{figure*}
	\includegraphics[width=1\linewidth]{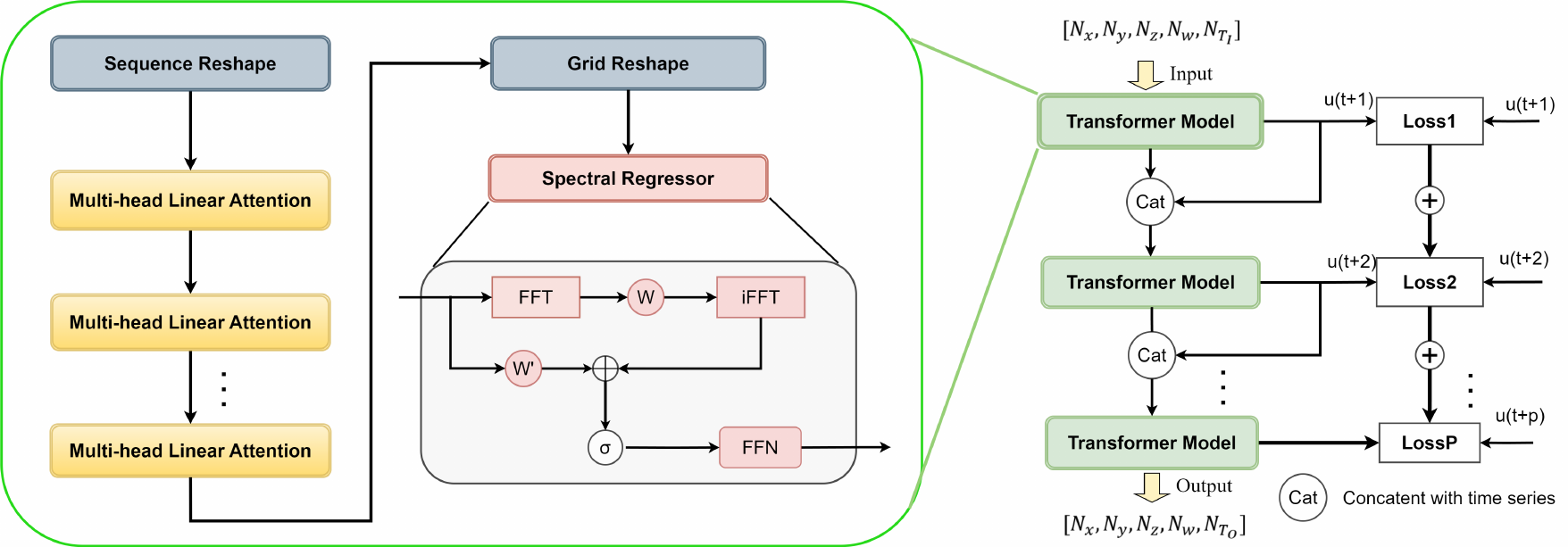}
	\caption{The transformer-based neural operator (TNO) architecture.}
	\label{TNO}
	\end{figure*}

Similar to FNO model, the flow field information of the first $N_{T_I}$ time steps for the original input tensor is defined as $a(x)$, with a size of $a(x)=[N_x, N_y, N_z, N_w, N_{T_I}]$. Here, to simplify notation, we have omitted the batch size dimension and utilized a broadcasting mechanism for parallel computation along this dimension. Thus, only the remaining dimensions need to be considered. The transformer-based model is shown in Fig.~\ref{TNO}. 

%--------------------sequence reshape and grid reshape
The sequence reshape block $\mathcal{R}(a)$ transforms a three-dimensional tensor of spatial data into a sequence, converting the temporal and physical field components into features. It can be represented as $\mathcal{R}(a):=[N_x, N_y, N_z, N_w, N_{T_I}] \rightarrow [N_s, N_c]$. Here, the sequence $ N_s = N_x \cdot N_y \cdot N_z $ and the features $ N_c = N_w \cdot N_{T_I} $. The grid reshape block can be defined as $\mathcal{R}^{-1}(a):=[{N_x \cdot N_y \cdot N_z}, {N_w }] \rightarrow [N_x, N_y, N_z, N_w]$.

%-------------------------------------------attention
	\begin{figure*}
	\includegraphics[width=1\linewidth]{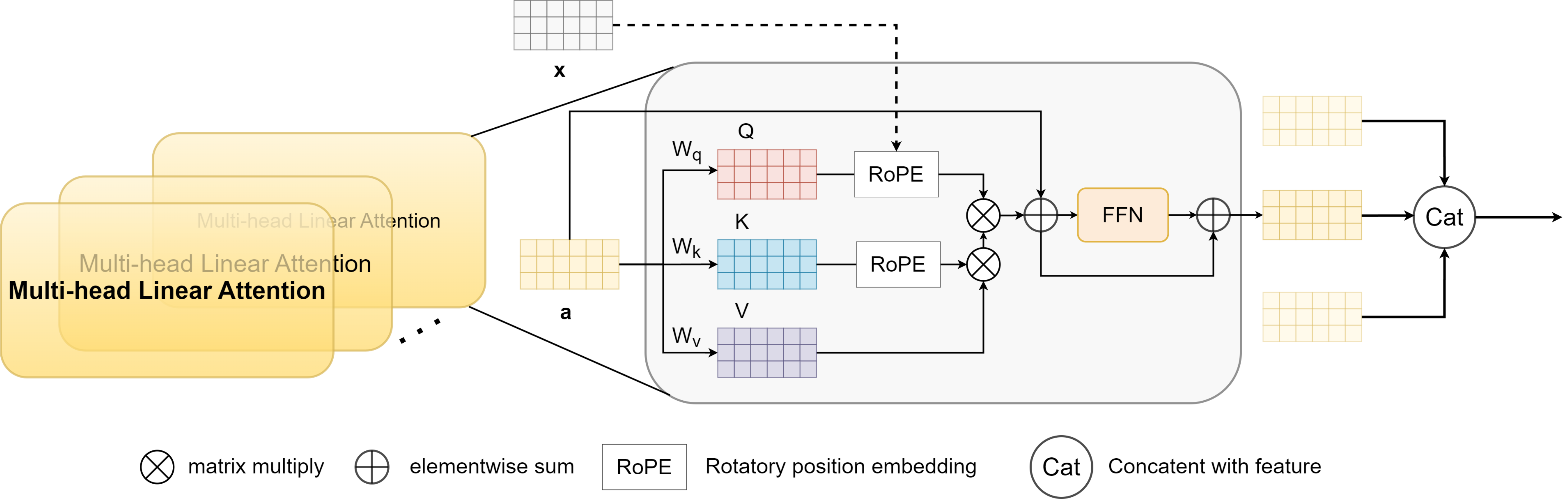}
	\caption{The architecture of multi-head linear attention.}
	\label{MHLattention}
	\end{figure*}

Figure.~\ref{MHLattention} illustrates a schematic diagram of the multi-head attention mechanism. For the input $\mathbf{a}$, it is initially divided into $N_h$ submatrices, corresponding to $N_h$ heads. It can be represented as $\mathbf{a}=\left[\mathbf{a}^1, \mathbf{a}^2, \ldots, \mathbf{a}^{N_h}\right]$. The embedding of the query, key, and value vectors $\mathbf{Q}, \mathbf{K}, \mathbf{V}$ for each head can be expressed as
\begin{equation}
	\mathbf{Q}^i=\mathbf{W}_{\mathbf{q}} \mathbf{a}^i, \mathbf{K}^i=\mathbf{W}_{\mathbf{k}} \mathbf{a}^i, \mathbf{V}^i=\mathbf{W}_{\mathbf{v}} \mathbf{a}^i.
	\label{eq-tno-d}
\end{equation}
Here, the matrices $\mathbf{W}_{\mathbf{q}}, \mathbf{W}_{\mathbf{k}}, \mathbf{W}_{\mathbf{v}}$ are learnable and shared across all heads. The dimensions of $\mathbf{Q}$, $\mathbf{K}$, and $\mathbf{V}$ are $[N_s, N_d]$, where $N_s$ represents the number of samples and $N_d$ denotes the dimensionality. Each $\mathbf{Q}^i$ for each head is of size $N_d/N_h$. It can be written as
\begin{equation}
	\quad \mathbf{Q}=\left[\mathbf{Q}^1, \ldots \mathbf{Q}^{N_h}\right] ; \mathbf{K}=\left[\mathbf{K}^1, \ldots \mathbf{K}^{N_h}\right] ; \mathbf{V}=\left[\mathbf{V}^1, \ldots \mathbf{V}^{N_h}\right].
	\label{eq-tno-e}
\end{equation}

By utilizing $\operatorname{Attention}(\cdot)$ to denote the general notion of attention calculation, it can be defined as
\begin{equation}
	\mathbf{z} = \operatorname{Attention}(\mathbf{Q}, \mathbf{K}, \mathbf{V}) = \operatorname{sim}(\mathbf{Q}\mathbf{K}^T)\mathbf{V}.
	\label{eq-tno-f}
\end{equation}
Here, the function $sim(\cdot)$ represents the similarity calculation function (including the normalization process), and $(\cdot)^T$ denotes matrix transpose. The computational complexity defined by Eq.~\ref{eq-tno-f} is $\mathcal{O}(N_s^2N_d)$, where $N_s$ represents the length of the grid node sequence and $N_d$ denotes the dimensionality of the embedding. This method poses significant computational challenges, particularly in the case of predicting physical fields where $N_s$ is typically proportional to the number of grid nodes. In the case of predicting three-dimensional turbulence, the scale of nodes can reach magnitudes of $10^6$ to $10^7$, imposing a substantial computational burden.

To maintain consistency with the notation commonly used in attention literature, we represent the $i$-th row in the sequence as $q_i, k_i, v_i$, where $0 < i < d$. Similarly, we denote the vector formed by the $j$-th column in the sequence as $q^j, k^j, v^j$, where $0 < j < N_s$. It can be rewritten as
\begin{equation}
	\left(\mathbf{z}_i\right)^j=\sum_{j=1}^{N_s} w_s\left(q_i \cdot k_l\right) v_l \approx \int_{\Omega} \kappa\left(x_i, \xi\right) v^j(\xi) d \xi.
	\label{eq-tno-g}
\end{equation}
Here, the operation between the attention matrix and the values can be considered as an integration operation of a kernel function. By considering the values as the basis functions, selecting the Fourier basis functions as the set of kernel functions, utilizing the inverse Fourier transform basis functions for the values, and incorporating learnable parameters and mode truncation method, the aforementioned operation is equivalent to the fourier neural operator.

By employing linear integral attention to approximate the Eq.~\ref{eq-tno-g}, we can derive the following equation
\begin{equation}
	\quad\left(\mathbf{z}^j\right)_i=\sum_{l=1}^d \frac{\left(\mathbf{k}^l \cdot \mathbf{v}^j\right)}{n}\left(\mathbf{q}^l\right)_i \approx \sum_{l=1}^d\left(\int_{\Omega}\left(k^l(\xi) v^j(\xi)\right) d\xi\right) q_l\left(x_i\right).
	\label{eq-tno-h}
\end{equation}

By adopting this linear attention mechanism, the computational complexity can be reduced to $\mathcal{O}(N_sN_d^2)$, leading to a significant reduction in computational workload when dealing with millions of grid nodes\cite{peng2023linear}.

The Relative Positional Encoding(RoPE) is a relative position encoding method with promising theoretical properties. For a given three-dimensional spatial coordinate $\mathbf{x}$, it can be represented as $ \mathbf{x}_i = (\alpha_i, \beta_i, \gamma_i), 0<=i<=N_s $. Correspondingly, $\mathbf{q}_i$ can be split along the feature dimension into three sub-matrices $[q^\alpha_i, q^\beta_i, q^\gamma_i]$, where $q^\alpha_i$ represents the q-matrix for the first coordinate, with a length of $[N_s, N_d/3]$. For one of the coordinates, taking $\alpha_i$ as an example, the RoPE can be expressed as\cite{su2024roformer}
\begin{equation}
\begin{gathered}
	\psi\left(\mathbf{q}_i, \mathbf{x}_i\right)=\boldsymbol{\Theta}\left(\mathbf{x}_i\right) \mathbf{q}_i, \\
	\boldsymbol{\Theta}\left(\mathbf{x}_i\right)=\operatorname{Diag}\left(\boldsymbol{\Theta}\left(\alpha_i\right), \boldsymbol{\Theta}\left(\beta_i\right), \boldsymbol{\Theta}\left(\gamma_i\right)\right), \\
	\boldsymbol{\Theta}\left(\alpha_i\right)=\operatorname{Diag}\left(\mathbf{R}_1^{\alpha_i}, \mathbf{R}_2^{\alpha_i}, \ldots, \mathbf{R}_{d / 6}^{\alpha_i}\right), \\
	\mathbf{R}_l^{\alpha_i}=\left[\begin{array}{cc}
		\cos \left(\lambda \alpha_i \theta_l\right) & -\sin \left(\lambda \alpha_i \theta_l\right) \\
		\sin \left(\lambda \alpha_i \theta_l\right) & \cos \left(\lambda \alpha_i \theta_l\right)
	\end{array}\right].
\end{gathered}
	\label{eq-tno-6}
\end{equation}
Here, $\operatorname{Diag(\cdot)}$ denotes stacking sub-matrices along the diagonal, $\lambda$ is wavelength of the spatial domain(e.g. $\lambda$ = 128 for a spatial dimension discretized by 128 equi-distant points in one dimension). $\theta_l=10000^{-2(l-1) / d}, l \in\{1,2, \ldots, d / 2\}$.

It can be proven that the RoPE defined for the 3D sequence satisfies the following properties
\begin{equation}
	\psi(\mathbf{q}_i, \mathbf{x}_i) \cdot \psi(\mathbf{k}_j, \mathbf{x}_j) = \left(\mathbf{\Theta}\left(x_i\right) \mathbf{q}_i\right) \cdot\left(\mathbf{\Theta}\left(x_j\right) \mathbf{k}_j\right) = \mathbf{\Theta}(\mathbf{x}_i-\mathbf{x}_j)\left(\mathbf{q}_i \cdot \mathbf{k}_j\right).
	\label{eq-tno-7}
\end{equation}
Here, ``$\cdot$'' represents the dot product of vectors.

%--------------------------------------------FFN
The Feed Forward Neural network (FFN) in the spectral regressor consists of two multilayer perceptron (MLP) layers and is mainly used to adjust the number of feature channels. The FFN is defined as
\begin{equation}
	\operatorname{FFN}(a):=W_2\sigma(W_1 a+b_1),
	\label{eq-tno-2}
\end{equation}
where, $W_1$, $b_1$, and $W_2$ are the learnable parameters of fully connected networks. $\sigma$ is nonlinear activation functions.

%--------------------------------------------Spectral Regressor
For the spectral regressor module, the regressor can be expressed as a function of the given input function $v(x)$. it can be defined as
\begin{equation}
	\operatorname{FNO}(v):=\sigma\left(\mathcal{F}^{-1}\left(W(k) * \mathcal{F}\left(v\right)(k)\right)+W^{\prime} v+b\right)
	\label{eq-tno-1}
\end{equation}
Here, $*$ denotes convolution operation, $\mathcal{F}$ and $\mathcal{F}^{-1}$ represents the fast Fourier transform and its inverse, respectively. The weight matrix function $W(k)$, which is applied in the Fourier space, is a learnable function that depends on the wavenumber $k$. The weight matrix $W'$ is a learnable matrix applied in the physical space. $b$ and $\sigma$ denote learnable bias terms and nonlinear activation functions, respectively. The spectral regressor module can be seen as an extension of the Fourier layer in the FNO model introduced in Chapter \ref{sec:3}, combined with an additional FFN module. For detailed information on the Fourier layer, please refer to Chapter \ref{sec:3}.

%-----------------------------------------------------------
%网络预测部分
When using the network for prediction, the input of the velocity field at the previous $N_{T_I}$ time steps to predict the velocity field at the next time step can be represented as
\begin{equation}
	\hat{\mathbf{u}}^{N_{T_I}+1} = \mathcal{G} ( \mathbf{x}, [\mathbf{u}^{1}, \mathbf{u}^{2}, ... , \mathbf{u}^{N_{T_I}-1} , \mathbf{u}^{N_{T_I}}]).
	\label{eq-tno-b}
\end{equation}

When making predictions for the second time step using the same approach, it can be represented as
\begin{equation}
	\hat{\mathbf{u}}^{N_{T_I}+2} = \mathcal{G} ( \mathbf{x}, [\mathbf{u}^{2}, \mathbf{u}^{3}, ... , \mathbf{u}^{N_{T_I}} , \hat{\mathbf{u}}^{N_{T_I}+1}]).
	\label{eq-tno-c}
\end{equation}
Here, we concatenate the results obtained from the first prediction $\hat{\mathbf{u}}^{N_{T_I}+1}$ with the input tensor for the second prediction, while considering only the previous $N_{T_I}$ steps that are relevant to the target prediction time step. This recursive prediction approach can be employed for long-term forecasting of physical fields. When the prediction time is greater than $N_{T_I}$, all input tensors have transformed into the predicted tensors of the model.

%------------------------------------------------损失函数计算
The general loss function for data-driven methods of velocity fields can be expressed as
\begin{equation}
	\|\hat{\mathbf{u}}-\mathbf{u}\|_{\boldsymbol{\Omega}, \mathbf{u}}=\int_{\mathrm{u}} \int_{\boldsymbol{\Omega}} \sum_{m=1}^{N_w}\left\|\hat{\mathbf{u}}^m(\mathbf{x}, t)-\mathbf{u}^m(\mathbf{x}, t)\right\| d \boldsymbol{\Omega} d \mathbf{T}.
	\label{eq-tno-8}
\end{equation}
Here, $\hat{\mathbf{u}}$ is the predicted velocity field, $\mathbf{u}$ is ground truth velocity field. $\boldsymbol{\Omega}$ and $\mathbf{T}$ are the spatial domain and temporal domain, respectively.

Given the iterative computation of velocity fields at different time steps in a recurrent manner, the temporal dimension can be discretized into $N_t$ time steps. Consequently, the loss function can be discretized as 
\begin{equation}
	\int_{\mathrm{T}} \int_{\boldsymbol{\Omega}} \sum_{m=1}^{N_w}\left\|\hat{\mathbf{u}}^m(\mathbf{x}, t)-\mathbf{u}^m(\mathbf{x}, t)\right\| d \boldsymbol{\Omega} d \mathbf{T} \approx \frac{1}{N_t} \sum_{i=1}^{N_p} \| \hat{ \mathbf{u}}^{im} (\mathbf{x})  -  \mathbf{u}^{im} (\mathbf{x}) \| _ {\boldsymbol{\Omega}}.
	\label{eq-tno-9}
\end{equation}
Here, $\hat{\mathbf{u}}^i$ is the predicted velocity field, $\mathbf{u}^i$ denotes ground truth, and $N_p$ represents the number of time steps that need to be iterated.

Considering the uniform spatial distribution of the points where the velocity field is evaluated, it can be further discretized into a high-dimensional tensor representation using the L2 norm.
\begin{equation}
	\| \hat{ \mathbf{u} }^i (\mathbf{x}) -  \mathbf{u}^i (\mathbf{x}) \| _ {\boldsymbol{\Omega}} \approx \frac{1}{N_s} \sum_{j,k,l=1}^{N_x,N_y,N_z} \left( \hat{ \mathbf{u} }^{im}_{jkl} -  \mathbf{u}^{im}_{jkl} \right) ^ 2.
	\label{eq-tno-10}
\end{equation}
Therefore, the final loss function can be rewritten as
\begin{equation}
	\mathcal{L(\hat{\mathbf{u}}, \mathbf{u})} = \frac{1}{N_p} \sum_{i,m=1}^{N_p,N_w} \frac{1}{N_s} \sum_{j,k,l=1}^{N_x,N_y,N_z} \left( \hat{ \mathbf{u} }^{im}_{jkl} -  \mathbf{u}^{im}_{jkl} \right) ^ 2.
	\label{eq-tno-11}
\end{equation}

%-------------------------------------------------------------------------------------------------------------
\section{\label{sec:5}Numerical experiments}
In this chapter, we will systematically analyze and compare the performance between the FNO model, commonly used traditional sub-grid models such as the dynamic Smagorinsky model (DSM) and dynamic mixed model (DMM), and our proposed TNO model. The numerical experiments will be conducted in two different flow types: statistically steady forced homogeneous isotropic turbulence (HIT) and unsteady free-shear turbulence. By showcasing the data generation and training processes, comparing a \textit{posterior} results, and evaluating computational efficiency, we aim to compare the performance of different models comprehensively.

In different flow cases, all models underwent ten independent repeated numerical experiments, each evolving from a different random initial field. Then, the \textit{a posterior} statistical results from each model's ten runs were averaged separately to reduce the influence of experimental randomness and facilitate a more meaningful comparison.

%--------------------------------------------------------------------------------------------------
\subsection{\label{sec:5.1} Performance on forced homogeneous isotropic turbulence}

\subsubsection{\label{sec:5.1.1} Dataset description}
The direct numerical simulation (DNS) of forced homogeneous isotropic turbulence is performed in a $(2\pi)^3$ cubic box using $256^3$ uniform grid, with periodic boundary conditions and a second-order two-step Adams–Bashforth time scheme\cite{wyp2022,yuan2020deconvolutional}. The pseudo-spectral method is adopted, and the aliasing error caused by nonlinear advection terms is eliminated by the two-thirds rule\cite{hussaini1987spectral}. The kinematic viscosity is $\nu=0.00625$, and the Taylor Reynolds number is $Re_\lambda\approx100$\cite{li2022fourier}. To ensure the turbulence remains in a statistically steady state, we apply large-scale forcing and save the data after an extended period\cite{yuan2020deconvolutional}(more than $10\tau$, here $\tau={L_I}/{u_{\text{rms}}}\approx1.0$ is large-eddy turnover times). 

In the present study, the DNS data is filtered into large-scale flow fields at grid resolutions of $64^3$ by the sharp spectral filter $\hat{G}(\boldsymbol{k})=$ $H\left(k_c-|\boldsymbol{k}|\right)$ in Fourier space for homogeneous isotropic turbulence\cite{pope2000}. Here, the cutoff wavenumber $k_c=\pi / \Delta = 21$, and $\Delta$ denotes the filter width. The Heaviside step function $H(x)=1$ if $x \geq 0$; otherwise $H(x)=0$\cite{pope2000,chang2022}. 

\begin{figure*}
	\includegraphics[width=1\linewidth]{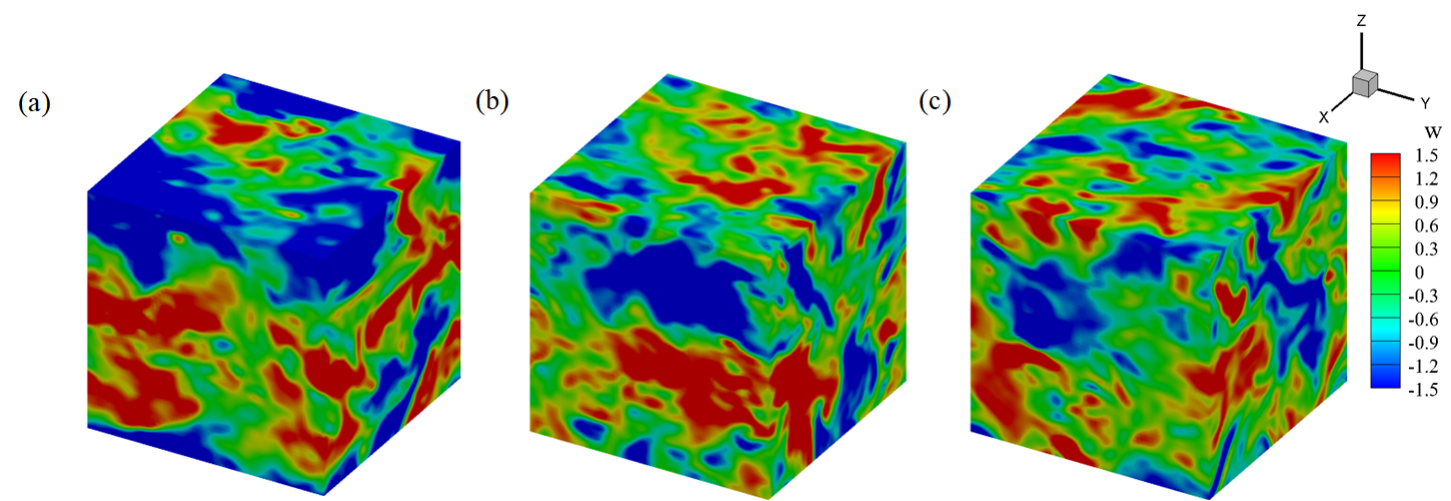}
	\caption{Velocity contours of different initial fields}
	\label{initial}
\end{figure*}

The DNS data is recorded with a time step of 0.001, and snapshots of the numerical solution are captured at intervals of 200 steps as a time node. Therefore, each prediction step performed by the neural operator is equivalent to 200 steps of DNS computation, corresponding to $0.2\tau$. We employ 100 different random fields as initial conditions, and for each set of computations, we save 600 time nodes. Therefore, the DNS data with the tensor size of $[100\times 600\times 256\times 256\times 256\times 3]$ can be obtained, and the filtered direct numerical simulation (fDNS) data with the tensor size of $[100\times 600\times 64\times 64\times 64\times 3]$ will be served as a training and testing dataset\cite{li2022fourier}. Specifically, the dataset we use to train the neural operator model consists of 100 groups, each group has 600 time nodes, and each time node denotes a filtered velocity field of $64^3$ with three directions. For these 100 groups of HIT DNS data, each group would  take approximately 4.3 hours to parallel computing on a 32-core CPU. Figure.~\ref{initial} demonstrates the velocity contours of three random initial fields, so that the differences between different initial fields can be seen more intuitively.

We employ a consistent training approach for both FNO and TNO models. We use the data from the previous five time steps as inputs to the neural operator models, aiming to learn the mapping between these inputs and the corresponding data at the sixth time step\cite{li2022fourier}. Denotes the $i$-th time-node filtered velocity field as $U_i$. Therefore, the input tensor is $[U_i, U_{i+1}, U_{i+2}, U_{i+3}, U_{i+4}]$, and the output tensor is $[U_{i+5}]$, which is taken as input-output pairs\cite{li2022fourier}. By utilizing 100 different initial conditions and capturing 600 time nodes in each group. These samples are split into 80\% for training and 20\% for testing\cite{li2022fourier}. Once the model training is completed, it will be able to make fast predictions for the evolution of the filtered velocity field by directly mapping the relationship between inputs and outputs.

\begin{figure*}
	\includegraphics[width=0.8\linewidth]{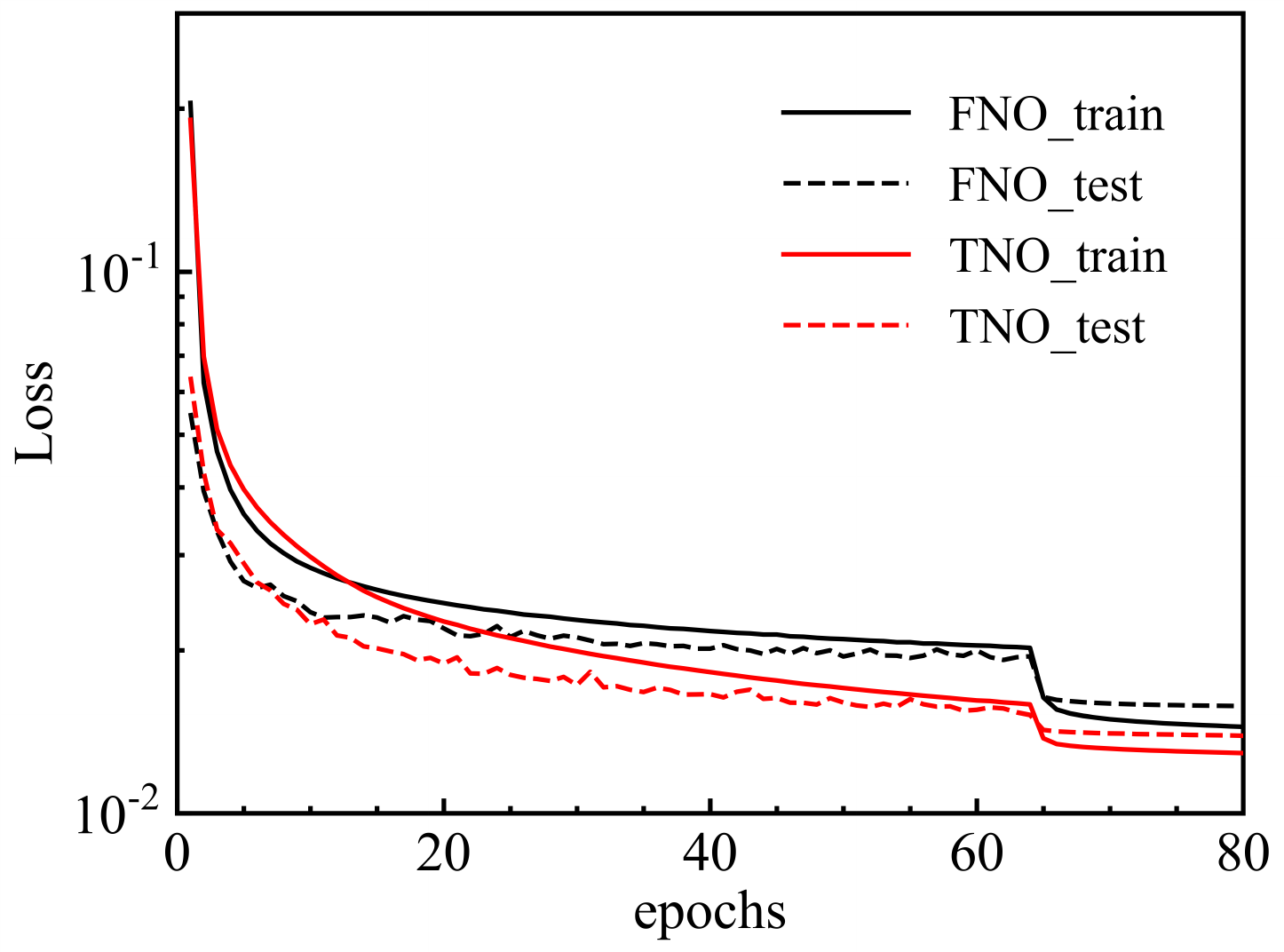}
	\caption{The learning curves of the FNO and TNO models for LES of 3D HIT.}
	\label{HIT_learning}
\end{figure*}
To ensure a fair comparison, the hyperparameters are set to be identical. The learning curves of the FNO and TNO models for 3D HIT is shown in Fig.~\ref{HIT_learning}. The FNO model and TNO model take 14.6 hours and 11.2 hours to train the network, respectively. Both FNO and TNO models employed in this study utilize the same number of Fourier modes, specifically 16, and share a common initial learning rate of $10^{-3}$\cite{peng2023linear}. The channel width of $P$ is configured as 64, and the channel width of $Q$ is 128. The AdamW optimizer is chosen as optimization with $\beta =(0.9,0.999)$\cite{zhuang2022understanding,loshchilov2018fixing}. The GELU function is used for the activation function\cite{hendrycks2016gaussian}.
% The loss function is defined as
%\begin{equation}
%	\textit{Loss}=\frac{\|u^*-u\|_2}{\|u\|_2}, \text { where }\|\mathbf{A}\|_2=\frac{1}{n} \sqrt{\sum_{k=1}^n\left|\mathbf{A}_{\mathbf{k}}\right|^2}.
%	\label{eqloss}
%\end{equation}
%Here, $u^*$ denotes the prediction of velocity fields and $u$ is the ground truth.

To avoid overfitting in the models, we generated and used an additional ten independent groups of data from different initial fields for the \textit{a posteriori} evaluation. In the \textit{a posteriori} study, fDNS data is utilized as a benchmark to evaluate various models. Moreover, all the neural operator models and classical SGS models used for comparison are initialized with the same initial field. More specifically, while the DSM and DMM models utilize the flow field data of the fifth step as the initial conditions, the TNO and FNO models employ the first five steps of the flow field data of the new initial field for the inference process.

%--------------------------------------------------------
\subsubsection{\label{sec:5.1.2}A \textit{posteriori} study}
\begin{figure*}
	\includegraphics[width=1\linewidth]{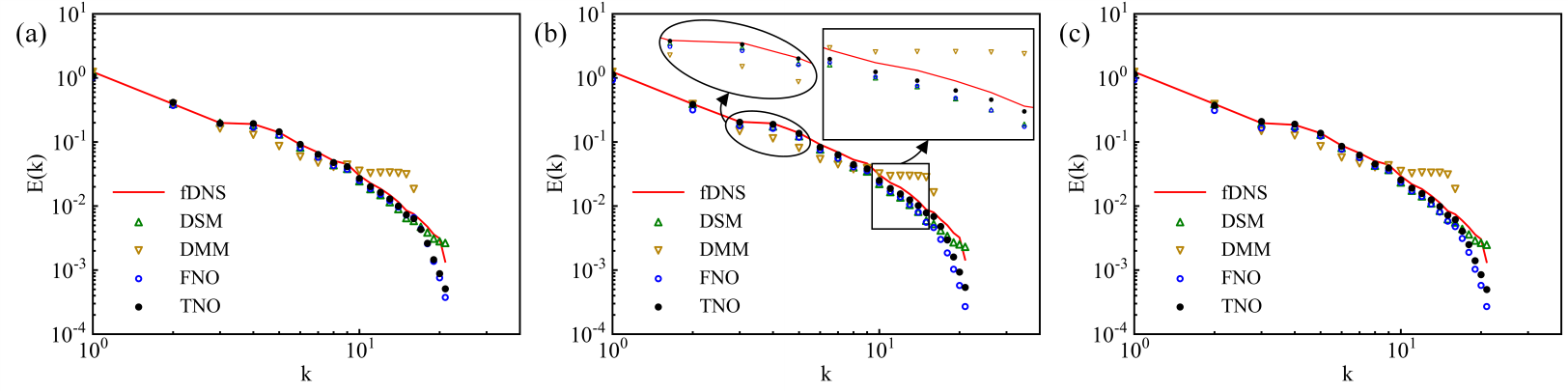}
	\caption{Comparisons of velocity spectrum for LES in the forced HIT at different time instants: (a)$t/\tau\approx6.0$; (b)$t/\tau\approx18$; (c)$t/\tau\approx40$.}
	\label{HIT_spectrum}
\end{figure*}

The velocity spectrum predicted by various models at different time instants is shown in Fig.~\ref{HIT_spectrum}. It can be seen that the velocity spectrum, as predicted by both classical LES models and data-driven models, demonstrates persistent consistency with fDNS result in both short-term and long-term predictions. Here, the large-eddy turnover time $\tau$ is provided in Eq.~(\ref{eq4}). As observed in Fig.~\ref{HIT_spectrum}, the DMM model underestimates the velocity spectrum compared to the fDNS result at low wavenumbers, and significantly overestimates the spectrum at high wavenumbers. The figure also reveals that the TNO and FNO models exhibit similar overall trends in predicting the velocity spectrum. However, the TNO model demonstrates a slightly more accuracy than the FNO model at both high and low wavenumbers regions, giving a closer alignment with the fDNS. Furthermore, it can be seen from Fig.~\ref{HIT_spectrum}(b) that the velocity spectrum predicted by the TNO model are more accurate than those of the DSM model within the wavenumber $ 3\leq k \leq17$, while the DSM model demonstrates greater accuracy when the wavenumber $k \geq 18$. Therefore, both the TNO and DSM models show similar performance in predicting the velocity spectrum in the HIT case, achieving good agreement with the fDNS.

\begin{figure*}
	\includegraphics[width=1\linewidth]{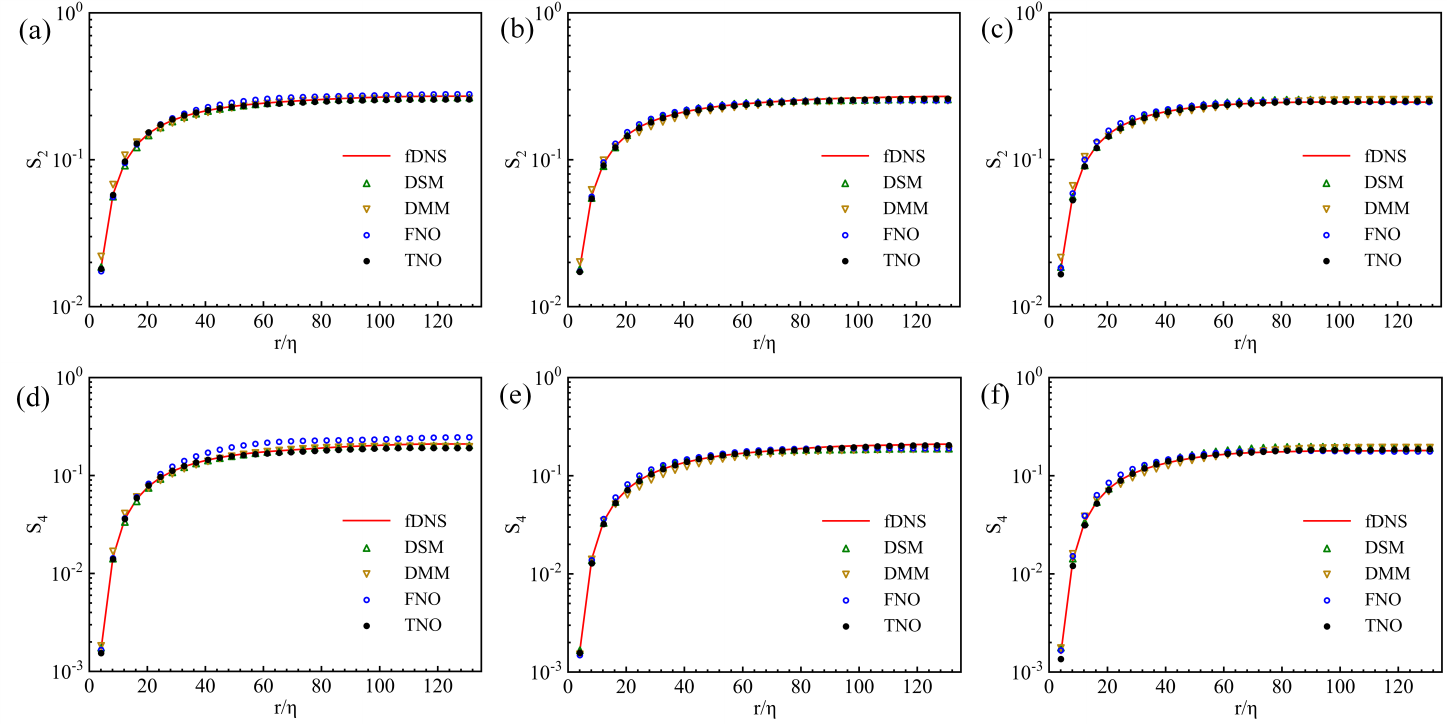}
	\caption{Structure functions of the velocity for LES in the forced HIT at different time instants: (a)$S_2$, $t/\tau\approx6.0$; (b)$S_2$, $t/\tau\approx18$; (c)$S_2$, $t/\tau\approx40$; (d)$S_4$, $t/\tau\approx6.0$; (e)$S_4$, $t/\tau\approx18$; (f)$S_4$, $t/\tau\approx40$.}
	\label{HIT_structure}
\end{figure*}
In order to further compare the performance of different models in predicting multi-scale properties of turbulence, we calculate the longitudinal structure functions of the filtered velocity. These structure functions are defined by\cite{xie2018modified,xie2020approximate}
\begin{equation}
	S_n(r)=\left\langle\left|\frac{\delta_r \bar{u}}{\bar{u}_{\text{rms}}}\right|^n\right\rangle,
	\label{eq20}
\end{equation}
where $n$ is the order of structure function, $\bar{u}_{\text{rms}}=\sqrt{\left\langle \bar{u}_i \bar{u}_i\right\rangle}$, and $\delta_r \bar{u}=[\overline{\mathbf{u}}(\mathbf{x}+\mathbf{r})-\overline{\mathbf{u}}(\mathbf{x})] \cdot \hat{\mathbf{r}}$ represents the longitudinal increment of the velocity at the separation $\mathbf{r}$. Here, $\hat{\mathbf{r}}=\mathbf{r} /|\mathbf{r}|$ denotes the unit distance vector. 

Fig.~\ref{HIT_structure} compares the structure functions of the filtered velocity for different models with fDNS data at different time instants. Here, the Kolmogorov scale are used to normalize the distance. It can be seen that the DMM model exhibits an overestimation of the structure functions at short distances in comparison to those of the fDNS data. As can be observed from Fig.~\ref{HIT_structure}(a) and (d), the FNO model exhibits an increasing deviation in predicting the structure functions at $t/\tau\approx6.0$ as the distance $r/\eta$ increases, with the predicted results overestimating those obtained from fDNS. In contrast, the TNO and DSM models can always accurately predict the structure functions at both short-term and long-term prediction.

\begin{figure*}
	\includegraphics[width=1\linewidth]{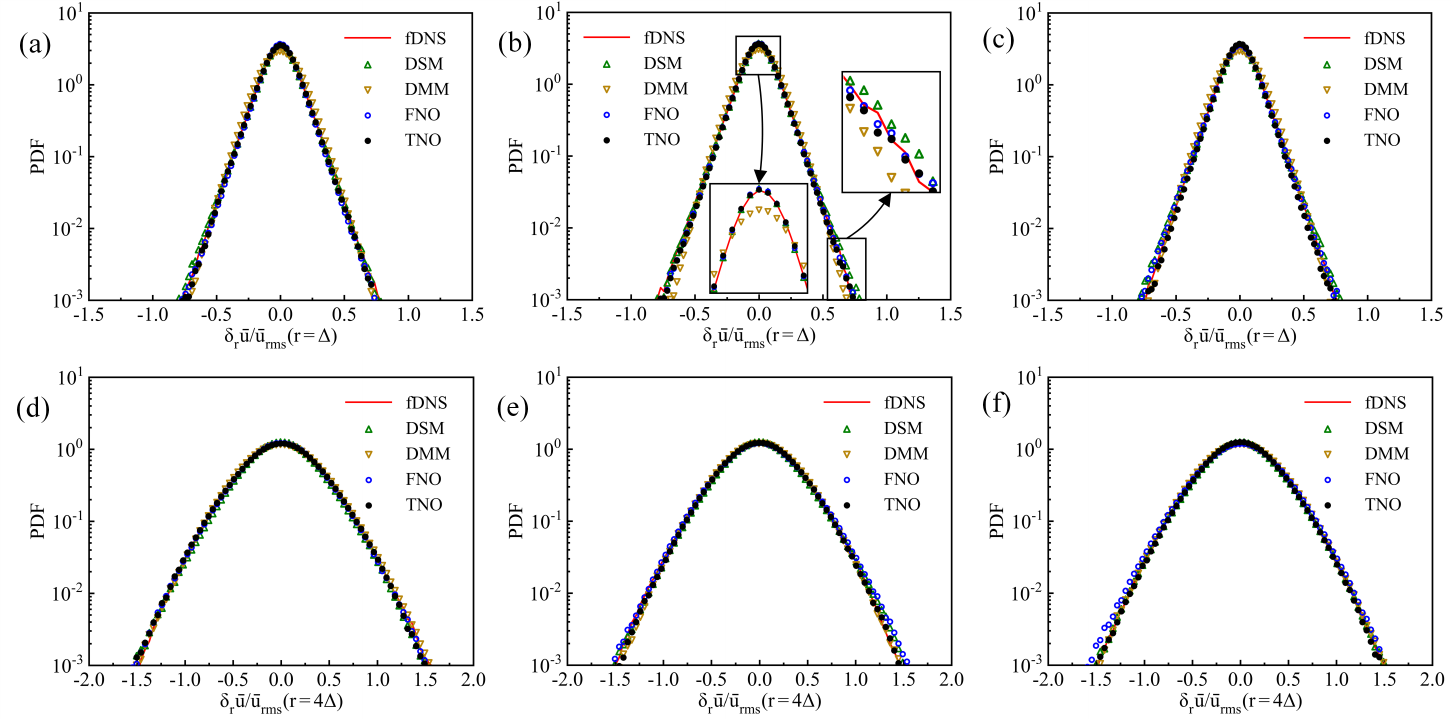}
	\caption{The PDFs of the normalized velocity increments for LES in the forced HIT at different time instants: (a)$r/\Delta=1$, $t/\tau\approx6.0$; (b)$r/\Delta=1$, $t/\tau\approx18$; (c)$r/\Delta=1$, $t/\tau\approx40$; (d)$r/\Delta=4$, $t/\tau\approx6.0$; (e)$r/\Delta=4$, $t/\tau\approx18$; (f)$r/\Delta=4$, $t/\tau\approx40$.}
	\label{HIT_inc14}
\end{figure*}
Furthermore, we compare PDFs of the normalized velocity increments $\delta_r \bar{u}/\bar{u}^{rms}$ with distance $r=\Delta$ and $r=4\Delta$ at different time instants in Fig.~\ref{HIT_inc14}. It can be seen that the PDFs of the normalized velocity increments predicted by all models show a good agreement with the fDNS data. From Fig.~\ref{HIT_inc14}(b), it is seen that the predictions of the velocity increments from the TNO model and DSM model are very close to fDNS in the region near zero velocity increment. However, the predictions from the DSM model and the DMM model deviate after $\delta_r \bar{u}/\bar{u}_{\text{rms}}=0.5$, while the TNO model maintains a good consistency with fDNS. This further demonstrates the TNO model's superior accuracy in comparison to the other models.

\begin{figure*}
	\includegraphics[width=1\linewidth]{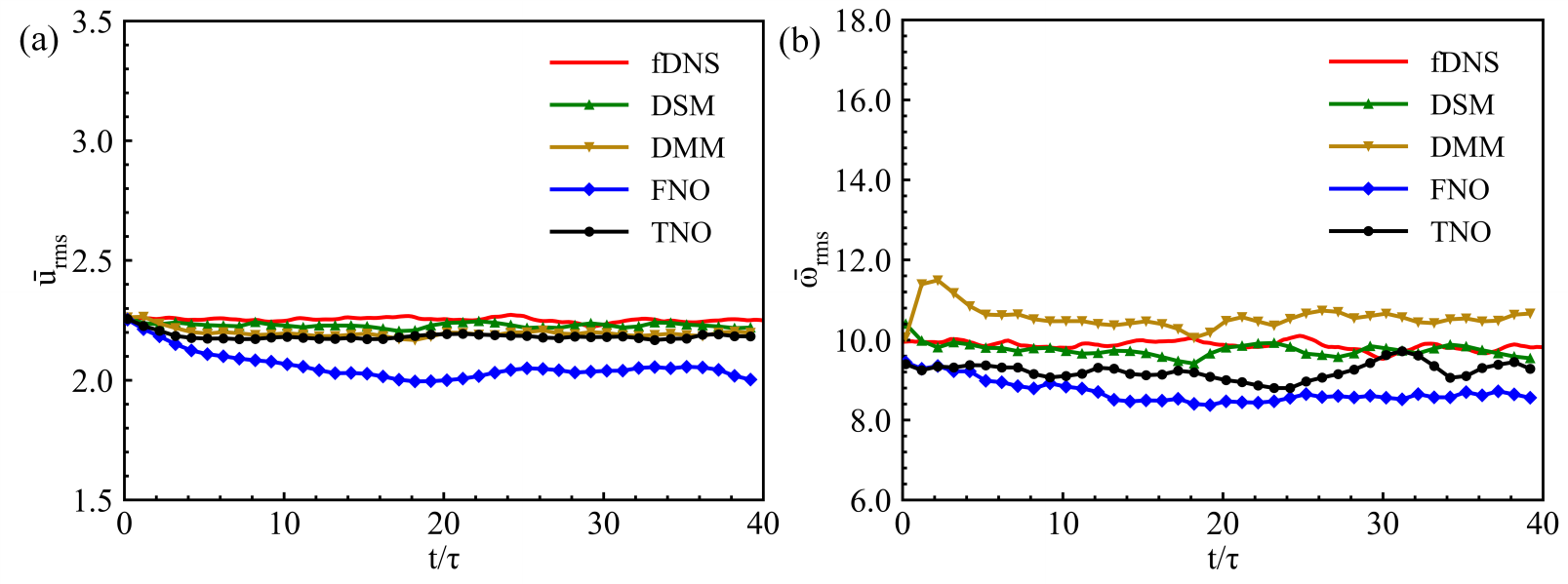}
	\caption{Temporal evolutions of the velocity rms value and vorticity rms value for LES in the forced HIT.}
	\label{HIT_uwRMS}
\end{figure*}
In order to demonstrate the stability of the different models, we depict the temporal evolution of the root mean square (rms) values of velocity and vorticity in Fig.~\ref{HIT_uwRMS}. It can be observed that both DSM and DMM models, as well as the enhanced FNO and TNO models, exhibit relatively stable predictive performance. The stability of classical LES models can be attributed to their inherent dissipative characteristics\cite{smagorinsky1963,kleissl2006numerical}. The improved data-driven models, FNO and TNO models, can provide stable predictions for simple flow cases such as HIT by effectively learning from vast of data. It can be seen that the rms values of velocity and vorticity predicted by DSM model are closer to those of fDNS when comparing with other models. This is also due to the fact that the DSM model applies large-scale forces identical to fDNS at the two largest wavenumbers during numerical simulation. Among the two data-driven models, the TNO model exhibits a more precise performance than the FNO model.

\begin{figure*}
	\includegraphics[width=1\linewidth]{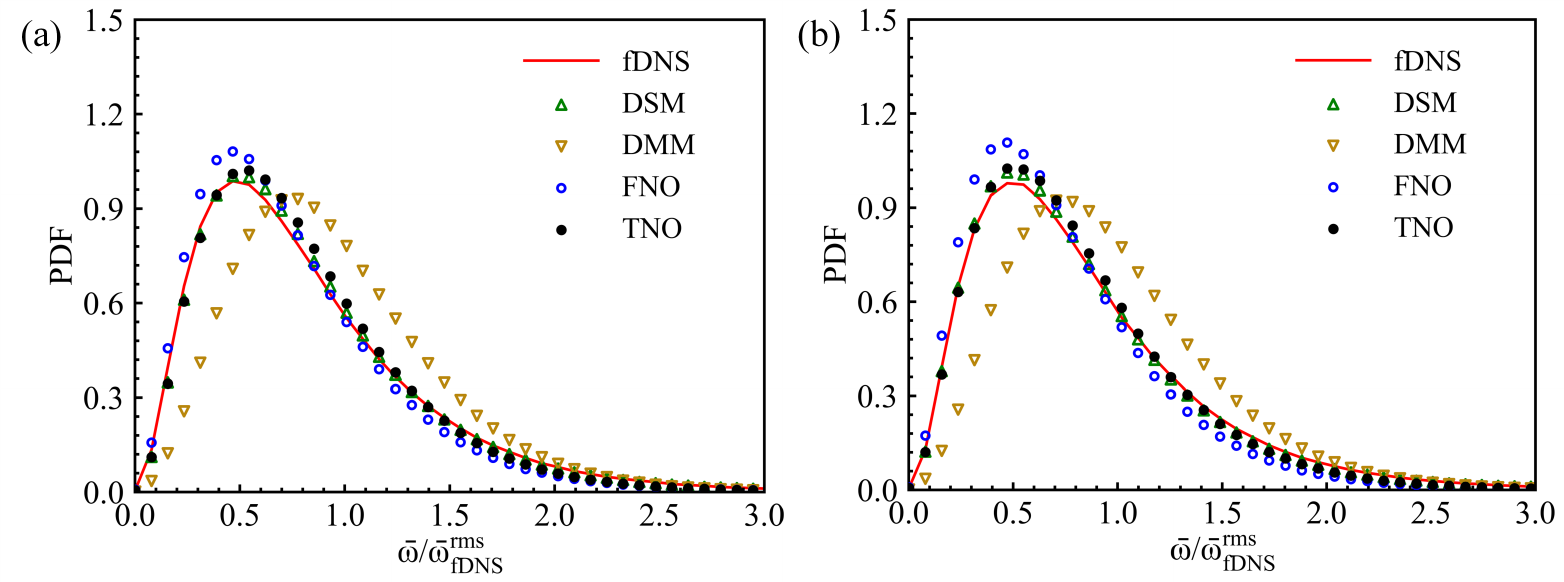}
	\caption{The PDFs of the normalized vorticity for LES in the forced HIT at different time instants: (a)$t/\tau\approx6.0$; (b)$t/\tau\approx40$.}
	\label{HIT_vort}
\end{figure*}
Furthermore, the PDFs of the normalized vorticity magnitude at different time instants are shown in Fig.~\ref{HIT_vort}. Here, the vorticity is normalized by the root-mean-square values of the vorticity obtained from the fDNS data. It is observed that the PDFs predicted by DSM and TNO models have a good agreement with those of fDNS in both short-term and long-term prediction. However, both FNO and DMM models exhibit a significant deviation from fDNS.

\begin{figure*}
	\includegraphics[width=1\linewidth]{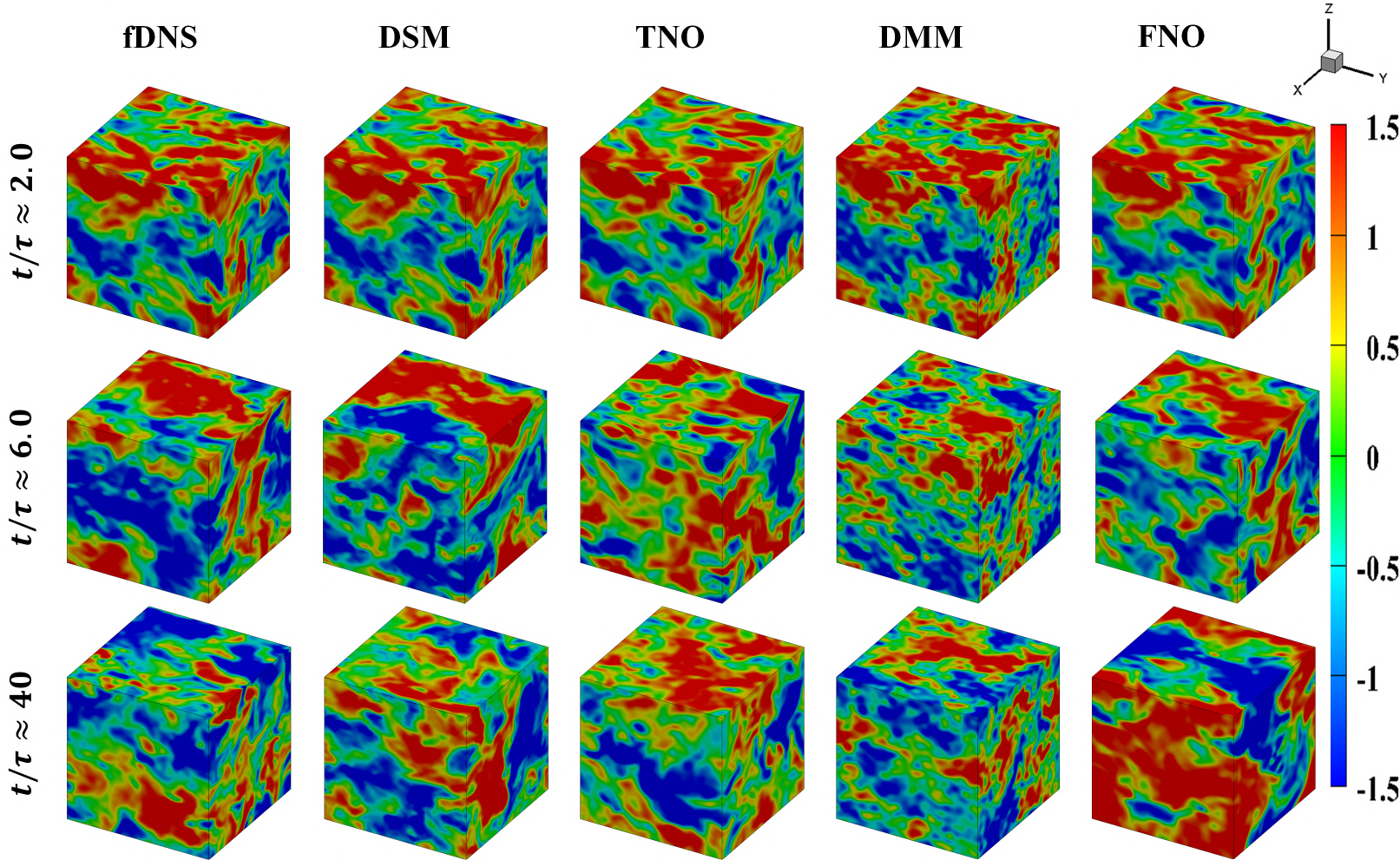}
	\caption{Contours of the predicted velocity field in the z-direction for LES in the forced HIT at different time instants.}
	\label{HIT_vel3D}
\end{figure*}
We demonstrate contours of the predicted velocity fields in the z-direction for LES in the forced HIT at several time instants in Fig.~\ref{HIT_vel3D}. Here, due to the characteristics of homogeneous isotropic turbulence, we have arbitrarily chosen to present the velocity in the z-direction. Additionally, we randomly selected one case from the 10 different cases with different initial fields to generate the contours. It is observed that the velocity fields predicted by TNO and DSM models exhibit a closer tendency to the fDNS results within a short-term forecasting $t/\tau\approx2.0$. It becomes challenging to maintain complete similarity in the flow field after a long-term evolution $t/\tau\approx40$. Therefore, for long-term predictions, our primary focus is on observing the overall statistical trends of the flow field. We can observe that the LES with DSM and TNO models exhibit a closer tendency to the fDNS results.

%-----------------------------------------------
\subsubsection{Computational efficiency}
\begin{table*}
	\caption{\label{CE}Computational efficiency of different approaches on forced HIT.}
	\begin{ruledtabular}
		\begin{tabular}{cccc}
			Method& Total parameters($\times10^6$)& GPU$\cdot$s& CPU$\cdot$s\\
			\hline
			DSM& N/A & N/A & 105.6 \\
			DMM& N/A & N/A & 166.4 \\
			FNO& 563.9 & 0.206 & 4.366 \\
			TNO& 268.6 & 1.021 & 12.48 \\ 
		\end{tabular}
	\end{ruledtabular}
\end{table*}
Here, we compare the computational efficiency among different models in Table.~\ref{CE} The recorded time represents the time required by different models to predict each time node in the forced HIT. We conduct the neural network models using Pytorch and train the models on Nvidia Tesla V100 GPU, where the CPU type is Intel(R) Xeon(R) Gold 6240 CPU @2.60GHz. The DSM simulations are implemented on a computing cluster, where the type of CPU is Intel Xeon Gold 6148 with 16 cores each @2.40 GHz. It can be observed that the TNO model exhibits a prediction speed that is three times slower than the FNO model. However, it achieves a notable reduction in the number of network parameters when compared to FNO. It can be seen that the TNO model requires only 12.48s of CPU time to predict a single time step, while the LES with DSM and DMM models require 105.6s and 166.4s, respectively. Furthermore, if the TNO model is run on a GPU, the time can be further reduced to 1s. Compared to LES with DSM and DMM models, the trained TNO model demonstrates a significant advantage in computational efficiency. Although it is important to consider the longer training time required for the TNO model, its ability to achieve faster flow field predictions is a notable advantage.

%------------------------------------

\subsubsection{Generalization on higher Taylor Reynolds numbers}
Moreover, the generalization ability of the 3D TNO model is tested. The well-trained TNO model can be directly used to predict the fDNS data at higher Taylor Reynolds number cases without additional training or modifications. In this study, we generated four sets of forced HIT data with different initial fields at higher Taylor Reynolds numbers $Re_\lambda \approx 160$ and $Re_\lambda \approx 250$, respectively, to assess the model's generalization performance during the inference process. By employing the same prediction methodology as in the forced HIT with low Taylor Reynolds number, we input the first five time steps of $Re_\lambda \approx 160$ forced HIT velocity field data into the well-trained data-driven models to forecast the further evolution of the flow field. Here, we perform individual predictions for four sets of data with distinct initial conditions and subsequently average the statistical quantities in the \textit{a posterior} analysis.

\begin{figure*}
	\includegraphics[width=1\linewidth]{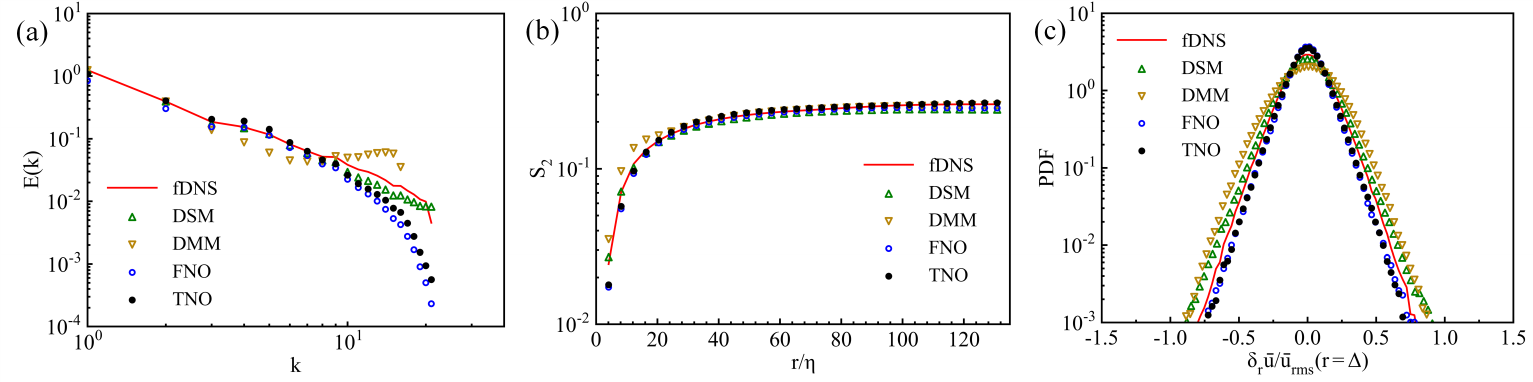}
	\caption{Statistical analysis of prediction when Taylor Reynolds number $Re_\lambda \approx 160$: (a)Velocity spectrum; (b)Second-order structure functions; (c)PDFs of the normalized velocity increments.}
	\label{HIT_Re400}
\end{figure*}
Figure.~\ref{HIT_Re400} presents the statistical quantities, including the velocity spectrum, second-order structure functions, and PDFs of the normalized velocity increments, predicted by different models in forced HIT with Taylor Reynolds numbers $Re_\lambda \approx 160$ at $t/\tau_m$ $\approx$ 28.6. Here, the large-eddy turnover time is $\tau_m \approx 0.7$. It can be seen from Fig.~\ref{HIT_Re400}(a) that the velocity spectrum predicted by the DSM model demonstrates higher accuracy and closer agreement with those of fDNS compared to other models. However, Fig.~\ref{HIT_Re400}(b) and (c) reveal that the TNO model exhibits superior accuracy in predicting the second-order structure functions and PDFs of velocity increments compared to LES with DSM and DMM models.

\begin{figure*}
	\includegraphics[width=1\linewidth]{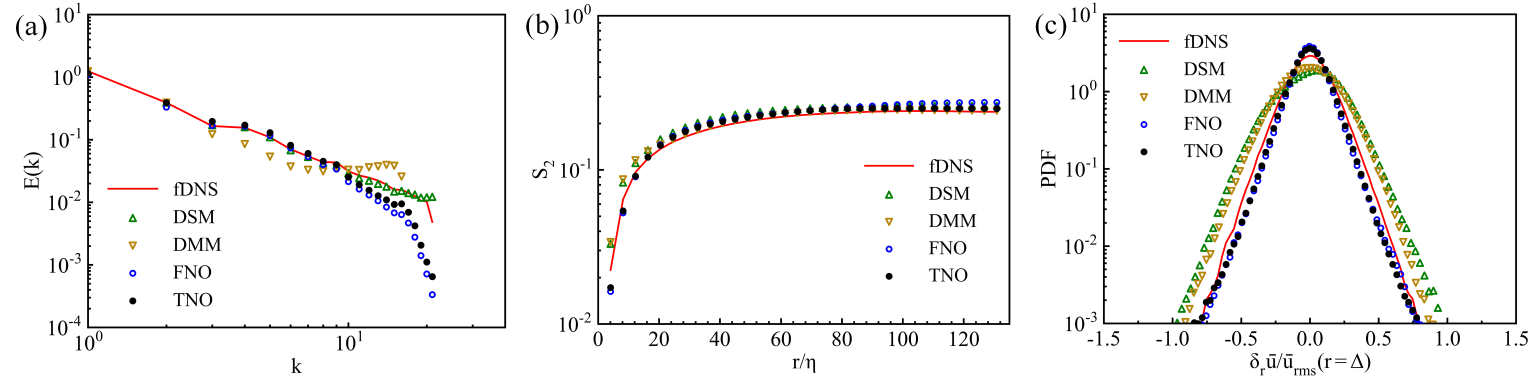}
	\caption{Statistical analysis of prediction when Taylor Reynolds number $Re_\lambda \approx 250$: (a)Velocity spectrum; (b)Second-order structure functions; (c)PDFs of the normalized velocity increments.}
	\label{HIT_Re1000}
\end{figure*}

Moreover, we further test different models at a higher Taylor Reynolds number $Re_\lambda \approx 250$. The average statistical quantities predicted by different models in forced HIT with Taylor Reynolds numbers $Re_\lambda \approx 250$ at $t/\tau_m$ $\approx$ 26.7 are illustrated in Fig.~\ref{HIT_Re1000}. Here, the large-eddy turnover time is $\tau_h \approx 0.6$. It can be observed that the velocity spectrum predicted by the DSM model demonstrates higher accuracy compared to other models in general. However, in the prediction of second-order structure functions and PDFs of velocity increments, the TNO model exhibits superior precision compared to LES with DSM and DMM models.

Furthermore, it can be observed from Fig.~\ref{HIT_Re400}(c) and Fig.~\ref{HIT_Re1000}(c) that as the Taylor Reynolds number increases, the DSM and DMM models exhibit deviations in predicting the PDFs of velocity increments. In contrast, the PDFs of velocity increments predicted by the TNO model demonstrates better consistency with those of the fDNS data. Overall, both the TNO and FNO models exhibit generalization capabilities to higher Taylor Reynolds numbers, producing predictions that are comparable to those of the DSM model and closely consistent with the fDNS data.
	
When dealing with higher Taylor Reynolds number cases, both the DSM and DMM models require higher temporal resolution, resulting in an increased computational time. However, the TNO model can be directly applied to higher Taylor Reynolds number cases and achieve accurate predictions while maintaining the computational efficiency. Since turbulent flow data for high Taylor Reynolds numbers is scarce, it is important to train the TNO model using data with lower Reynolds numbers and apply the well-trained model to predict turbulence with higher Taylor Reynolds numbers. Considering the self-similarity of turbulence, the large-scale statistical features and flow structures in some regions are insensitive to the Reynolds numbers, which provides a theoretical foundation for developing such a model.

%--------------------------------------------------------------------------------------------------
\subsection{\label{sec:5.2}Performance on free-shear turbulence}
Apart from assessing the performance of data-driven models and classical subgrid-scale (SGS) models on a three-dimensional (3D) forced homogeneous isotropic turbulence (HIT), we also evaluate their capabilities in a more challenging simulation task: a 3D turbulent mixing layer with free-shear.

\subsubsection{\label{sec:5.2.1} Dataset description}
\begin{figure}
	\includegraphics[width=1\columnwidth]{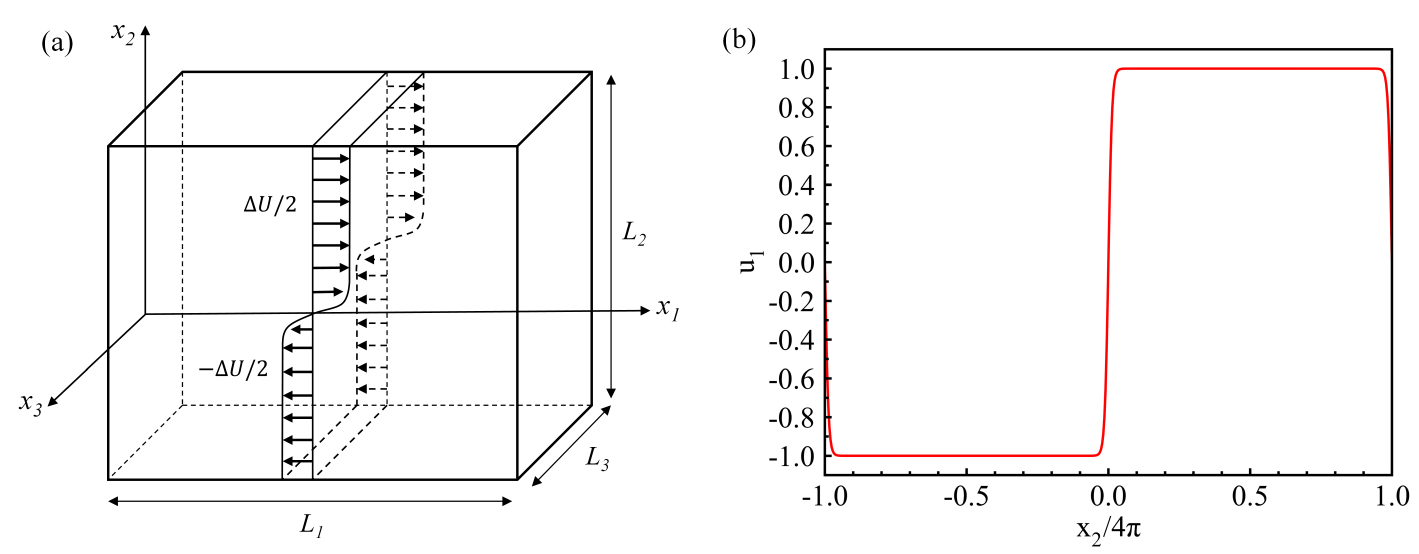}
	\caption{Diagram of the temporally evolving mixing layer with the velocity profile: (a)schematic of the mixing layer;(b)streamwise velocity profile along the normal direction.}
	\label{FS_schem}
\end{figure}

The dynamics of the free-shear turbulent mixing layer are described by the same Navier-Stokes equations (Eqs.~\ref{eq1} and \ref{eq2}) but without a forcing term. Fig.~\ref{FS_schem} illustrates the diagram of the flow configuration for the temporally evolving turbulent mixing layer with the initial hyperbolic tangent streamwise velocity profile. The mixing layer is performed in a cuboid domain with lengths $L_1\times L_2\times L_3=8\pi\times 8\pi\times 4\pi$ using a uniform grid resolution of $N_1\times N_2\times N_3=256\times 256\times 128$ where $x_1 \in\left[-L_1 / 2, L_1 / 2\right]$, $x_2 \in\left[-L_2 / 2, L_2 / 2\right]$ and $x_3 \in\left[-L_3 / 2, L_3 / 2\right]$ denote the streamwise, normal, and spanwise directions, respectively. The initial streamwise velocity is given by\cite{wyp2022,wang2022compressibility,sharan2019turbulent}
\begin{equation}
	 u_1=\frac{\Delta U}{2}\left[\tanh \left(\frac{x_2}{2 \delta_\theta^0}\right)-\tanh \left(\frac{x_2+L_2 / 2}{2 \delta_\theta^0}\right)-\tanh \left(\frac{x_2-L_2 / 2}{2 \delta_\theta^0}\right)\right] + \lambda_1, 
	\label{eq21}
\end{equation}
where, $\delta_\theta^0=0.08$ is the initial momentum thickness and $\Delta U = U_2 - U_1=2$ represents the velocity contrast between two identical and opposing free streams across the shear layer\cite{wyp2022,yuan2023adjoint}. The momentum thickness measures the extent of the turbulent region within the mixing layer, which is given by\cite{sharan2019turbulent,rogers1994direct,yuan2023adjoint}
\begin{equation}
	\delta_\theta=\int_{-L_2 / 4}^{L_2 / 4}\left[\frac{1}{4}-\left(\frac{\left\langle\bar{u}_1\right\rangle}{\Delta U}\right)^2\right] d x_2.
	\label{eqxita}
\end{equation}

The initial normal and spanwise velocities are given as $u_2=\lambda_2$, and $u_3=\lambda_3$, respectively. Here, $\lambda_1, \lambda_2, \lambda_3 \sim$ $\mathcal{N}\left(\mu,\sigma^2\right)$, i.e., $\lambda_1, \lambda_2, \lambda_3$ satisfy the Gaussian random distribution\cite{wang2022compressibility}. The expectation of the distribution is $\mu=0$ and the variance of the distribution is $\sigma^2=0.01$. $\tau_\theta=\delta_\theta^0/\Delta U=20dt$, and $dt=0.002$ represents the time step used in the DNS simulation. In order to minimize the influence of the upper and lower boundaries on the central mixing layer, two numerical diffusion buffer zones are introduced at the vertical edges of the computational domain\cite{wang2022compressibility,yuan2023adjoint,wyp2022}. Periodic boundary conditions are employed in all three directions. The spatial discretization is performed using the pseudo-spectral method with the two-thirds dealiasing rule, while the time-advancing scheme is implemented using an explicit two-step Adam-Bashforth scheme.

The DNS data is subjected to explicit filtering using a widely used Gaussian filter, which is defined by\cite{pope2000,sagaut2006}
\begin{equation}
	G(\mathbf{r} ; \bar{\Delta})=\left(\frac{6}{\pi \bar{\Delta}^2}\right)^{1 / 2} \exp \left(-\frac{6 \mathbf{r}^2}{\bar{\Delta}^2}\right).
	\label{eq22}
\end{equation}
Here, the filter scale $\bar{\Delta}=4h_{DNS}$ is chosen for the free-shear turbulent mixing layer, where $h_{DNS}$ is the grid spacing of DNS. The filter-to-grid ratio FGR=$\bar{\Delta}/h_{LES}$=1 is utilized and the corresponding grid resolution of LES: $64\times 64\times 32$ can be obtained\cite{chang2022,wyp2022}. 

\begin{figure}
	\includegraphics[width=0.8\columnwidth]{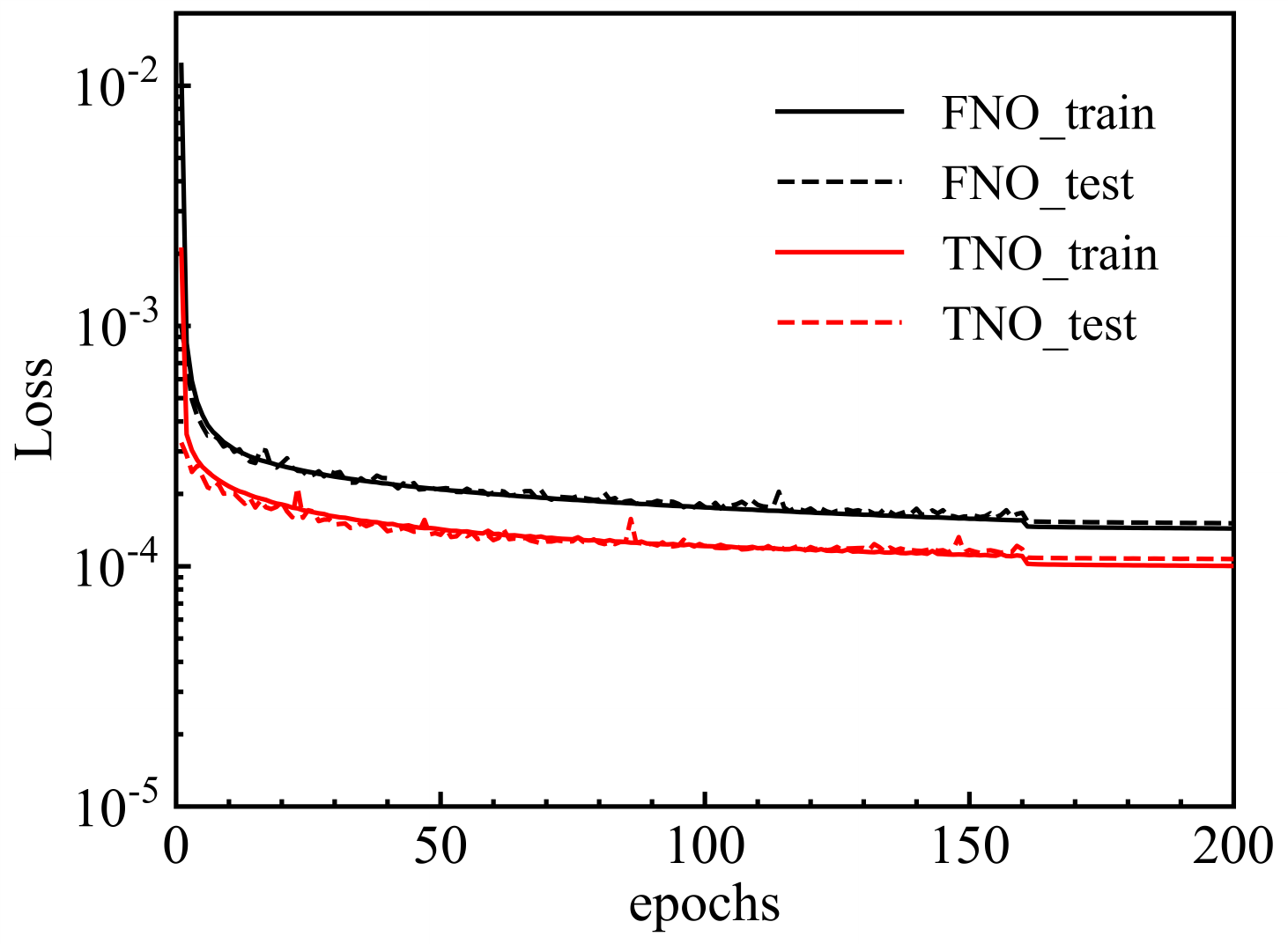}
	\caption{The learning curves of the FNO and TNO models for 3D free-shear turbulent mixing layer.}
	\label{FS_loss}
\end{figure}

With the same method, we conduct numerical simulations for 200 different sets of initial fields and save the data for 90 time snapshots for each set of initial fields. Each snapshot is taken at a time interval of $200dt$. Therefore, the data of size $[200\times 90\times 64\times 64\times 32 \times 3]$ can be obtained as training and testing sets. For these 200 groups of turbulent mixing layer DNS data, each group would  take approximately 1.1 hours to parallel computing on a 32-core CPU. Similar to Section\ref{sec:5.1}, 80\% of data is used for training, and 20\% is used for testing. In this case, employing the same network parameters, the study sets the Fourier modes to 16, the initial learning rate to $10^{-3}$, the channel width of $P$ to 64, and configures the channel width of $Q$ as 128. The AdamW optimizer is used, and the GELU function is employed as the activation function. The learning curves of the FNO and TNO models for 3D free-shear turbulent mixing layer is shown in Fig.~\ref{FS_loss}. The FNO model and TNO model take 15.2 hours and 10.6 hours to train the network, respectively. It can be seen that the training and testing loss of the TNO model are lower than those of the FNO model.

In order to mitigate overfitting in the models, we generated and employed ten extra independent datasets originating from diverse initial fields to conduct the \textit{a posteriori} evaluation. In the a \textit{posteriori} study, fDNS data is utilized as a baseline to evaluate various models. Moreover, all neural operators and classical models used for comparison are initialized with the same initial field.

%--------------------------------------------------
\subsubsection{A \textit{posteriori} study}
\begin{figure}
	\includegraphics[width=0.8\columnwidth]{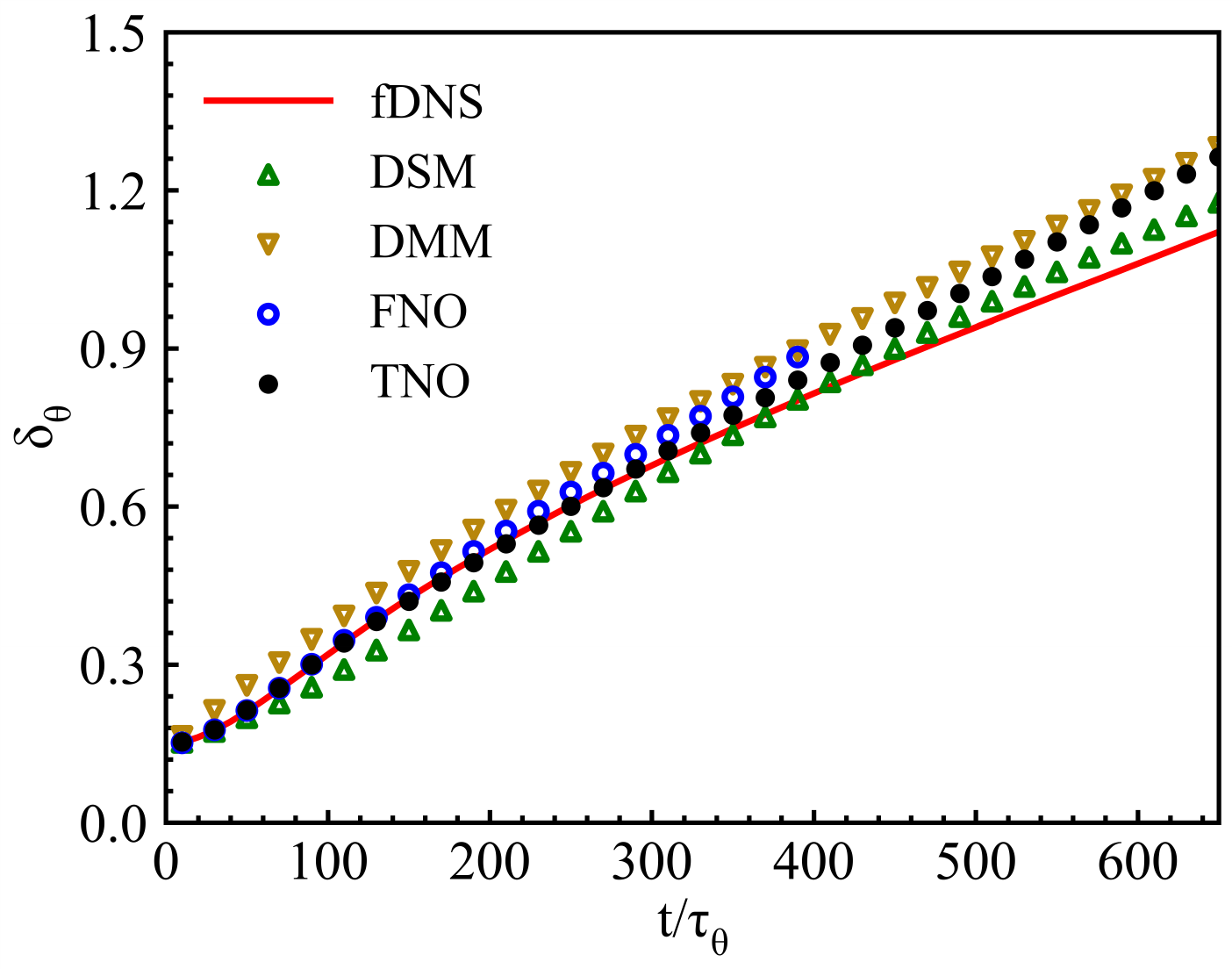}
	\caption{Temporal evolutions of the momentum thickness $\delta_\theta$ for LES in the free-shear turbulent mixing layer.}
	\label{FS_theta}
\end{figure}
The temporal evolutions of the momentum thickness $\delta_\theta$ for LES in the free-shear turbulent mixing layer are shown in Fig.~\ref{FS_theta}. It can be seen that the DMM model tends to overestimate the momentum thickness in comparison to the fDNS model. The DSM model tends to underestimate the momentum thickness during the initial stages of the transition region. However, it tends to overestimate the momentum thickness in the linear growth region. The FNO model demonstrates a strong capability to accurately capture the growth rate of momentum thickness in the early stages of temporal development. Nevertheless, its predictive ability becomes unreliable after reaching a period of 400 units $(t/\tau_\theta \geqslant 400)$. In contrast, the predictions of TNO model show a good agreement with fDNS in the transition region and maintain relatively stable results in the linear growth regions.

\begin{figure*}
	\includegraphics[width=1\linewidth]{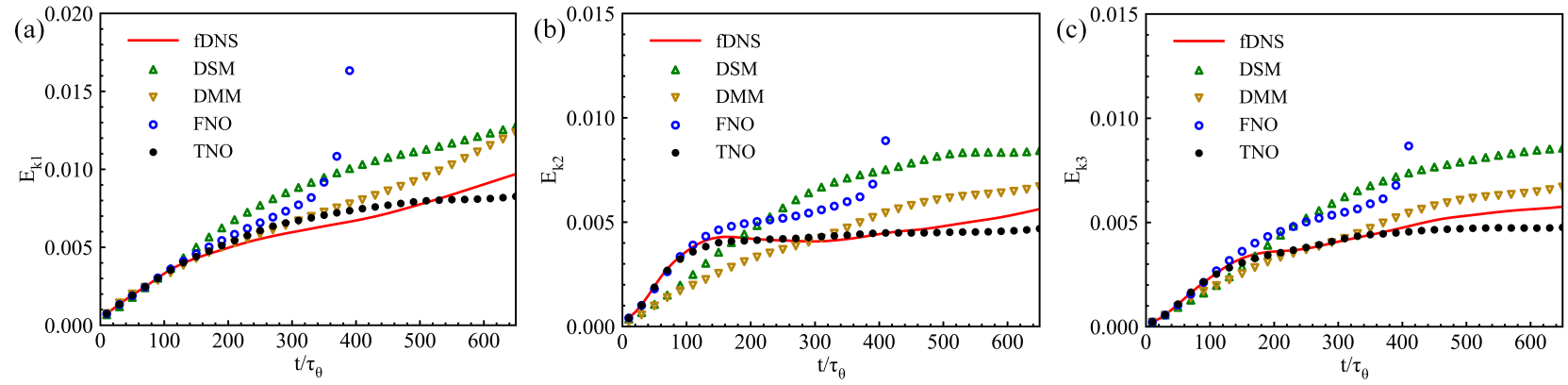}
	\caption{Temporal evolutions of the turbulent kinetic energy for LES in the free-shear turbulent mixing layer: (a)streamwise; (b)normal; (c)spanwise.}
	\label{FS_Ek123}
\end{figure*}
Moreover, the temporal evolutions of the streamwise turbulent kinetic energy $E_{k1}=\frac{1}{2}\left(\sqrt{\left\langle u_1 u_1\right\rangle}\right)^2$, normal  turbulent kinetic energy $E_{k2}=\frac{1}{2}\left(\sqrt{\left\langle u_2 u_2\right\rangle}\right)^2$ and spanwise turbulent kinetic energy $E_{k3}=\frac{1}{2}\left(\sqrt{\left\langle u_3 u_3\right\rangle}\right)^2$ are displayed in Fig.~\ref{FS_Ek123}. Here, $\langle\cdot\rangle$ is a spatial average over the whole computational domain. During the development of the shear layer in fDNS, there is a gradual increase in the turbulent kinetic energy across various directions. The DSM and DMM models underestimate the turbulent kinetic energy at the beginning, and overestimate the result in the linear growth region. The FNO model predicts reasonable results during the first 200 time units $(t/\tau_\theta\leq200)$, after that the results diverge quickly. By contrast, the TNO model closely matches the fDNS data by accurately predicting the kinetic energy in all three directions throughout the entire development of the shear layer.

\begin{figure*}
	\includegraphics[width=1\linewidth]{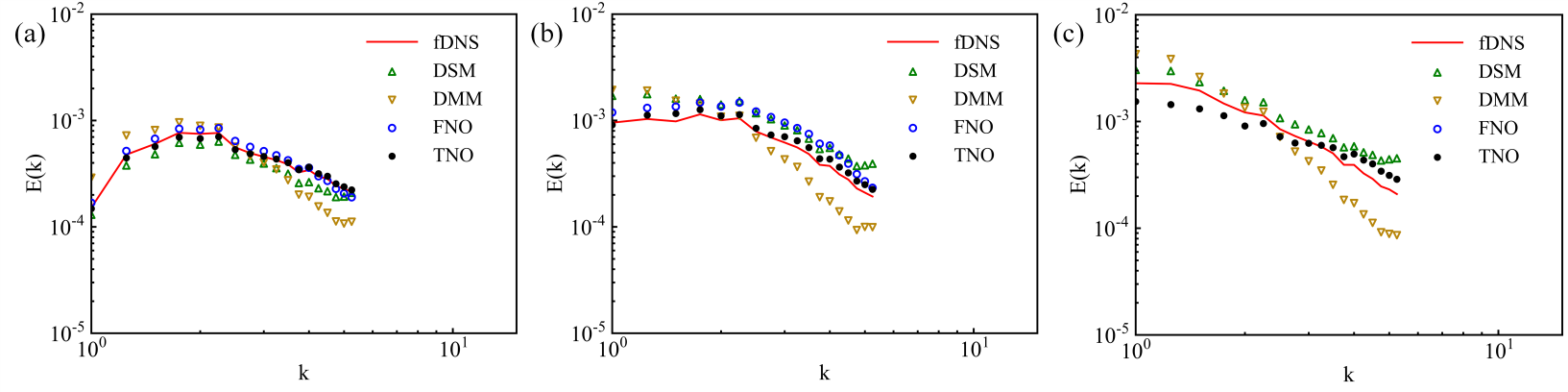}
	\caption{Velocity spectrum for LES in the free-shear turbulent mixing layer at different time instants: (a)$t/\tau_\theta\approx 100$; (b)$t/\tau_\theta\approx 300$; (b)$t/\tau_\theta\approx 650$.}
	\label{FS_spectrum}
\end{figure*}
The velocity spectrum of different models at time instants  $t/\tau_\theta\approx 100$, $t/\tau_\theta\approx 300$ and $t/\tau_\theta\approx 650$ are shown in Fig.~\ref{FS_spectrum}. The velocity spectrum predicted by the DSM model is observed to be underestimated at  $t/\tau_\theta\approx 100$, whereas it is overestimated for the subsequent time period starting from $t/\tau_\theta\approx 300$. The DMM model overestimates the velocity spectrum at low wavenumbers, and underestimates the velocity spectrum at high wavenumbers when compared to those of the fDNS. The FNO model is capable of providing reasonable predictions in the short-term, but the deviation from the fDNS results becomes more significant for longer time. In comparison, the TNO model demonstrates superior performance compared to other models at $t/\tau_\theta\approx 100$ and $t/\tau_\theta\approx 300$, with predictions closely matching the results of fDNS. The velocity spectrum predicted by all models exhibit a certain degree of deviation from fDNS at $t/\tau_\theta\approx 650$, while the TNO model slightly outperforms the other models in terms of prediction accuracy.

\begin{figure*}
	\includegraphics[width=1\linewidth]{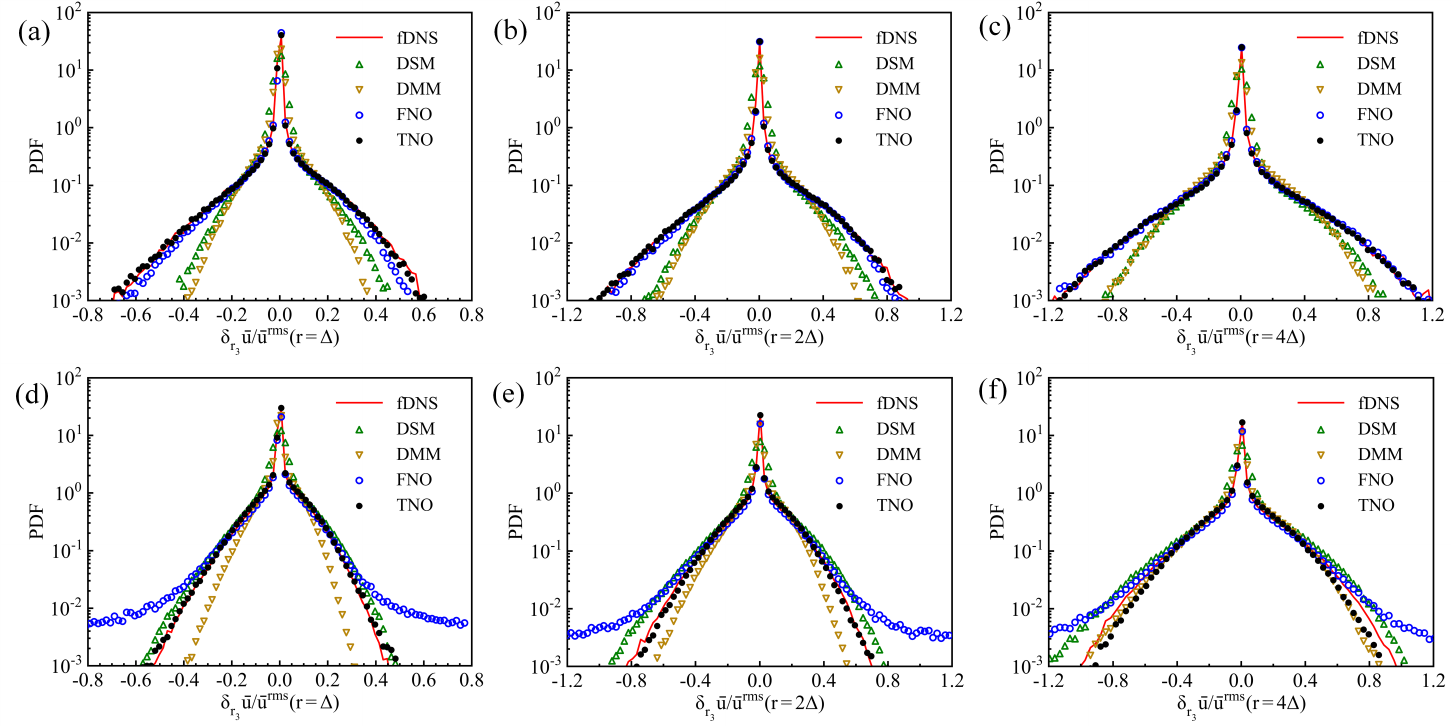}
	\caption{The PDFs of the spanwise velocity increment for LES in the free-shear turbulent mixing layer at different time instants: (a)$r=\Delta$, $t/\tau_\theta\approx 100$; (b)$r=2\Delta$, $t/\tau_\theta\approx 100$; (c)$r=4\Delta$, $t/\tau_\theta\approx 100$; (d)$r=\Delta$, $t/\tau_\theta\approx 650$; (e)$r=2\Delta$, $t/\tau_\theta\approx 650$ (f)$r=4\Delta$, $t/\tau_\theta\approx 650$.}
	\label{FS_inc124}
\end{figure*}
We then illustrate the PDFs of velocity increment in the spanwise direction, as shown in Fig.~\ref{FS_inc124}. The spanwise velocity increment is defined by $\delta_{r_3}\bar{u}=[\overline{\mathbf{u}}(\mathbf{x}+\mathbf{r})-\overline{\mathbf{u}}(\mathbf{x})] \cdot \hat{\mathbf{e}_3}$, where $\hat{\mathbf{e}_3}$ is the unit vector in the spanwise direction, and the velocity increments are normalized by the rms values of velocity $\bar{u}^{rms}$. The sharp peak of PDF can be attributed to the non-turbulent regions where the velocity increment is close to zero in the spanwise direction.\cite{wyp2022} Conversely, the regions characterized by non-zero velocity increments are predominantly influenced by turbulence. It is observed that both DSM and DMM models exhibit deviations from the fDNS results in both short-term and long-term predictions. The FNO model provides reasonably accurate predictions at $t/\tau_\theta\approx 100$, but the deviations become unacceptable as the prediction time increases. In contrast, the velocity increment given by the TNO model shows an excellent agreement with those of the fDNS for both short-term and long-term predictions.

\begin{figure*}
	\includegraphics[width=1\linewidth]{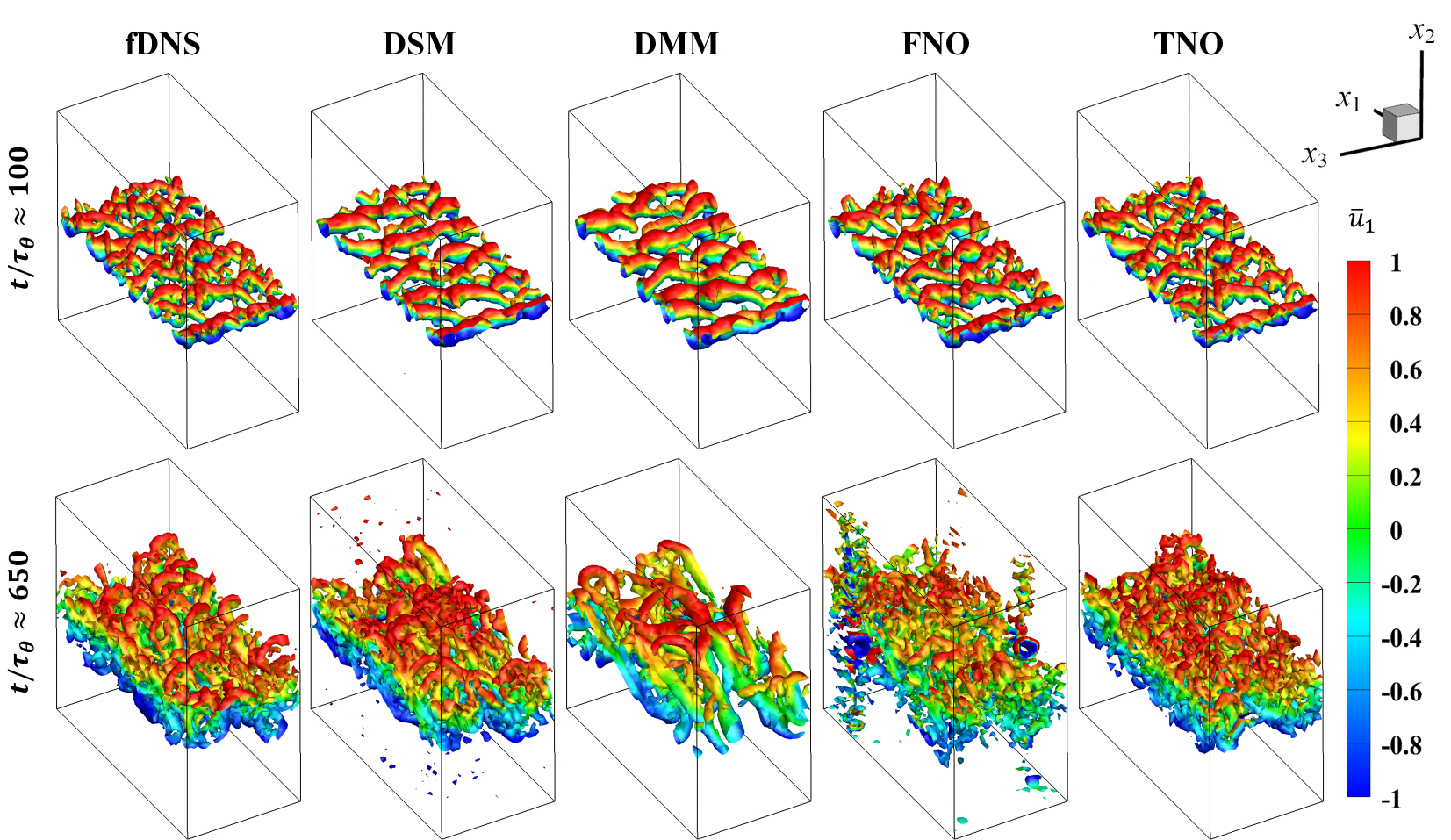}
	\caption{The iso-surface of the Q-criterion at $Q=0.2$ colored by the streamwise velocity at $t/\tau_\theta\approx 100$ and $t/\tau_\theta\approx 650$ in the free-shear turbulent mixing layer.}
	\label{FS_Q}
\end{figure*}
The instantaneous isosurfaces of $Q=0.2$ at $t/\tau_\theta\approx 100$ and $t/\tau_\theta\approx 650$ colored by the streamwise velocity in the free-shear turbulent mixing layer are shown in Fig.~\ref{FS_Q}. The Q-criterion has been widely used for visualizing vortex structures in turbulent flows and is defined by\cite{dubief2000coherent,zhan2019comparison}  
\begin{equation}
	Q=\frac{1}{2}\left(\bar{\Omega}_{i j} \bar{\Omega}_{i j}-\bar{S}_{i j} \bar{S}_{i j}\right),
	\label{eq23}
\end{equation}
where $\bar{\Omega}_{i j}=\left(\partial \bar{u}_i / \partial x_j-\partial \bar{u}_j / \partial x_i\right)/2$ denotes the filtered rotation-rate tensor. It is observed that the LES with DSM and DMM models predict relatively larger vortex structures compared to the fDNS result. The vortex structures predicted by the FNO model become unreliable after $t/\tau_\theta\approx 650$. On the contrary, the TNO model demonstrates a higher level of agreement with fDNS results, particularly in terms of reconstructing the small vortex structures, highlighting its advantage in improving the accuracy of LES.

\begin{figure*}
	\includegraphics[width=1\linewidth]{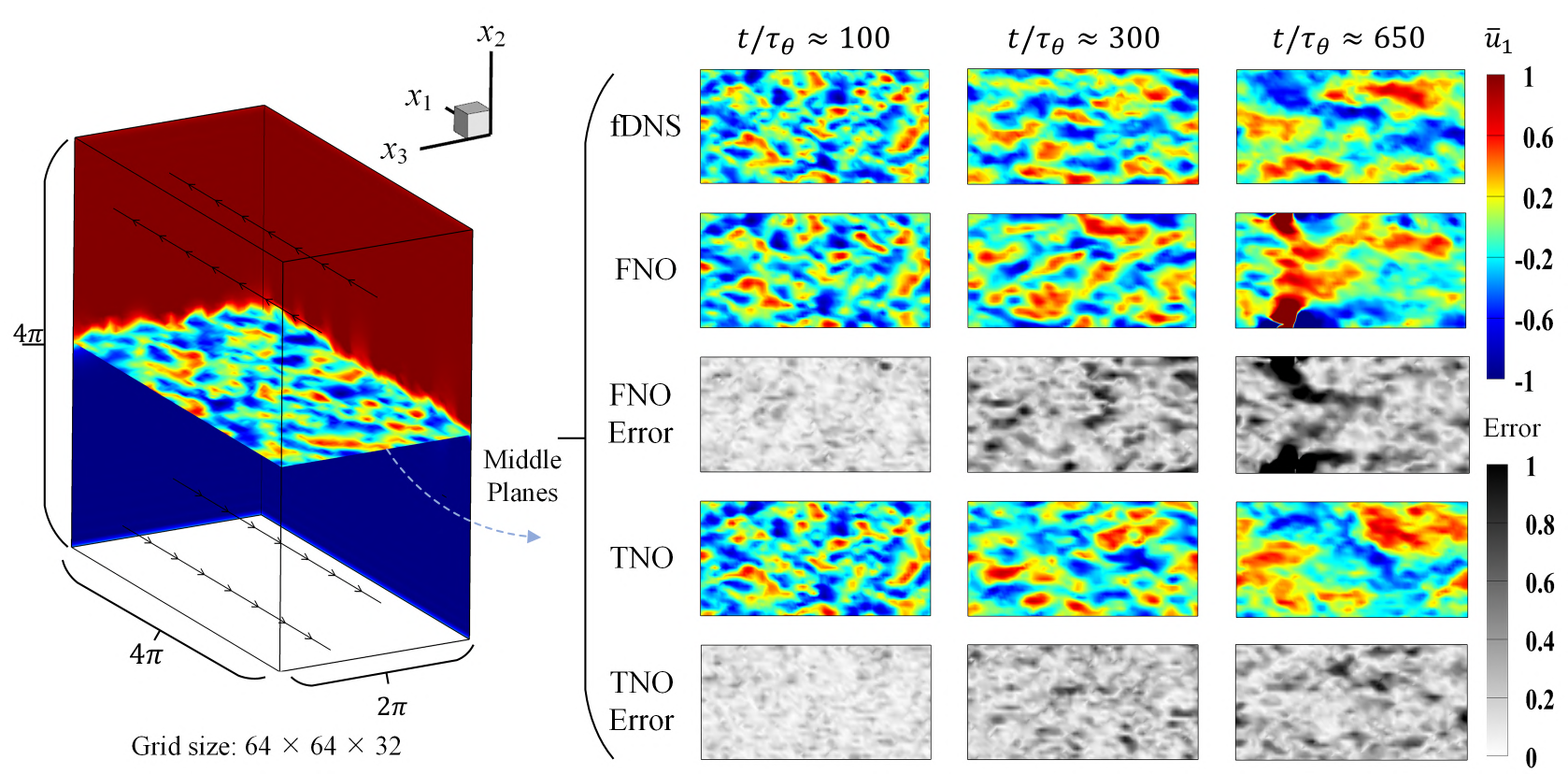}
	\caption{Velocity profile in the free-shear turbulent mixing layer at different time instants.}
	\label{FS_vel_profile}
\end{figure*}
Finally, we present the velocity profiles and absolute errors of the predicted streamwise velocities for both FNO and TNO models at different time instants in Fig.~\ref{FS_vel_profile}. It can be observed that the velocity profiles predicted by both the FNO and TNO models demonstrate a good agreement with those obtained from the fDNS results at $t/\tau_\theta\approx 100$. However, as time advances, a significant increase in prediction error is observed for the FNO model at $t/\tau_\theta\approx 650$. The TNO model demonstrates smaller errors in predicting the streamwise velocity compared to the fDNS results, highlighting its superior accuracy over the FNO model.

%--------------------------------------------------------------------------------------------------
%-----------------------------------------------

\subsubsection{Computational efficiency}
Here, we compare the computational efficiency among different models in Table.~\ref{CE_FS} The recorded time represents the time required by different models to predict each time node in free-shear turbulent mixing layer. We train the FNO and TNO models on Nvidia Tesla V100 GPU.The LES with DSM and DMM models are implemented on a computing cluster, where the type of CPU is Intel Xeon Gold 6148 with 16 cores each @2.40 GHz. It can be observed that the TNO model achieves a notable reduction in the number of network parameters when compared to FNO model. It can be seen that the TNO model requires only 20.18s of CPU time to predict a single time step, while the LES with DSM and DMM models require 83.55s and 97.86s, respectively. Furthermore, if the TNO model is run on a GPU, the time can be further reduced to 1.691s. This also demonstrates the potential of the TNO model in terms of fast predictions.
	\begin{table*}
	\caption{\label{CE_FS}Computational efficiency of different approaches on free-shear turbulent mixing layer.}
	\begin{ruledtabular}
		\begin{tabular}{cccc}
			Method& Total parameters($\times10^6$)& GPU$\cdot$s& CPU$\cdot$s\\
			\hline
			DSM& N/A & N/A & 83.55 \\
			DMM& N/A & N/A & 97.86 \\
			FNO& 563.9 & 0.426 & 7.912 \\
			TNO& 268.6 & 1.691 & 20.18 \\ 
		\end{tabular}
	\end{ruledtabular}
	\end{table*}

%===============================================================================================
\section{\label{sec:6}Conclusion}

%limitation是怎样，后续改进的方向是怎样。

Simulating 3D nonlinear PDEs is crucial in engineering applications, but data-driven approaches for fast 3D PDEs simulations are scarce compared to their success in 1D and 2D PDEs. Accurately modeling the complex non-linear interactions in 3D PDEs requires significant model complexity and a large number of parameters, posing a major challenge. Compared to the FNO model, transformer-based models are more suitable for developing accurate large-scale models due to their adaptability to larger datasets. Recently, the transformer-based models have demonstrated remarkable effectiveness as surrogate models for solving PDEs, highlighting their considerable potential in addressing 3D nonlinear problems. 

In this work, we developed a transformer-based neural operator (TNO) model to effectively predict the large-scale dynamics of 3D turbulence. The velocity fields of the previous time steps are taken as inputs of the model, aiming to directly map them to the velocity field of the next time step. By training the model with a large amount of data, TNO learns the evolution rules of the fluid dynamics from the data, enabling faster and more accurate predictions of the flow field evolution. The proposed model is tested in the LES of two types of 3D turbulence: forced homogeneous isotropic turbulence and free-shear turbulent mixing layer. The classical SGS models, including the DSM and DMM models, are utilized as comparison models in LES, with fDNS results serving as the benchmark.

A more systematic evaluation method is employed to validate the performance of different models. The numerical simulations comprehensively evaluate the performance of these models on a variety of flow statistics, including the velocity spectrum, structure functions, the PDFs of vorticity, the PDFs of velocity increments, momentum thickness, turbulent kinetic energy, and the iso-surface of the Q-criterion. The results demonstrate that the TNO model performs similarly to the LES with DSM model, and outperforms the FNO model and the LES using DMM model in HIT. Moreover, in the turbulent mixing layer, the TNO model exhibits superior accuracy compared to the other models. Moreover, compared to the FNO model, the TNO model not only achieves higher prediction accuracy while significantly reducing the number of network parameters, but it also demonstrates the ability to provide long-term stable predictions in the turbulent mixing layer where the FNO model fails. Meanwhile, the well-trained TNO model is much more efficient than traditional LES with DSM and DMM models. Moreover, the trained TNO model can be well generalized to the situations of higher Taylor–Reynolds numbers. Therefore, TNO has great potential in developing advanced neural network models to solve 3D nonlinear problems in engineering applications. 

One limitation of the proposed model is that it has only been evaluated on simple flow cases, while real-world engineering applications typically involve much more complex flows. Due to the influence of boundary conditions, further improvements and modifications to the model are required when extending its application to more complex flow cases. Recently, more advanced FNO-based and transformer-based model have been proposed\cite{guibas2021adaptive,kurth2022fourcastnet,alkin2024universal,zhao2023pinnsformer}, but their evaluation has been primarily limited to 2D problems. Li et al. proposed a geo-FNO model to handle PDEs on irregular geometries by mapping the input physical domain into a uniform latent space using a deformation function\cite{li2022fouriergeo}. This approach provides valuable insights for dealing with 3D turbulence on non-uniform grids and non-periodic boundary conditions. In future work, we will also consider the attention mechanism in the temporal dimension to enhance prediction accuracy and generalization abilities. The data-driven TNO model requires separate training for each distinct flow configuration. In the follow-up study, we will try to develop large models so that they can be better generalized to different flow types. Moreover, we will integrate various improvement approaches to explore the application of the proposed model in more complex engineering flow problems.

%--------------------------------------------------------------------------------------------------------------
\begin{acknowledgments}
This work was supported by the National Natural Science Foundation of China (NSFC Grant Nos. 12172161, 12161141017, 92052301 and 91952104), by the NSFC Basic Science Center Program (grant no. 11988102), by the Shenzhen Science and Technology Program (Grants No.KQTD20180411143441009), by Key Special Project for Introduced Talents Team of Southern Marine Science and Engineering Guangdong Laboratory (Guangzhou) (Grant No. GML2019ZD0103), and by Department of Science and Technology of Guangdong Province (Grants No. 2019B21203001, 2020B1212030001, 2023B1212060001). This work was also supported by Center for Computational Science and Engineering of Southern University of Science and Technology, and by Special Funds for the Cultivation of Guangdong College Students’ Scientific and Technological Innovation. (“Climbing Program” Special Fund pdjh2024c20702). Additionally, I would like to thank Siqi Ding for his support on this paper.
\end{acknowledgments}

\section*{AUTHOR DECLARATIONS}
\subsection*{Conflict of Interest}
The authors have no conflicts to disclose.

\section*{Data Availability}
The data that support the findings of this study are available from the corresponding author upon reasonable request.

%-------------------------------------------------------------------------------------------------------
%\appendix
%\section{Appendixes}

\bibliography{aipsamp}% Produces the bibliography via BibTeX.

\providecommand{\noopsort}[1]{}\providecommand{\singleletter}[1]{#1}%
\begin{thebibliography}{117}%
\makeatletter
\providecommand \@ifxundefined [1]{%
 \@ifx{#1\undefined}
}%
\providecommand \@ifnum [1]{%
 \ifnum #1\expandafter \@firstoftwo
 \else \expandafter \@secondoftwo
 \fi
}%
\providecommand \@ifx [1]{%
 \ifx #1\expandafter \@firstoftwo
 \else \expandafter \@secondoftwo
 \fi
}%
\providecommand \natexlab [1]{#1}%
\providecommand \enquote  [1]{``#1''}%
\providecommand \bibnamefont  [1]{#1}%
\providecommand \bibfnamefont [1]{#1}%
\providecommand \citenamefont [1]{#1}%
\providecommand \href@noop [0]{\@secondoftwo}%
\providecommand \href [0]{\begingroup \@sanitize@url \@href}%
\providecommand \@href[1]{\@@startlink{#1}\@@href}%
\providecommand \@@href[1]{\endgroup#1\@@endlink}%
\providecommand \@sanitize@url [0]{\catcode `\\12\catcode `\$12\catcode
  `\&12\catcode `\#12\catcode `\^12\catcode `\_12\catcode `\%12\relax}%
\providecommand \@@startlink[1]{}%
\providecommand \@@endlink[0]{}%
\providecommand \url  [0]{\begingroup\@sanitize@url \@url }%
\providecommand \@url [1]{\endgroup\@href {#1}{\urlprefix }}%
\providecommand \urlprefix  [0]{URL }%
\providecommand \Eprint [0]{\href }%
\providecommand \doibase [0]{http://dx.doi.org/}%
\providecommand \selectlanguage [0]{\@gobble}%
\providecommand \bibinfo  [0]{\@secondoftwo}%
\providecommand \bibfield  [0]{\@secondoftwo}%
\providecommand \translation [1]{[#1]}%
\providecommand \BibitemOpen [0]{}%
\providecommand \bibitemStop [0]{}%
\providecommand \bibitemNoStop [0]{.\EOS\space}%
\providecommand \EOS [0]{\spacefactor3000\relax}%
\providecommand \BibitemShut  [1]{\csname bibitem#1\endcsname}%
\let\auto@bib@innerbib\@empty
%</preamble>
\bibitem [{\citenamefont {Smagorinsky}(1963)}]{smagorinsky1963}%
  \BibitemOpen
  \bibfield  {author} {\bibinfo {author} {\bibfnamefont {J.}~\bibnamefont
  {Smagorinsky}},\ }\bibfield  {title} {\enquote {\bibinfo {title} {{General
  circulation experiments with the primitive equations: I. The basic
  experiment}},}\ }\href@noop {} {\bibfield  {journal} {\bibinfo  {journal}
  {Mon Weather Rev}\ }\textbf {\bibinfo {volume} {91}},\ \bibinfo {pages}
  {99--164} (\bibinfo {year} {1963})}\BibitemShut {NoStop}%
\bibitem [{\citenamefont {Lilly}(1967)}]{lilly1967}%
  \BibitemOpen
  \bibfield  {author} {\bibinfo {author} {\bibfnamefont {D.~K.}\ \bibnamefont
  {Lilly}},\ }\bibfield  {title} {\enquote {\bibinfo {title} {{The
  representation of small-scale turbulence in numerical simulation
  experiments}},}\ }\href@noop {} {\bibfield  {journal} {\bibinfo  {journal}
  {IBM Form}\ ,\ \bibinfo {pages} {195--210}} (\bibinfo {year}
  {1967})}\BibitemShut {NoStop}%
\bibitem [{\citenamefont {Germano}(1992)}]{germano1992}%
  \BibitemOpen
  \bibfield  {author} {\bibinfo {author} {\bibfnamefont {M.}~\bibnamefont
  {Germano}},\ }\bibfield  {title} {\enquote {\bibinfo {title} {{Turbulence:
  the filtering approach}},}\ }\href@noop {} {\bibfield  {journal} {\bibinfo
  {journal} {J. Fluid Mech.}\ }\textbf {\bibinfo {volume} {238}},\ \bibinfo
  {pages} {325--336} (\bibinfo {year} {1992})}\BibitemShut {NoStop}%
\bibitem [{\citenamefont {Garnier}, \citenamefont {Adams},\ and\ \citenamefont
  {Sagaut}(2009)}]{garnier2009large}%
  \BibitemOpen
  \bibfield  {author} {\bibinfo {author} {\bibfnamefont {E.}~\bibnamefont
  {Garnier}}, \bibinfo {author} {\bibfnamefont {N.}~\bibnamefont {Adams}}, \
  and\ \bibinfo {author} {\bibfnamefont {P.}~\bibnamefont {Sagaut}},\
  }\href@noop {} {\emph {\bibinfo {title} {{Large eddy simulation for
  compressible flows}}}}\ (\bibinfo  {publisher} {Springer Science \& Business
  Media},\ \bibinfo {year} {2009})\BibitemShut {NoStop}%
\bibitem [{\citenamefont {Deardorff}(1970)}]{deardorff1970numerical}%
  \BibitemOpen
  \bibfield  {author} {\bibinfo {author} {\bibfnamefont {J.~W.}\ \bibnamefont
  {Deardorff}},\ }\bibfield  {title} {\enquote {\bibinfo {title} {{A numerical
  study of three-dimensional turbulent channel flow at large Reynolds
  numbers}},}\ }\href@noop {} {\bibfield  {journal} {\bibinfo  {journal} {J.
  Fluid Mech.}\ }\textbf {\bibinfo {volume} {41}},\ \bibinfo {pages} {453--480}
  (\bibinfo {year} {1970})}\BibitemShut {NoStop}%
\bibitem [{\citenamefont {Durbin}(2018)}]{durbin2018some}%
  \BibitemOpen
  \bibfield  {author} {\bibinfo {author} {\bibfnamefont {P.~A.}\ \bibnamefont
  {Durbin}},\ }\bibfield  {title} {\enquote {\bibinfo {title} {{Some recent
  developments in turbulence closure modeling}},}\ }\href@noop {} {\bibfield
  {journal} {\bibinfo  {journal} {Annu. Rev. Fluid Mech.}\ }\textbf {\bibinfo
  {volume} {50}},\ \bibinfo {pages} {77--103} (\bibinfo {year}
  {2018})}\BibitemShut {NoStop}%
\bibitem [{\citenamefont {Pope}(1975)}]{pope1975more}%
  \BibitemOpen
  \bibfield  {author} {\bibinfo {author} {\bibfnamefont {S.~B.}\ \bibnamefont
  {Pope}},\ }\bibfield  {title} {\enquote {\bibinfo {title} {{A more general
  effective-viscosity hypothesis}},}\ }\href@noop {} {\bibfield  {journal}
  {\bibinfo  {journal} {J. Fluid Mech.}\ }\textbf {\bibinfo {volume} {72}},\
  \bibinfo {pages} {331--340} (\bibinfo {year} {1975})}\BibitemShut {NoStop}%
\bibitem [{\citenamefont {Choi}\ and\ \citenamefont
  {Moin}(2012)}]{choi2012grid}%
  \BibitemOpen
  \bibfield  {author} {\bibinfo {author} {\bibfnamefont {H.}~\bibnamefont
  {Choi}}\ and\ \bibinfo {author} {\bibfnamefont {P.}~\bibnamefont {Moin}},\
  }\bibfield  {title} {\enquote {\bibinfo {title} {{Grid-point requirements for
  large eddy simulation: Chapman’s estimates revisited}},}\ }\href@noop {}
  {\bibfield  {journal} {\bibinfo  {journal} {Phys. Fluids}\ }\textbf {\bibinfo
  {volume} {24}} (\bibinfo {year} {2012})}\BibitemShut {NoStop}%
\bibitem [{\citenamefont {Yang}\ and\ \citenamefont
  {Griffin}(2021)}]{yang2021grid}%
  \BibitemOpen
  \bibfield  {author} {\bibinfo {author} {\bibfnamefont {X.~I.}\ \bibnamefont
  {Yang}}\ and\ \bibinfo {author} {\bibfnamefont {K.~P.}\ \bibnamefont
  {Griffin}},\ }\bibfield  {title} {\enquote {\bibinfo {title} {{Grid-point and
  time-step requirements for direct numerical simulation and large-eddy
  simulation}},}\ }\href@noop {} {\bibfield  {journal} {\bibinfo  {journal}
  {Phys. Fluids}\ }\textbf {\bibinfo {volume} {33}} (\bibinfo {year}
  {2021})}\BibitemShut {NoStop}%
\bibitem [{\citenamefont {Bin}\ \emph {et~al.}(2023)\citenamefont {Bin},
  \citenamefont {Huang}, \citenamefont {Kunz},\ and\ \citenamefont
  {Yang}}]{bin2023constrained}%
  \BibitemOpen
  \bibfield  {author} {\bibinfo {author} {\bibfnamefont {Y.}~\bibnamefont
  {Bin}}, \bibinfo {author} {\bibfnamefont {G.}~\bibnamefont {Huang}}, \bibinfo
  {author} {\bibfnamefont {R.}~\bibnamefont {Kunz}}, \ and\ \bibinfo {author}
  {\bibfnamefont {X.~I.}\ \bibnamefont {Yang}},\ }\bibfield  {title} {\enquote
  {\bibinfo {title} {{Constrained Recalibration of Reynolds-Averaged
  Navier--Stokes Models}},}\ }\href@noop {} {\bibfield  {journal} {\bibinfo
  {journal} {AIAA Journal}\ ,\ \bibinfo {pages} {1--13}} (\bibinfo {year}
  {2023})}\BibitemShut {NoStop}%
\bibitem [{\citenamefont {Meneveau}\ and\ \citenamefont
  {Katz}(2000)}]{meneveau2000}%
  \BibitemOpen
  \bibfield  {author} {\bibinfo {author} {\bibfnamefont {C.}~\bibnamefont
  {Meneveau}}\ and\ \bibinfo {author} {\bibfnamefont {J.}~\bibnamefont
  {Katz}},\ }\bibfield  {title} {\enquote {\bibinfo {title} {{Scale-invariance
  and turbulence models for large-eddy simulation}},}\ }\href@noop {}
  {\bibfield  {journal} {\bibinfo  {journal} {Annu. Rev. Fluid Mech.}\ }\textbf
  {\bibinfo {volume} {32}},\ \bibinfo {pages} {1--32} (\bibinfo {year}
  {2000})}\BibitemShut {NoStop}%
\bibitem [{\citenamefont {Fan}\ \emph {et~al.}(2023)\citenamefont {Fan},
  \citenamefont {Yuan}, \citenamefont {Wang},\ and\ \citenamefont
  {Wang}}]{fan2023eddy}%
  \BibitemOpen
  \bibfield  {author} {\bibinfo {author} {\bibfnamefont {B.}~\bibnamefont
  {Fan}}, \bibinfo {author} {\bibfnamefont {Z.}~\bibnamefont {Yuan}}, \bibinfo
  {author} {\bibfnamefont {Y.}~\bibnamefont {Wang}}, \ and\ \bibinfo {author}
  {\bibfnamefont {J.}~\bibnamefont {Wang}},\ }\bibfield  {title} {\enquote
  {\bibinfo {title} {{Eddy viscosity enhanced temporal direct deconvolution
  models for temporal large-eddy simulation of turbulence}},}\ }\href@noop {}
  {\bibfield  {journal} {\bibinfo  {journal} {Phys. Fluids}\ }\textbf {\bibinfo
  {volume} {35}} (\bibinfo {year} {2023})}\BibitemShut {NoStop}%
\bibitem [{\citenamefont {Chen}\ \emph {et~al.}(2012)\citenamefont {Chen},
  \citenamefont {Xia}, \citenamefont {Pei}, \citenamefont {Wang}, \citenamefont
  {Yang}, \citenamefont {Xiao},\ and\ \citenamefont {Shi}}]{chen2012reynolds}%
  \BibitemOpen
  \bibfield  {author} {\bibinfo {author} {\bibfnamefont {S.}~\bibnamefont
  {Chen}}, \bibinfo {author} {\bibfnamefont {Z.}~\bibnamefont {Xia}}, \bibinfo
  {author} {\bibfnamefont {S.}~\bibnamefont {Pei}}, \bibinfo {author}
  {\bibfnamefont {J.}~\bibnamefont {Wang}}, \bibinfo {author} {\bibfnamefont
  {Y.}~\bibnamefont {Yang}}, \bibinfo {author} {\bibfnamefont {Z.}~\bibnamefont
  {Xiao}}, \ and\ \bibinfo {author} {\bibfnamefont {Y.}~\bibnamefont {Shi}},\
  }\bibfield  {title} {\enquote {\bibinfo {title} {{Reynolds-stress-constrained
  large-eddy simulation of wall-bounded turbulent flows}},}\ }\href@noop {}
  {\bibfield  {journal} {\bibinfo  {journal} {J. Fluid Mech.}\ }\textbf
  {\bibinfo {volume} {703}},\ \bibinfo {pages} {1--28} (\bibinfo {year}
  {2012})}\BibitemShut {NoStop}%
\bibitem [{\citenamefont {Moin}\ \emph {et~al.}(1991)\citenamefont {Moin},
  \citenamefont {Squires}, \citenamefont {Cabot},\ and\ \citenamefont
  {Lee}}]{moin1991dynamic}%
  \BibitemOpen
  \bibfield  {author} {\bibinfo {author} {\bibfnamefont {P.}~\bibnamefont
  {Moin}}, \bibinfo {author} {\bibfnamefont {K.}~\bibnamefont {Squires}},
  \bibinfo {author} {\bibfnamefont {W.}~\bibnamefont {Cabot}}, \ and\ \bibinfo
  {author} {\bibfnamefont {S.}~\bibnamefont {Lee}},\ }\bibfield  {title}
  {\enquote {\bibinfo {title} {A dynamic subgrid-scale model for compressible
  turbulence and scalar transport},}\ }\href@noop {} {\bibfield  {journal}
  {\bibinfo  {journal} {Phys. Fluids A: Fluid Dynamics}\ }\textbf {\bibinfo
  {volume} {3}},\ \bibinfo {pages} {2746--2757} (\bibinfo {year}
  {1991})}\BibitemShut {NoStop}%
\bibitem [{\citenamefont {Germano}\ \emph {et~al.}(1991)\citenamefont
  {Germano}, \citenamefont {Piomelli}, \citenamefont {Moin},\ and\
  \citenamefont {Cabot}}]{germano1991}%
  \BibitemOpen
  \bibfield  {author} {\bibinfo {author} {\bibfnamefont {M.}~\bibnamefont
  {Germano}}, \bibinfo {author} {\bibfnamefont {U.}~\bibnamefont {Piomelli}},
  \bibinfo {author} {\bibfnamefont {P.}~\bibnamefont {Moin}}, \ and\ \bibinfo
  {author} {\bibfnamefont {W.~H.}\ \bibnamefont {Cabot}},\ }\bibfield  {title}
  {\enquote {\bibinfo {title} {{A dynamic subgrid-scale eddy viscosity
  model}},}\ }\href@noop {} {\bibfield  {journal} {\bibinfo  {journal} {Phys.
  Fluids A}\ }\textbf {\bibinfo {volume} {3}},\ \bibinfo {pages} {1760--1765}
  (\bibinfo {year} {1991})}\BibitemShut {NoStop}%
\bibitem [{\citenamefont {Shi}, \citenamefont {Xiao},\ and\ \citenamefont
  {Chen}(2008)}]{shi2008constrained}%
  \BibitemOpen
  \bibfield  {author} {\bibinfo {author} {\bibfnamefont {Y.}~\bibnamefont
  {Shi}}, \bibinfo {author} {\bibfnamefont {Z.}~\bibnamefont {Xiao}}, \ and\
  \bibinfo {author} {\bibfnamefont {S.}~\bibnamefont {Chen}},\ }\bibfield
  {title} {\enquote {\bibinfo {title} {{Constrained subgrid-scale stress model
  for large eddy simulation}},}\ }\href@noop {} {\bibfield  {journal} {\bibinfo
   {journal} {Phys. Fluids}\ }\textbf {\bibinfo {volume} {20}} (\bibinfo {year}
  {2008})}\BibitemShut {NoStop}%
\bibitem [{\citenamefont {Erlebacher}\ \emph {et~al.}(1992)\citenamefont
  {Erlebacher}, \citenamefont {Hussaini}, \citenamefont {Speziale},\ and\
  \citenamefont {Zang}}]{erlebacher1992toward}%
  \BibitemOpen
  \bibfield  {author} {\bibinfo {author} {\bibfnamefont {G.}~\bibnamefont
  {Erlebacher}}, \bibinfo {author} {\bibfnamefont {M.~Y.}\ \bibnamefont
  {Hussaini}}, \bibinfo {author} {\bibfnamefont {C.~G.}\ \bibnamefont
  {Speziale}}, \ and\ \bibinfo {author} {\bibfnamefont {T.~A.}\ \bibnamefont
  {Zang}},\ }\bibfield  {title} {\enquote {\bibinfo {title} {{Toward the
  large-eddy simulation of compressible turbulent flows}},}\ }\href@noop {}
  {\bibfield  {journal} {\bibinfo  {journal} {J. Fluid Mech.}\ }\textbf
  {\bibinfo {volume} {238}},\ \bibinfo {pages} {155--185} (\bibinfo {year}
  {1992})}\BibitemShut {NoStop}%
\bibitem [{\citenamefont {Clark}, \citenamefont {Ferziger},\ and\ \citenamefont
  {Reynolds}(1979)}]{clark1979evaluation}%
  \BibitemOpen
  \bibfield  {author} {\bibinfo {author} {\bibfnamefont {R.~A.}\ \bibnamefont
  {Clark}}, \bibinfo {author} {\bibfnamefont {J.~H.}\ \bibnamefont {Ferziger}},
  \ and\ \bibinfo {author} {\bibfnamefont {W.~C.}\ \bibnamefont {Reynolds}},\
  }\bibfield  {title} {\enquote {\bibinfo {title} {{Evaluation of subgrid-scale
  models using an accurately simulated turbulent flow}},}\ }\href@noop {}
  {\bibfield  {journal} {\bibinfo  {journal} {J. Fluid Mech.}\ }\textbf
  {\bibinfo {volume} {91}},\ \bibinfo {pages} {1--16} (\bibinfo {year}
  {1979})}\BibitemShut {NoStop}%
\bibitem [{\citenamefont {Maulik}\ \emph {et~al.}(2018)\citenamefont {Maulik},
  \citenamefont {San}, \citenamefont {Rasheed},\ and\ \citenamefont
  {Vedula}}]{maulik2018data}%
  \BibitemOpen
  \bibfield  {author} {\bibinfo {author} {\bibfnamefont {R.}~\bibnamefont
  {Maulik}}, \bibinfo {author} {\bibfnamefont {O.}~\bibnamefont {San}},
  \bibinfo {author} {\bibfnamefont {A.}~\bibnamefont {Rasheed}}, \ and\
  \bibinfo {author} {\bibfnamefont {P.}~\bibnamefont {Vedula}},\ }\bibfield
  {title} {\enquote {\bibinfo {title} {{Data-driven deconvolution for large
  eddy simulations of Kraichnan turbulence}},}\ }\href@noop {} {\bibfield
  {journal} {\bibinfo  {journal} {Phys. Fluids}\ }\textbf {\bibinfo {volume}
  {30}},\ \bibinfo {pages} {125109} (\bibinfo {year} {2018})}\BibitemShut
  {NoStop}%
\bibitem [{\citenamefont {Yuan}\ \emph {et~al.}(2021)\citenamefont {Yuan},
  \citenamefont {Wang}, \citenamefont {Xie},\ and\ \citenamefont
  {Wang}}]{yuan2021dynamic}%
  \BibitemOpen
  \bibfield  {author} {\bibinfo {author} {\bibfnamefont {Z.}~\bibnamefont
  {Yuan}}, \bibinfo {author} {\bibfnamefont {Y.}~\bibnamefont {Wang}}, \bibinfo
  {author} {\bibfnamefont {C.}~\bibnamefont {Xie}}, \ and\ \bibinfo {author}
  {\bibfnamefont {J.}~\bibnamefont {Wang}},\ }\bibfield  {title} {\enquote
  {\bibinfo {title} {{Dynamic iterative approximate deconvolution models for
  large-eddy simulation of turbulence}},}\ }\href@noop {} {\bibfield  {journal}
  {\bibinfo  {journal} {Phys. Fluids}\ }\textbf {\bibinfo {volume} {33}},\
  \bibinfo {pages} {085125} (\bibinfo {year} {2021})}\BibitemShut {NoStop}%
\bibitem [{\citenamefont {Brunton}, \citenamefont {Noack},\ and\ \citenamefont
  {Koumoutsakos}(2020)}]{brunton2020machine}%
  \BibitemOpen
  \bibfield  {author} {\bibinfo {author} {\bibfnamefont {S.~L.}\ \bibnamefont
  {Brunton}}, \bibinfo {author} {\bibfnamefont {B.~R.}\ \bibnamefont {Noack}},
  \ and\ \bibinfo {author} {\bibfnamefont {P.}~\bibnamefont {Koumoutsakos}},\
  }\bibfield  {title} {\enquote {\bibinfo {title} {{Machine learning for fluid
  mechanics}},}\ }\href@noop {} {\bibfield  {journal} {\bibinfo  {journal}
  {Annu. Rev. Fluid Mech.}\ }\textbf {\bibinfo {volume} {52}},\ \bibinfo
  {pages} {477--508} (\bibinfo {year} {2020})}\BibitemShut {NoStop}%
\bibitem [{\citenamefont {Duraisamy}, \citenamefont {Iaccarino},\ and\
  \citenamefont {Xiao}(2019)}]{duraisamy2019turbulence}%
  \BibitemOpen
  \bibfield  {author} {\bibinfo {author} {\bibfnamefont {K.}~\bibnamefont
  {Duraisamy}}, \bibinfo {author} {\bibfnamefont {G.}~\bibnamefont
  {Iaccarino}}, \ and\ \bibinfo {author} {\bibfnamefont {H.}~\bibnamefont
  {Xiao}},\ }\bibfield  {title} {\enquote {\bibinfo {title} {{Turbulence
  modeling in the age of data}},}\ }\href@noop {} {\bibfield  {journal}
  {\bibinfo  {journal} {Annu. Rev. Fluid Mech.}\ }\textbf {\bibinfo {volume}
  {51}},\ \bibinfo {pages} {357--377} (\bibinfo {year} {2019})}\BibitemShut
  {NoStop}%
\bibitem [{\citenamefont {Lienen}\ \emph {et~al.}(2023)\citenamefont {Lienen},
  \citenamefont {Hansen-Palmus}, \citenamefont {L{\"u}dke},\ and\ \citenamefont
  {G{\"u}nnemann}}]{lienen2023generative}%
  \BibitemOpen
  \bibfield  {author} {\bibinfo {author} {\bibfnamefont {M.}~\bibnamefont
  {Lienen}}, \bibinfo {author} {\bibfnamefont {J.}~\bibnamefont
  {Hansen-Palmus}}, \bibinfo {author} {\bibfnamefont {D.}~\bibnamefont
  {L{\"u}dke}}, \ and\ \bibinfo {author} {\bibfnamefont {S.}~\bibnamefont
  {G{\"u}nnemann}},\ }\bibfield  {title} {\enquote {\bibinfo {title}
  {{Generative Diffusion for 3D Turbulent Flows}},}\ }\href@noop {} {\bibfield
  {journal} {\bibinfo  {journal} {arXiv preprint arXiv:2306.01776}\ } (\bibinfo
  {year} {2023})}\BibitemShut {NoStop}%
\bibitem [{\citenamefont {LeCun}, \citenamefont {Bengio},\ and\ \citenamefont
  {Hinton}(2015)}]{lecun2015deep}%
  \BibitemOpen
  \bibfield  {author} {\bibinfo {author} {\bibfnamefont {Y.}~\bibnamefont
  {LeCun}}, \bibinfo {author} {\bibfnamefont {Y.}~\bibnamefont {Bengio}}, \
  and\ \bibinfo {author} {\bibfnamefont {G.}~\bibnamefont {Hinton}},\
  }\bibfield  {title} {\enquote {\bibinfo {title} {{Deep learning}},}\
  }\href@noop {} {\bibfield  {journal} {\bibinfo  {journal} {nature}\ }\textbf
  {\bibinfo {volume} {521}},\ \bibinfo {pages} {436--444} (\bibinfo {year}
  {2015})}\BibitemShut {NoStop}%
\bibitem [{\citenamefont {Wang}\ \emph {et~al.}(2021)\citenamefont {Wang},
  \citenamefont {Yuan}, \citenamefont {Xie},\ and\ \citenamefont
  {Wang}}]{wang2021artificial}%
  \BibitemOpen
  \bibfield  {author} {\bibinfo {author} {\bibfnamefont {Y.}~\bibnamefont
  {Wang}}, \bibinfo {author} {\bibfnamefont {Z.}~\bibnamefont {Yuan}}, \bibinfo
  {author} {\bibfnamefont {C.}~\bibnamefont {Xie}}, \ and\ \bibinfo {author}
  {\bibfnamefont {J.}~\bibnamefont {Wang}},\ }\bibfield  {title} {\enquote
  {\bibinfo {title} {{Artificial neural network-based spatial gradient models
  for large-eddy simulation of turbulence}},}\ }\href@noop {} {\bibfield
  {journal} {\bibinfo  {journal} {AIP Adv.}\ }\textbf {\bibinfo {volume}
  {11}},\ \bibinfo {pages} {055216} (\bibinfo {year} {2021})}\BibitemShut
  {NoStop}%
\bibitem [{\citenamefont {Beck}, \citenamefont {Flad},\ and\ \citenamefont
  {Munz}(2019)}]{beck2019deep}%
  \BibitemOpen
  \bibfield  {author} {\bibinfo {author} {\bibfnamefont {A.}~\bibnamefont
  {Beck}}, \bibinfo {author} {\bibfnamefont {D.}~\bibnamefont {Flad}}, \ and\
  \bibinfo {author} {\bibfnamefont {C.-D.}\ \bibnamefont {Munz}},\ }\bibfield
  {title} {\enquote {\bibinfo {title} {{Deep neural networks for data-driven
  turbulence models}},}\ }in\ \href@noop {} {\emph {\bibinfo {booktitle}
  {Proceedings of the APS Division of Fluid Dynamics Meeting Abstracts}}}\
  (\bibinfo {year} {2019})\ p.\ \bibinfo {pages} {G16}\BibitemShut {NoStop}%
\bibitem [{\citenamefont {Xie}\ \emph {et~al.}(2019)\citenamefont {Xie},
  \citenamefont {Wang}, \citenamefont {Li},\ and\ \citenamefont
  {Ma}}]{xie2019artificial}%
  \BibitemOpen
  \bibfield  {author} {\bibinfo {author} {\bibfnamefont {C.}~\bibnamefont
  {Xie}}, \bibinfo {author} {\bibfnamefont {J.}~\bibnamefont {Wang}}, \bibinfo
  {author} {\bibfnamefont {K.}~\bibnamefont {Li}}, \ and\ \bibinfo {author}
  {\bibfnamefont {C.}~\bibnamefont {Ma}},\ }\bibfield  {title} {\enquote
  {\bibinfo {title} {{Artificial neural network approach to large-eddy
  simulation of compressible isotropic turbulence}},}\ }\href@noop {}
  {\bibfield  {journal} {\bibinfo  {journal} {Phys. Rev. E}\ }\textbf {\bibinfo
  {volume} {99}},\ \bibinfo {pages} {053113} (\bibinfo {year}
  {2019})}\BibitemShut {NoStop}%
\bibitem [{\citenamefont {Wang}\ \emph {et~al.}(2018)\citenamefont {Wang},
  \citenamefont {Luo}, \citenamefont {Li}, \citenamefont {Tan},\ and\
  \citenamefont {Fan}}]{wang2018investigations}%
  \BibitemOpen
  \bibfield  {author} {\bibinfo {author} {\bibfnamefont {Z.}~\bibnamefont
  {Wang}}, \bibinfo {author} {\bibfnamefont {K.}~\bibnamefont {Luo}}, \bibinfo
  {author} {\bibfnamefont {D.}~\bibnamefont {Li}}, \bibinfo {author}
  {\bibfnamefont {J.}~\bibnamefont {Tan}}, \ and\ \bibinfo {author}
  {\bibfnamefont {J.}~\bibnamefont {Fan}},\ }\bibfield  {title} {\enquote
  {\bibinfo {title} {{Investigations of data-driven closure for subgrid-scale
  stress in large-eddy simulation}},}\ }\href@noop {} {\bibfield  {journal}
  {\bibinfo  {journal} {Phys. Fluids}\ }\textbf {\bibinfo {volume} {30}},\
  \bibinfo {pages} {125101} (\bibinfo {year} {2018})}\BibitemShut {NoStop}%
\bibitem [{\citenamefont {Gamahara}\ and\ \citenamefont
  {Hattori}(2017)}]{gamahara2017searching}%
  \BibitemOpen
  \bibfield  {author} {\bibinfo {author} {\bibfnamefont {M.}~\bibnamefont
  {Gamahara}}\ and\ \bibinfo {author} {\bibfnamefont {Y.}~\bibnamefont
  {Hattori}},\ }\bibfield  {title} {\enquote {\bibinfo {title} {{Searching for
  turbulence models by artificial neural network}},}\ }\href@noop {} {\bibfield
   {journal} {\bibinfo  {journal} {Phys. Rev. Fluid}\ }\textbf {\bibinfo
  {volume} {2}},\ \bibinfo {pages} {054604} (\bibinfo {year}
  {2017})}\BibitemShut {NoStop}%
\bibitem [{\citenamefont {Zhou}\ \emph {et~al.}(2019)\citenamefont {Zhou},
  \citenamefont {He}, \citenamefont {Wang},\ and\ \citenamefont
  {Jin}}]{zhou2019subgrid}%
  \BibitemOpen
  \bibfield  {author} {\bibinfo {author} {\bibfnamefont {Z.}~\bibnamefont
  {Zhou}}, \bibinfo {author} {\bibfnamefont {G.}~\bibnamefont {He}}, \bibinfo
  {author} {\bibfnamefont {S.}~\bibnamefont {Wang}}, \ and\ \bibinfo {author}
  {\bibfnamefont {G.}~\bibnamefont {Jin}},\ }\bibfield  {title} {\enquote
  {\bibinfo {title} {{Subgrid-scale model for large-eddy simulation of
  isotropic turbulent flows using an artificial neural network}},}\ }\href@noop
  {} {\bibfield  {journal} {\bibinfo  {journal} {Comput Fluids}\ }\textbf
  {\bibinfo {volume} {195}},\ \bibinfo {pages} {104319} (\bibinfo {year}
  {2019})}\BibitemShut {NoStop}%
\bibitem [{\citenamefont {Li}\ \emph {et~al.}(2021{\natexlab{a}})\citenamefont
  {Li}, \citenamefont {Zhao}, \citenamefont {Wang},\ and\ \citenamefont
  {Sandberg}}]{li2021data}%
  \BibitemOpen
  \bibfield  {author} {\bibinfo {author} {\bibfnamefont {H.}~\bibnamefont
  {Li}}, \bibinfo {author} {\bibfnamefont {Y.}~\bibnamefont {Zhao}}, \bibinfo
  {author} {\bibfnamefont {J.}~\bibnamefont {Wang}}, \ and\ \bibinfo {author}
  {\bibfnamefont {R.~D.}\ \bibnamefont {Sandberg}},\ }\bibfield  {title}
  {\enquote {\bibinfo {title} {{Data-driven model development for large-eddy
  simulation of turbulence using gene-expression programing}},}\ }\href@noop {}
  {\bibfield  {journal} {\bibinfo  {journal} {Phys. Fluids}\ }\textbf {\bibinfo
  {volume} {33}},\ \bibinfo {pages} {125127} (\bibinfo {year}
  {2021}{\natexlab{a}})}\BibitemShut {NoStop}%
\bibitem [{\citenamefont {Ling}, \citenamefont {Kurzawski},\ and\ \citenamefont
  {Templeton}(2016)}]{ling2016reynolds}%
  \BibitemOpen
  \bibfield  {author} {\bibinfo {author} {\bibfnamefont {J.}~\bibnamefont
  {Ling}}, \bibinfo {author} {\bibfnamefont {A.}~\bibnamefont {Kurzawski}}, \
  and\ \bibinfo {author} {\bibfnamefont {J.}~\bibnamefont {Templeton}},\
  }\bibfield  {title} {\enquote {\bibinfo {title} {{Reynolds averaged
  turbulence modelling using deep neural networks with embedded invariance}},}\
  }\href@noop {} {\bibfield  {journal} {\bibinfo  {journal} {J. Fluid Mech.}\
  }\textbf {\bibinfo {volume} {807}},\ \bibinfo {pages} {155--166} (\bibinfo
  {year} {2016})}\BibitemShut {NoStop}%
\bibitem [{\citenamefont {Guan}\ \emph {et~al.}(2022)\citenamefont {Guan},
  \citenamefont {Chattopadhyay}, \citenamefont {Subel},\ and\ \citenamefont
  {Hassanzadeh}}]{guan2022stable}%
  \BibitemOpen
  \bibfield  {author} {\bibinfo {author} {\bibfnamefont {Y.}~\bibnamefont
  {Guan}}, \bibinfo {author} {\bibfnamefont {A.}~\bibnamefont {Chattopadhyay}},
  \bibinfo {author} {\bibfnamefont {A.}~\bibnamefont {Subel}}, \ and\ \bibinfo
  {author} {\bibfnamefont {P.}~\bibnamefont {Hassanzadeh}},\ }\bibfield
  {title} {\enquote {\bibinfo {title} {{Stable a posteriori LES of 2D
  turbulence using convolutional neural networks: Backscattering analysis and
  generalization to higher Re via transfer learning}},}\ }\href@noop {}
  {\bibfield  {journal} {\bibinfo  {journal} {J. Comput. Phys.}\ }\textbf
  {\bibinfo {volume} {458}},\ \bibinfo {pages} {111090} (\bibinfo {year}
  {2022})}\BibitemShut {NoStop}%
\bibitem [{\citenamefont {Han}\ \emph {et~al.}(2019)\citenamefont {Han},
  \citenamefont {Wang}, \citenamefont {Zhang},\ and\ \citenamefont
  {Chen}}]{han2019novel}%
  \BibitemOpen
  \bibfield  {author} {\bibinfo {author} {\bibfnamefont {R.}~\bibnamefont
  {Han}}, \bibinfo {author} {\bibfnamefont {Y.}~\bibnamefont {Wang}}, \bibinfo
  {author} {\bibfnamefont {Y.}~\bibnamefont {Zhang}}, \ and\ \bibinfo {author}
  {\bibfnamefont {G.}~\bibnamefont {Chen}},\ }\bibfield  {title} {\enquote
  {\bibinfo {title} {{A novel spatial-temporal prediction method for unsteady
  wake flows based on hybrid deep neural network}},}\ }\href@noop {} {\bibfield
   {journal} {\bibinfo  {journal} {Phys. Fluids}\ }\textbf {\bibinfo {volume}
  {31}} (\bibinfo {year} {2019})}\BibitemShut {NoStop}%
\bibitem [{\citenamefont {Cai}\ \emph {et~al.}(2021)\citenamefont {Cai},
  \citenamefont {Mao}, \citenamefont {Wang}, \citenamefont {Yin},\ and\
  \citenamefont {Karniadakis}}]{cai2021physics}%
  \BibitemOpen
  \bibfield  {author} {\bibinfo {author} {\bibfnamefont {S.}~\bibnamefont
  {Cai}}, \bibinfo {author} {\bibfnamefont {Z.}~\bibnamefont {Mao}}, \bibinfo
  {author} {\bibfnamefont {Z.}~\bibnamefont {Wang}}, \bibinfo {author}
  {\bibfnamefont {M.}~\bibnamefont {Yin}}, \ and\ \bibinfo {author}
  {\bibfnamefont {G.~E.}\ \bibnamefont {Karniadakis}},\ }\bibfield  {title}
  {\enquote {\bibinfo {title} {{Physics-informed neural networks (PINNs) for
  fluid mechanics: A review}},}\ }\href@noop {} {\bibfield  {journal} {\bibinfo
   {journal} {Acta Mech Sin}\ }\textbf {\bibinfo {volume} {37}},\ \bibinfo
  {pages} {1727--1738} (\bibinfo {year} {2021})}\BibitemShut {NoStop}%
\bibitem [{\citenamefont {Lanthaler}, \citenamefont {Mishra},\ and\
  \citenamefont {Karniadakis}(2022)}]{lanthaler2022error}%
  \BibitemOpen
  \bibfield  {author} {\bibinfo {author} {\bibfnamefont {S.}~\bibnamefont
  {Lanthaler}}, \bibinfo {author} {\bibfnamefont {S.}~\bibnamefont {Mishra}}, \
  and\ \bibinfo {author} {\bibfnamefont {G.~E.}\ \bibnamefont {Karniadakis}},\
  }\bibfield  {title} {\enquote {\bibinfo {title} {{Error estimates for
  deeponets: A deep learning framework in infinite dimensions}},}\ }\href@noop
  {} {\bibfield  {journal} {\bibinfo  {journal} {Transactions of Mathematics
  and Its Applications}\ }\textbf {\bibinfo {volume} {6}},\ \bibinfo {pages}
  {tnac001} (\bibinfo {year} {2022})}\BibitemShut {NoStop}%
\bibitem [{\citenamefont {Karniadakis}\ \emph {et~al.}(2021)\citenamefont
  {Karniadakis}, \citenamefont {Kevrekidis}, \citenamefont {Lu}, \citenamefont
  {Perdikaris}, \citenamefont {Wang},\ and\ \citenamefont
  {Yang}}]{karniadakis2021physics}%
  \BibitemOpen
  \bibfield  {author} {\bibinfo {author} {\bibfnamefont {G.~E.}\ \bibnamefont
  {Karniadakis}}, \bibinfo {author} {\bibfnamefont {I.~G.}\ \bibnamefont
  {Kevrekidis}}, \bibinfo {author} {\bibfnamefont {L.}~\bibnamefont {Lu}},
  \bibinfo {author} {\bibfnamefont {P.}~\bibnamefont {Perdikaris}}, \bibinfo
  {author} {\bibfnamefont {S.}~\bibnamefont {Wang}}, \ and\ \bibinfo {author}
  {\bibfnamefont {L.}~\bibnamefont {Yang}},\ }\bibfield  {title} {\enquote
  {\bibinfo {title} {{Physics-informed machine learning}},}\ }\href@noop {}
  {\bibfield  {journal} {\bibinfo  {journal} {Nat. Rev. Phys.}\ }\textbf
  {\bibinfo {volume} {3}},\ \bibinfo {pages} {422--440} (\bibinfo {year}
  {2021})}\BibitemShut {NoStop}%
\bibitem [{\citenamefont {Raissi}, \citenamefont {Perdikaris},\ and\
  \citenamefont {Karniadakis}(2017)}]{raissi2017physics}%
  \BibitemOpen
  \bibfield  {author} {\bibinfo {author} {\bibfnamefont {M.}~\bibnamefont
  {Raissi}}, \bibinfo {author} {\bibfnamefont {P.}~\bibnamefont {Perdikaris}},
  \ and\ \bibinfo {author} {\bibfnamefont {G.~E.}\ \bibnamefont
  {Karniadakis}},\ }\bibfield  {title} {\enquote {\bibinfo {title} {{Physics
  informed deep learning (part i): Data-driven solutions of nonlinear partial
  differential equations}},}\ }\href@noop {} {\bibfield  {journal} {\bibinfo
  {journal} {arXiv preprint arXiv:1711.10561}\ } (\bibinfo {year}
  {2017})}\BibitemShut {NoStop}%
\bibitem [{\citenamefont {Yang}\ \emph {et~al.}(2019)\citenamefont {Yang},
  \citenamefont {Zafar}, \citenamefont {Wang},\ and\ \citenamefont
  {Xiao}}]{yang2019predictive}%
  \BibitemOpen
  \bibfield  {author} {\bibinfo {author} {\bibfnamefont {X.}~\bibnamefont
  {Yang}}, \bibinfo {author} {\bibfnamefont {S.}~\bibnamefont {Zafar}},
  \bibinfo {author} {\bibfnamefont {J.-X.}\ \bibnamefont {Wang}}, \ and\
  \bibinfo {author} {\bibfnamefont {H.}~\bibnamefont {Xiao}},\ }\bibfield
  {title} {\enquote {\bibinfo {title} {{Predictive large-eddy-simulation wall
  modeling via physics-informed neural networks}},}\ }\href@noop {} {\bibfield
  {journal} {\bibinfo  {journal} {Phys. Rev. Fluid}\ }\textbf {\bibinfo
  {volume} {4}},\ \bibinfo {pages} {034602} (\bibinfo {year}
  {2019})}\BibitemShut {NoStop}%
\bibitem [{\citenamefont {Raissi}, \citenamefont {Perdikaris},\ and\
  \citenamefont {Karniadakis}(2019)}]{raissi2019physics}%
  \BibitemOpen
  \bibfield  {author} {\bibinfo {author} {\bibfnamefont {M.}~\bibnamefont
  {Raissi}}, \bibinfo {author} {\bibfnamefont {P.}~\bibnamefont {Perdikaris}},
  \ and\ \bibinfo {author} {\bibfnamefont {G.~E.}\ \bibnamefont
  {Karniadakis}},\ }\bibfield  {title} {\enquote {\bibinfo {title}
  {{Physics-informed neural networks: A deep learning framework for solving
  forward and inverse problems involving nonlinear partial differential
  equations}},}\ }\href@noop {} {\bibfield  {journal} {\bibinfo  {journal} {J.
  Comput. Phys.}\ }\textbf {\bibinfo {volume} {378}},\ \bibinfo {pages}
  {686--707} (\bibinfo {year} {2019})}\BibitemShut {NoStop}%
\bibitem [{\citenamefont {Chen}\ \emph {et~al.}(2021)\citenamefont {Chen},
  \citenamefont {Huang}, \citenamefont {Zhang}, \citenamefont {Zeng},
  \citenamefont {Wang}, \citenamefont {Zhang},\ and\ \citenamefont
  {Yan}}]{chen2021theory}%
  \BibitemOpen
  \bibfield  {author} {\bibinfo {author} {\bibfnamefont {Y.}~\bibnamefont
  {Chen}}, \bibinfo {author} {\bibfnamefont {D.}~\bibnamefont {Huang}},
  \bibinfo {author} {\bibfnamefont {D.}~\bibnamefont {Zhang}}, \bibinfo
  {author} {\bibfnamefont {J.}~\bibnamefont {Zeng}}, \bibinfo {author}
  {\bibfnamefont {N.}~\bibnamefont {Wang}}, \bibinfo {author} {\bibfnamefont
  {H.}~\bibnamefont {Zhang}}, \ and\ \bibinfo {author} {\bibfnamefont
  {J.}~\bibnamefont {Yan}},\ }\bibfield  {title} {\enquote {\bibinfo {title}
  {{Theory-guided hard constraint projection (HCP): A knowledge-based
  data-driven scientific machine learning method}},}\ }\href@noop {} {\bibfield
   {journal} {\bibinfo  {journal} {J. Comput. Phys.}\ }\textbf {\bibinfo
  {volume} {445}},\ \bibinfo {pages} {110624} (\bibinfo {year}
  {2021})}\BibitemShut {NoStop}%
\bibitem [{\citenamefont {Jin}\ \emph {et~al.}(2021)\citenamefont {Jin},
  \citenamefont {Cai}, \citenamefont {Li},\ and\ \citenamefont
  {Karniadakis}}]{jin2021nsfnets}%
  \BibitemOpen
  \bibfield  {author} {\bibinfo {author} {\bibfnamefont {X.}~\bibnamefont
  {Jin}}, \bibinfo {author} {\bibfnamefont {S.}~\bibnamefont {Cai}}, \bibinfo
  {author} {\bibfnamefont {H.}~\bibnamefont {Li}}, \ and\ \bibinfo {author}
  {\bibfnamefont {G.~E.}\ \bibnamefont {Karniadakis}},\ }\bibfield  {title}
  {\enquote {\bibinfo {title} {{NSFnets (Navier-Stokes flow nets):
  Physics-informed neural networks for the incompressible Navier-Stokes
  equations}},}\ }\href@noop {} {\bibfield  {journal} {\bibinfo  {journal} {J.
  Comput. Phys.}\ }\textbf {\bibinfo {volume} {426}},\ \bibinfo {pages}
  {109951} (\bibinfo {year} {2021})}\BibitemShut {NoStop}%
\bibitem [{\citenamefont {Wu}\ and\ \citenamefont {Xiu}(2020)}]{wu2020data}%
  \BibitemOpen
  \bibfield  {author} {\bibinfo {author} {\bibfnamefont {K.}~\bibnamefont
  {Wu}}\ and\ \bibinfo {author} {\bibfnamefont {D.}~\bibnamefont {Xiu}},\
  }\bibfield  {title} {\enquote {\bibinfo {title} {{Data-driven deep learning
  of partial differential equations in modal space}},}\ }\href@noop {}
  {\bibfield  {journal} {\bibinfo  {journal} {J. Comput. Phys.}\ }\textbf
  {\bibinfo {volume} {408}},\ \bibinfo {pages} {109307} (\bibinfo {year}
  {2020})}\BibitemShut {NoStop}%
\bibitem [{\citenamefont {Xu}, \citenamefont {Zhang},\ and\ \citenamefont
  {Zeng}(2021)}]{xu2021deep}%
  \BibitemOpen
  \bibfield  {author} {\bibinfo {author} {\bibfnamefont {H.}~\bibnamefont
  {Xu}}, \bibinfo {author} {\bibfnamefont {D.}~\bibnamefont {Zhang}}, \ and\
  \bibinfo {author} {\bibfnamefont {J.}~\bibnamefont {Zeng}},\ }\bibfield
  {title} {\enquote {\bibinfo {title} {{Deep-learning of parametric partial
  differential equations from sparse and noisy data}},}\ }\href@noop {}
  {\bibfield  {journal} {\bibinfo  {journal} {Phys. Fluids}\ }\textbf {\bibinfo
  {volume} {33}},\ \bibinfo {pages} {037132} (\bibinfo {year}
  {2021})}\BibitemShut {NoStop}%
\bibitem [{\citenamefont {Lu}\ \emph {et~al.}(2022)\citenamefont {Lu},
  \citenamefont {Meng}, \citenamefont {Cai}, \citenamefont {Mao}, \citenamefont
  {Goswami}, \citenamefont {Zhang},\ and\ \citenamefont
  {Karniadakis}}]{lu2022comprehensive}%
  \BibitemOpen
  \bibfield  {author} {\bibinfo {author} {\bibfnamefont {L.}~\bibnamefont
  {Lu}}, \bibinfo {author} {\bibfnamefont {X.}~\bibnamefont {Meng}}, \bibinfo
  {author} {\bibfnamefont {S.}~\bibnamefont {Cai}}, \bibinfo {author}
  {\bibfnamefont {Z.}~\bibnamefont {Mao}}, \bibinfo {author} {\bibfnamefont
  {S.}~\bibnamefont {Goswami}}, \bibinfo {author} {\bibfnamefont
  {Z.}~\bibnamefont {Zhang}}, \ and\ \bibinfo {author} {\bibfnamefont {G.~E.}\
  \bibnamefont {Karniadakis}},\ }\bibfield  {title} {\enquote {\bibinfo {title}
  {{A comprehensive and fair comparison of two neural operators (with practical
  extensions) based on fair data}},}\ }\href@noop {} {\bibfield  {journal}
  {\bibinfo  {journal} {Comput Methods Appl Mech Eng}\ }\textbf {\bibinfo
  {volume} {393}},\ \bibinfo {pages} {114778} (\bibinfo {year}
  {2022})}\BibitemShut {NoStop}%
\bibitem [{\citenamefont {Kovachki}\ \emph {et~al.}(2023)\citenamefont
  {Kovachki}, \citenamefont {Li}, \citenamefont {Liu}, \citenamefont
  {Azizzadenesheli}, \citenamefont {Bhattacharya}, \citenamefont {Stuart},\
  and\ \citenamefont {Anandkumar}}]{kovachki2023neural}%
  \BibitemOpen
  \bibfield  {author} {\bibinfo {author} {\bibfnamefont {N.}~\bibnamefont
  {Kovachki}}, \bibinfo {author} {\bibfnamefont {Z.}~\bibnamefont {Li}},
  \bibinfo {author} {\bibfnamefont {B.}~\bibnamefont {Liu}}, \bibinfo {author}
  {\bibfnamefont {K.}~\bibnamefont {Azizzadenesheli}}, \bibinfo {author}
  {\bibfnamefont {K.}~\bibnamefont {Bhattacharya}}, \bibinfo {author}
  {\bibfnamefont {A.}~\bibnamefont {Stuart}}, \ and\ \bibinfo {author}
  {\bibfnamefont {A.}~\bibnamefont {Anandkumar}},\ }\bibfield  {title}
  {\enquote {\bibinfo {title} {{Neural Operator: Learning Maps Between Function
  Spaces With Applications to PDEs}},}\ }\href@noop {} {\bibfield  {journal}
  {\bibinfo  {journal} {J Mach Learn Res}\ }\textbf {\bibinfo {volume} {24}},\
  \bibinfo {pages} {1--97} (\bibinfo {year} {2023})}\BibitemShut {NoStop}%
\bibitem [{\citenamefont {Goswami}\ \emph {et~al.}(2022)\citenamefont
  {Goswami}, \citenamefont {Kontolati}, \citenamefont {Shields},\ and\
  \citenamefont {Karniadakis}}]{goswami2022deep}%
  \BibitemOpen
  \bibfield  {author} {\bibinfo {author} {\bibfnamefont {S.}~\bibnamefont
  {Goswami}}, \bibinfo {author} {\bibfnamefont {K.}~\bibnamefont {Kontolati}},
  \bibinfo {author} {\bibfnamefont {M.~D.}\ \bibnamefont {Shields}}, \ and\
  \bibinfo {author} {\bibfnamefont {G.~E.}\ \bibnamefont {Karniadakis}},\
  }\bibfield  {title} {\enquote {\bibinfo {title} {{Deep transfer learning for
  partial differential equations under conditional shift with DeepONet}},}\
  }\href@noop {} {\bibfield  {journal} {\bibinfo  {journal} {arXiv preprint
  arXiv:2204.09810}\ } (\bibinfo {year} {2022})}\BibitemShut {NoStop}%
\bibitem [{\citenamefont {Li}\ \emph {et~al.}(2020{\natexlab{a}})\citenamefont
  {Li}, \citenamefont {Kovachki}, \citenamefont {Azizzadenesheli},
  \citenamefont {Liu}, \citenamefont {Bhattacharya}, \citenamefont {Stuart},\
  and\ \citenamefont {Anandkumar}}]{li2020fourier}%
  \BibitemOpen
  \bibfield  {author} {\bibinfo {author} {\bibfnamefont {Z.}~\bibnamefont
  {Li}}, \bibinfo {author} {\bibfnamefont {N.}~\bibnamefont {Kovachki}},
  \bibinfo {author} {\bibfnamefont {K.}~\bibnamefont {Azizzadenesheli}},
  \bibinfo {author} {\bibfnamefont {B.}~\bibnamefont {Liu}}, \bibinfo {author}
  {\bibfnamefont {K.}~\bibnamefont {Bhattacharya}}, \bibinfo {author}
  {\bibfnamefont {A.}~\bibnamefont {Stuart}}, \ and\ \bibinfo {author}
  {\bibfnamefont {A.}~\bibnamefont {Anandkumar}},\ }\bibfield  {title}
  {\enquote {\bibinfo {title} {{Fourier neural operator for parametric partial
  differential equations}},}\ }\href@noop {} {\bibfield  {journal} {\bibinfo
  {journal} {arXiv preprint arXiv:2010.08895}\ } (\bibinfo {year}
  {2020}{\natexlab{a}})}\BibitemShut {NoStop}%
\bibitem [{\citenamefont {Chen}, \citenamefont {Viquerat},\ and\ \citenamefont
  {Hachem}(2019)}]{chen2019u}%
  \BibitemOpen
  \bibfield  {author} {\bibinfo {author} {\bibfnamefont {J.}~\bibnamefont
  {Chen}}, \bibinfo {author} {\bibfnamefont {J.}~\bibnamefont {Viquerat}}, \
  and\ \bibinfo {author} {\bibfnamefont {E.}~\bibnamefont {Hachem}},\
  }\bibfield  {title} {\enquote {\bibinfo {title} {{U-net architectures for
  fast prediction of incompressible laminar flows}},}\ }\href@noop {}
  {\bibfield  {journal} {\bibinfo  {journal} {arXiv preprint arXiv:1910.13532}\
  } (\bibinfo {year} {2019})}\BibitemShut {NoStop}%
\bibitem [{\citenamefont {Wang}\ \emph {et~al.}(2020)\citenamefont {Wang},
  \citenamefont {Kashinath}, \citenamefont {Mustafa}, \citenamefont {Albert},\
  and\ \citenamefont {Yu}}]{wang2020towards}%
  \BibitemOpen
  \bibfield  {author} {\bibinfo {author} {\bibfnamefont {R.}~\bibnamefont
  {Wang}}, \bibinfo {author} {\bibfnamefont {K.}~\bibnamefont {Kashinath}},
  \bibinfo {author} {\bibfnamefont {M.}~\bibnamefont {Mustafa}}, \bibinfo
  {author} {\bibfnamefont {A.}~\bibnamefont {Albert}}, \ and\ \bibinfo {author}
  {\bibfnamefont {R.}~\bibnamefont {Yu}},\ }\bibfield  {title} {\enquote
  {\bibinfo {title} {Towards physics-informed deep learning for turbulent flow
  prediction},}\ }in\ \href@noop {} {\emph {\bibinfo {booktitle} {Proceedings
  of the 26th ACM SIGKDD International Conference on Knowledge Discovery \&
  Data Mining}}}\ (\bibinfo {year} {2020})\ pp.\ \bibinfo {pages}
  {1457--1466}\BibitemShut {NoStop}%
\bibitem [{\citenamefont {He}\ \emph {et~al.}(2016)\citenamefont {He},
  \citenamefont {Zhang}, \citenamefont {Ren},\ and\ \citenamefont
  {Sun}}]{he2016deep}%
  \BibitemOpen
  \bibfield  {author} {\bibinfo {author} {\bibfnamefont {K.}~\bibnamefont
  {He}}, \bibinfo {author} {\bibfnamefont {X.}~\bibnamefont {Zhang}}, \bibinfo
  {author} {\bibfnamefont {S.}~\bibnamefont {Ren}}, \ and\ \bibinfo {author}
  {\bibfnamefont {J.}~\bibnamefont {Sun}},\ }\bibfield  {title} {\enquote
  {\bibinfo {title} {{Deep residual learning for image recognition}},}\ }in\
  \href@noop {} {\emph {\bibinfo {booktitle} {Proceedings of the IEEE
  conference on computer vision and pattern recognition}}}\ (\bibinfo {year}
  {2016})\ pp.\ \bibinfo {pages} {770--778}\BibitemShut {NoStop}%
\bibitem [{\citenamefont {Wen}\ \emph {et~al.}(2022)\citenamefont {Wen},
  \citenamefont {Li}, \citenamefont {Azizzadenesheli}, \citenamefont
  {Anandkumar},\ and\ \citenamefont {Benson}}]{wen2022u}%
  \BibitemOpen
  \bibfield  {author} {\bibinfo {author} {\bibfnamefont {G.}~\bibnamefont
  {Wen}}, \bibinfo {author} {\bibfnamefont {Z.}~\bibnamefont {Li}}, \bibinfo
  {author} {\bibfnamefont {K.}~\bibnamefont {Azizzadenesheli}}, \bibinfo
  {author} {\bibfnamefont {A.}~\bibnamefont {Anandkumar}}, \ and\ \bibinfo
  {author} {\bibfnamefont {S.~M.}\ \bibnamefont {Benson}},\ }\bibfield  {title}
  {\enquote {\bibinfo {title} {{U-FNO—An enhanced Fourier neural
  operator-based deep-learning model for multiphase flow}},}\ }\href@noop {}
  {\bibfield  {journal} {\bibinfo  {journal} {Adv Water Resour}\ }\textbf
  {\bibinfo {volume} {163}},\ \bibinfo {pages} {104180} (\bibinfo {year}
  {2022})}\BibitemShut {NoStop}%
\bibitem [{\citenamefont {Choubineh}\ \emph {et~al.}(2023)\citenamefont
  {Choubineh}, \citenamefont {Chen}, \citenamefont {Wood}, \citenamefont
  {Coenen},\ and\ \citenamefont {Ma}}]{choubineh2023fourier}%
  \BibitemOpen
  \bibfield  {author} {\bibinfo {author} {\bibfnamefont {A.}~\bibnamefont
  {Choubineh}}, \bibinfo {author} {\bibfnamefont {J.}~\bibnamefont {Chen}},
  \bibinfo {author} {\bibfnamefont {D.~A.}\ \bibnamefont {Wood}}, \bibinfo
  {author} {\bibfnamefont {F.}~\bibnamefont {Coenen}}, \ and\ \bibinfo {author}
  {\bibfnamefont {F.}~\bibnamefont {Ma}},\ }\bibfield  {title} {\enquote
  {\bibinfo {title} {{Fourier neural operator for fluid flow in small-shape 2D
  simulated porous media dataset}},}\ }\href@noop {} {\bibfield  {journal}
  {\bibinfo  {journal} {Algorithms}\ }\textbf {\bibinfo {volume} {16}},\
  \bibinfo {pages} {24} (\bibinfo {year} {2023})}\BibitemShut {NoStop}%
\bibitem [{\citenamefont {Peng}, \citenamefont {Yuan},\ and\ \citenamefont
  {Wang}(2022)}]{peng2022attention}%
  \BibitemOpen
  \bibfield  {author} {\bibinfo {author} {\bibfnamefont {W.}~\bibnamefont
  {Peng}}, \bibinfo {author} {\bibfnamefont {Z.}~\bibnamefont {Yuan}}, \ and\
  \bibinfo {author} {\bibfnamefont {J.}~\bibnamefont {Wang}},\ }\bibfield
  {title} {\enquote {\bibinfo {title} {{Attention-enhanced neural network
  models for turbulence simulation}},}\ }\href@noop {} {\bibfield  {journal}
  {\bibinfo  {journal} {Phys. Fluids}\ }\textbf {\bibinfo {volume} {34}},\
  \bibinfo {pages} {025111} (\bibinfo {year} {2022})}\BibitemShut {NoStop}%
\bibitem [{\citenamefont {Li}\ \emph {et~al.}(2022{\natexlab{a}})\citenamefont
  {Li}, \citenamefont {Huang}, \citenamefont {Liu},\ and\ \citenamefont
  {Anandkumar}}]{li2022fouriergeo}%
  \BibitemOpen
  \bibfield  {author} {\bibinfo {author} {\bibfnamefont {Z.}~\bibnamefont
  {Li}}, \bibinfo {author} {\bibfnamefont {D.~Z.}\ \bibnamefont {Huang}},
  \bibinfo {author} {\bibfnamefont {B.}~\bibnamefont {Liu}}, \ and\ \bibinfo
  {author} {\bibfnamefont {A.}~\bibnamefont {Anandkumar}},\ }\bibfield  {title}
  {\enquote {\bibinfo {title} {{Fourier neural operator with learned
  deformations for pdes on general geometries}},}\ }\href@noop {} {\bibfield
  {journal} {\bibinfo  {journal} {arXiv preprint arXiv:2207.05209}\ } (\bibinfo
  {year} {2022}{\natexlab{a}})}\BibitemShut {NoStop}%
\bibitem [{\citenamefont {Jiang}\ \emph
  {et~al.}(2023{\natexlab{a}})\citenamefont {Jiang}, \citenamefont {Zhu},
  \citenamefont {Li}, \citenamefont {Li}, \citenamefont {Yuan},\ and\
  \citenamefont {Lu}}]{jiang2023fourier}%
  \BibitemOpen
  \bibfield  {author} {\bibinfo {author} {\bibfnamefont {Z.}~\bibnamefont
  {Jiang}}, \bibinfo {author} {\bibfnamefont {M.}~\bibnamefont {Zhu}}, \bibinfo
  {author} {\bibfnamefont {D.}~\bibnamefont {Li}}, \bibinfo {author}
  {\bibfnamefont {Q.}~\bibnamefont {Li}}, \bibinfo {author} {\bibfnamefont
  {Y.~O.}\ \bibnamefont {Yuan}}, \ and\ \bibinfo {author} {\bibfnamefont
  {L.}~\bibnamefont {Lu}},\ }\bibfield  {title} {\enquote {\bibinfo {title}
  {{Fourier-MIONet: Fourier-enhanced multiple-input neural operators for
  multiphase modeling of geological carbon sequestration}},}\ }\href@noop {}
  {\bibfield  {journal} {\bibinfo  {journal} {arXiv preprint arXiv:2303.04778}\
  } (\bibinfo {year} {2023}{\natexlab{a}})}\BibitemShut {NoStop}%
\bibitem [{\citenamefont {Tran}\ \emph {et~al.}(2021)\citenamefont {Tran},
  \citenamefont {Mathews}, \citenamefont {Xie},\ and\ \citenamefont
  {Ong}}]{tran2021factorized}%
  \BibitemOpen
  \bibfield  {author} {\bibinfo {author} {\bibfnamefont {A.}~\bibnamefont
  {Tran}}, \bibinfo {author} {\bibfnamefont {A.}~\bibnamefont {Mathews}},
  \bibinfo {author} {\bibfnamefont {L.}~\bibnamefont {Xie}}, \ and\ \bibinfo
  {author} {\bibfnamefont {C.~S.}\ \bibnamefont {Ong}},\ }\bibfield  {title}
  {\enquote {\bibinfo {title} {{Factorized Fourier neural operators}},}\
  }\href@noop {} {\bibfield  {journal} {\bibinfo  {journal} {arXiv preprint
  arXiv:2111.13802}\ } (\bibinfo {year} {2021})}\BibitemShut {NoStop}%
\bibitem [{\citenamefont {Renn}\ \emph {et~al.}(2023)\citenamefont {Renn},
  \citenamefont {Wang}, \citenamefont {Lale}, \citenamefont {Li}, \citenamefont
  {Anandkumar},\ and\ \citenamefont {Gharib}}]{renn2023forecasting}%
  \BibitemOpen
  \bibfield  {author} {\bibinfo {author} {\bibfnamefont {P.~I.}\ \bibnamefont
  {Renn}}, \bibinfo {author} {\bibfnamefont {C.}~\bibnamefont {Wang}}, \bibinfo
  {author} {\bibfnamefont {S.}~\bibnamefont {Lale}}, \bibinfo {author}
  {\bibfnamefont {Z.}~\bibnamefont {Li}}, \bibinfo {author} {\bibfnamefont
  {A.}~\bibnamefont {Anandkumar}}, \ and\ \bibinfo {author} {\bibfnamefont
  {M.}~\bibnamefont {Gharib}},\ }\bibfield  {title} {\enquote {\bibinfo {title}
  {{Forecasting subcritical cylinder wakes with Fourier Neural Operators}},}\
  }\href@noop {} {\bibfield  {journal} {\bibinfo  {journal} {arXiv preprint
  arXiv:2301.08290}\ } (\bibinfo {year} {2023})}\BibitemShut {NoStop}%
\bibitem [{\citenamefont {Li}\ \emph {et~al.}(2021{\natexlab{b}})\citenamefont
  {Li}, \citenamefont {Zheng}, \citenamefont {Kovachki}, \citenamefont {Jin},
  \citenamefont {Chen}, \citenamefont {Liu}, \citenamefont {Azizzadenesheli},\
  and\ \citenamefont {Anandkumar}}]{li2021physics}%
  \BibitemOpen
  \bibfield  {author} {\bibinfo {author} {\bibfnamefont {Z.}~\bibnamefont
  {Li}}, \bibinfo {author} {\bibfnamefont {H.}~\bibnamefont {Zheng}}, \bibinfo
  {author} {\bibfnamefont {N.}~\bibnamefont {Kovachki}}, \bibinfo {author}
  {\bibfnamefont {D.}~\bibnamefont {Jin}}, \bibinfo {author} {\bibfnamefont
  {H.}~\bibnamefont {Chen}}, \bibinfo {author} {\bibfnamefont {B.}~\bibnamefont
  {Liu}}, \bibinfo {author} {\bibfnamefont {K.}~\bibnamefont
  {Azizzadenesheli}}, \ and\ \bibinfo {author} {\bibfnamefont {A.}~\bibnamefont
  {Anandkumar}},\ }\bibfield  {title} {\enquote {\bibinfo {title}
  {{Physics-informed neural operator for learning partial differential
  equations}},}\ }\href@noop {} {\bibfield  {journal} {\bibinfo  {journal}
  {arXiv preprint arXiv:2111.03794}\ } (\bibinfo {year}
  {2021}{\natexlab{b}})}\BibitemShut {NoStop}%
\bibitem [{\citenamefont {Guibas}\ \emph {et~al.}(2021)\citenamefont {Guibas},
  \citenamefont {Mardani}, \citenamefont {Li}, \citenamefont {Tao},
  \citenamefont {Anandkumar},\ and\ \citenamefont
  {Catanzaro}}]{guibas2021adaptive}%
  \BibitemOpen
  \bibfield  {author} {\bibinfo {author} {\bibfnamefont {J.}~\bibnamefont
  {Guibas}}, \bibinfo {author} {\bibfnamefont {M.}~\bibnamefont {Mardani}},
  \bibinfo {author} {\bibfnamefont {Z.}~\bibnamefont {Li}}, \bibinfo {author}
  {\bibfnamefont {A.}~\bibnamefont {Tao}}, \bibinfo {author} {\bibfnamefont
  {A.}~\bibnamefont {Anandkumar}}, \ and\ \bibinfo {author} {\bibfnamefont
  {B.}~\bibnamefont {Catanzaro}},\ }\bibfield  {title} {\enquote {\bibinfo
  {title} {{Adaptive Fourier neural operators: Efficient token mixers for
  transformers}},}\ }\href@noop {} {\bibfield  {journal} {\bibinfo  {journal}
  {arXiv preprint arXiv:2111.13587}\ } (\bibinfo {year} {2021})}\BibitemShut
  {NoStop}%
\bibitem [{\citenamefont {Hao}\ \emph {et~al.}(2023)\citenamefont {Hao},
  \citenamefont {Ying}, \citenamefont {Wang}, \citenamefont {Su}, \citenamefont
  {Dong}, \citenamefont {Liu}, \citenamefont {Cheng}, \citenamefont {Zhu},\
  and\ \citenamefont {Song}}]{hao2023gnot}%
  \BibitemOpen
  \bibfield  {author} {\bibinfo {author} {\bibfnamefont {Z.}~\bibnamefont
  {Hao}}, \bibinfo {author} {\bibfnamefont {C.}~\bibnamefont {Ying}}, \bibinfo
  {author} {\bibfnamefont {Z.}~\bibnamefont {Wang}}, \bibinfo {author}
  {\bibfnamefont {H.}~\bibnamefont {Su}}, \bibinfo {author} {\bibfnamefont
  {Y.}~\bibnamefont {Dong}}, \bibinfo {author} {\bibfnamefont {S.}~\bibnamefont
  {Liu}}, \bibinfo {author} {\bibfnamefont {Z.}~\bibnamefont {Cheng}}, \bibinfo
  {author} {\bibfnamefont {J.}~\bibnamefont {Zhu}}, \ and\ \bibinfo {author}
  {\bibfnamefont {J.}~\bibnamefont {Song}},\ }\bibfield  {title} {\enquote
  {\bibinfo {title} {{GNOT: A General Neural Operator Transformer for Operator
  Learning}},}\ }\href@noop {} {\bibfield  {journal} {\bibinfo  {journal}
  {arXiv preprint arXiv:2302.14376}\ } (\bibinfo {year} {2023})}\BibitemShut
  {NoStop}%
\bibitem [{\citenamefont {Benitez}\ \emph {et~al.}(2023)\citenamefont
  {Benitez}, \citenamefont {Furuya}, \citenamefont {Faucher}, \citenamefont
  {Tricoche},\ and\ \citenamefont {de~Hoop}}]{benitez2023fine}%
  \BibitemOpen
  \bibfield  {author} {\bibinfo {author} {\bibfnamefont {J.~A.~L.}\
  \bibnamefont {Benitez}}, \bibinfo {author} {\bibfnamefont {T.}~\bibnamefont
  {Furuya}}, \bibinfo {author} {\bibfnamefont {F.}~\bibnamefont {Faucher}},
  \bibinfo {author} {\bibfnamefont {X.}~\bibnamefont {Tricoche}}, \ and\
  \bibinfo {author} {\bibfnamefont {M.~V.}\ \bibnamefont {de~Hoop}},\
  }\bibfield  {title} {\enquote {\bibinfo {title} {{Fine-tuning Neural-Operator
  architectures for training and generalization}},}\ }\href@noop {} {\bibfield
  {journal} {\bibinfo  {journal} {arXiv preprint arXiv:2301.11509}\ } (\bibinfo
  {year} {2023})}\BibitemShut {NoStop}%
\bibitem [{\citenamefont {Meng}\ \emph {et~al.}(2023)\citenamefont {Meng},
  \citenamefont {Zhu}, \citenamefont {Wang},\ and\ \citenamefont
  {Shi}}]{meng2023fast}%
  \BibitemOpen
  \bibfield  {author} {\bibinfo {author} {\bibfnamefont {D.}~\bibnamefont
  {Meng}}, \bibinfo {author} {\bibfnamefont {Y.}~\bibnamefont {Zhu}}, \bibinfo
  {author} {\bibfnamefont {J.}~\bibnamefont {Wang}}, \ and\ \bibinfo {author}
  {\bibfnamefont {Y.}~\bibnamefont {Shi}},\ }\bibfield  {title} {\enquote
  {\bibinfo {title} {{Fast flow prediction of airfoil dynamic stall based on
  Fourier neural operator}},}\ }\href@noop {} {\bibfield  {journal} {\bibinfo
  {journal} {Phys. Fluids}\ }\textbf {\bibinfo {volume} {35}} (\bibinfo {year}
  {2023})}\BibitemShut {NoStop}%
\bibitem [{\citenamefont {Deng}\ \emph {et~al.}(2023)\citenamefont {Deng},
  \citenamefont {Liu}, \citenamefont {Shi}, \citenamefont {Wang}, \citenamefont
  {Yu}, \citenamefont {Liu},\ and\ \citenamefont {Chen}}]{deng2023temporal}%
  \BibitemOpen
  \bibfield  {author} {\bibinfo {author} {\bibfnamefont {Z.}~\bibnamefont
  {Deng}}, \bibinfo {author} {\bibfnamefont {H.}~\bibnamefont {Liu}}, \bibinfo
  {author} {\bibfnamefont {B.}~\bibnamefont {Shi}}, \bibinfo {author}
  {\bibfnamefont {Z.}~\bibnamefont {Wang}}, \bibinfo {author} {\bibfnamefont
  {F.}~\bibnamefont {Yu}}, \bibinfo {author} {\bibfnamefont {Z.}~\bibnamefont
  {Liu}}, \ and\ \bibinfo {author} {\bibfnamefont {G.}~\bibnamefont {Chen}},\
  }\bibfield  {title} {\enquote {\bibinfo {title} {{Temporal predictions of
  periodic flows using a mesh transformation and deep learning-based
  strategy}},}\ }\href@noop {} {\bibfield  {journal} {\bibinfo  {journal}
  {Aerosp Sci Technol}\ }\textbf {\bibinfo {volume} {134}},\ \bibinfo {pages}
  {108081} (\bibinfo {year} {2023})}\BibitemShut {NoStop}%
\bibitem [{\citenamefont {Momenifar}\ \emph {et~al.}(2022)\citenamefont
  {Momenifar}, \citenamefont {Diao}, \citenamefont {Tarokh},\ and\
  \citenamefont {Bragg}}]{momenifar2022dimension}%
  \BibitemOpen
  \bibfield  {author} {\bibinfo {author} {\bibfnamefont {M.}~\bibnamefont
  {Momenifar}}, \bibinfo {author} {\bibfnamefont {E.}~\bibnamefont {Diao}},
  \bibinfo {author} {\bibfnamefont {V.}~\bibnamefont {Tarokh}}, \ and\ \bibinfo
  {author} {\bibfnamefont {A.~D.}\ \bibnamefont {Bragg}},\ }\bibfield  {title}
  {\enquote {\bibinfo {title} {{Dimension reduced turbulent flow data from deep
  vector quantisers}},}\ }\href@noop {} {\bibfield  {journal} {\bibinfo
  {journal} {J. Turbul.}\ }\textbf {\bibinfo {volume} {23}},\ \bibinfo {pages}
  {232--264} (\bibinfo {year} {2022})}\BibitemShut {NoStop}%
\bibitem [{\citenamefont {Mohan}\ \emph {et~al.}(2020)\citenamefont {Mohan},
  \citenamefont {Tretiak}, \citenamefont {Chertkov},\ and\ \citenamefont
  {Livescu}}]{mohan2020spatio}%
  \BibitemOpen
  \bibfield  {author} {\bibinfo {author} {\bibfnamefont {A.~T.}\ \bibnamefont
  {Mohan}}, \bibinfo {author} {\bibfnamefont {D.}~\bibnamefont {Tretiak}},
  \bibinfo {author} {\bibfnamefont {M.}~\bibnamefont {Chertkov}}, \ and\
  \bibinfo {author} {\bibfnamefont {D.}~\bibnamefont {Livescu}},\ }\bibfield
  {title} {\enquote {\bibinfo {title} {{Spatio-temporal deep learning models of
  3D turbulence with physics informed diagnostics}},}\ }\href@noop {}
  {\bibfield  {journal} {\bibinfo  {journal} {J. Turbul.}\ }\textbf {\bibinfo
  {volume} {21}},\ \bibinfo {pages} {484--524} (\bibinfo {year}
  {2020})}\BibitemShut {NoStop}%
\bibitem [{\citenamefont {Nakamura}\ \emph {et~al.}(2021)\citenamefont
  {Nakamura}, \citenamefont {Fukami}, \citenamefont {Hasegawa}, \citenamefont
  {Nabae},\ and\ \citenamefont {Fukagata}}]{nakamura2021convolutional}%
  \BibitemOpen
  \bibfield  {author} {\bibinfo {author} {\bibfnamefont {T.}~\bibnamefont
  {Nakamura}}, \bibinfo {author} {\bibfnamefont {K.}~\bibnamefont {Fukami}},
  \bibinfo {author} {\bibfnamefont {K.}~\bibnamefont {Hasegawa}}, \bibinfo
  {author} {\bibfnamefont {Y.}~\bibnamefont {Nabae}}, \ and\ \bibinfo {author}
  {\bibfnamefont {K.}~\bibnamefont {Fukagata}},\ }\bibfield  {title} {\enquote
  {\bibinfo {title} {{Convolutional neural network and long short-term memory
  based reduced order surrogate for minimal turbulent channel flow}},}\
  }\href@noop {} {\bibfield  {journal} {\bibinfo  {journal} {Phys. Fluids}\
  }\textbf {\bibinfo {volume} {33}},\ \bibinfo {pages} {025116} (\bibinfo
  {year} {2021})}\BibitemShut {NoStop}%
\bibitem [{\citenamefont {Li}\ \emph {et~al.}(2022{\natexlab{b}})\citenamefont
  {Li}, \citenamefont {Peng}, \citenamefont {Yuan},\ and\ \citenamefont
  {Wang}}]{li2022fourier}%
  \BibitemOpen
  \bibfield  {author} {\bibinfo {author} {\bibfnamefont {Z.}~\bibnamefont
  {Li}}, \bibinfo {author} {\bibfnamefont {W.}~\bibnamefont {Peng}}, \bibinfo
  {author} {\bibfnamefont {Z.}~\bibnamefont {Yuan}}, \ and\ \bibinfo {author}
  {\bibfnamefont {J.}~\bibnamefont {Wang}},\ }\bibfield  {title} {\enquote
  {\bibinfo {title} {{Fourier neural operator approach to large eddy simulation
  of three-dimensional turbulence}},}\ }\href@noop {} {\bibfield  {journal}
  {\bibinfo  {journal} {Theor. App. Mech. Lett.}\ }\textbf {\bibinfo {volume}
  {12}},\ \bibinfo {pages} {100389} (\bibinfo {year}
  {2022}{\natexlab{b}})}\BibitemShut {NoStop}%
\bibitem [{\citenamefont {Peng}\ \emph {et~al.}(2023)\citenamefont {Peng},
  \citenamefont {Yuan}, \citenamefont {Li},\ and\ \citenamefont
  {Wang}}]{peng2023linear}%
  \BibitemOpen
  \bibfield  {author} {\bibinfo {author} {\bibfnamefont {W.}~\bibnamefont
  {Peng}}, \bibinfo {author} {\bibfnamefont {Z.}~\bibnamefont {Yuan}}, \bibinfo
  {author} {\bibfnamefont {Z.}~\bibnamefont {Li}}, \ and\ \bibinfo {author}
  {\bibfnamefont {J.}~\bibnamefont {Wang}},\ }\bibfield  {title} {\enquote
  {\bibinfo {title} {{Linear attention coupled Fourier neural operator for
  simulation of three-dimensional turbulence}},}\ }\href@noop {} {\bibfield
  {journal} {\bibinfo  {journal} {Phys. Fluids}\ }\textbf {\bibinfo {volume}
  {35}},\ \bibinfo {pages} {015106} (\bibinfo {year} {2023})}\BibitemShut
  {NoStop}%
\bibitem [{\citenamefont {Li}\ \emph {et~al.}(2023{\natexlab{a}})\citenamefont
  {Li}, \citenamefont {Peng}, \citenamefont {Yuan},\ and\ \citenamefont
  {Wang}}]{li2023long}%
  \BibitemOpen
  \bibfield  {author} {\bibinfo {author} {\bibfnamefont {Z.}~\bibnamefont
  {Li}}, \bibinfo {author} {\bibfnamefont {W.}~\bibnamefont {Peng}}, \bibinfo
  {author} {\bibfnamefont {Z.}~\bibnamefont {Yuan}}, \ and\ \bibinfo {author}
  {\bibfnamefont {J.}~\bibnamefont {Wang}},\ }\bibfield  {title} {\enquote
  {\bibinfo {title} {{Long-term predictions of turbulence by implicit U-Net
  enhanced Fourier neural operator}},}\ }\href@noop {} {\bibfield  {journal}
  {\bibinfo  {journal} {Phys. Fluids}\ }\textbf {\bibinfo {volume} {35}}
  (\bibinfo {year} {2023}{\natexlab{a}})}\BibitemShut {NoStop}%
\bibitem [{\citenamefont {Li}\ \emph {et~al.}(2023{\natexlab{b}})\citenamefont
  {Li}, \citenamefont {Kovachki}, \citenamefont {Choy}, \citenamefont {Li},
  \citenamefont {Kossaifi}, \citenamefont {Otta}, \citenamefont {Nabian},
  \citenamefont {Stadler}, \citenamefont {Hundt}, \citenamefont
  {Azizzadenesheli} \emph {et~al.}}]{li2023geometry}%
  \BibitemOpen
  \bibfield  {author} {\bibinfo {author} {\bibfnamefont {Z.}~\bibnamefont
  {Li}}, \bibinfo {author} {\bibfnamefont {N.~B.}\ \bibnamefont {Kovachki}},
  \bibinfo {author} {\bibfnamefont {C.}~\bibnamefont {Choy}}, \bibinfo {author}
  {\bibfnamefont {B.}~\bibnamefont {Li}}, \bibinfo {author} {\bibfnamefont
  {J.}~\bibnamefont {Kossaifi}}, \bibinfo {author} {\bibfnamefont {S.~P.}\
  \bibnamefont {Otta}}, \bibinfo {author} {\bibfnamefont {M.~A.}\ \bibnamefont
  {Nabian}}, \bibinfo {author} {\bibfnamefont {M.}~\bibnamefont {Stadler}},
  \bibinfo {author} {\bibfnamefont {C.}~\bibnamefont {Hundt}}, \bibinfo
  {author} {\bibfnamefont {K.}~\bibnamefont {Azizzadenesheli}},  \emph
  {et~al.},\ }\bibfield  {title} {\enquote {\bibinfo {title}
  {{Geometry-informed neural operator for large-scale 3d pdes}},}\ }\href@noop
  {} {\bibfield  {journal} {\bibinfo  {journal} {arXiv preprint
  arXiv:2309.00583}\ } (\bibinfo {year} {2023}{\natexlab{b}})}\BibitemShut
  {NoStop}%
\bibitem [{\citenamefont {Vaswani}\ \emph {et~al.}(2017)\citenamefont
  {Vaswani}, \citenamefont {Shazeer}, \citenamefont {Parmar}, \citenamefont
  {Uszkoreit}, \citenamefont {Jones}, \citenamefont {Gomez}, \citenamefont
  {Kaiser},\ and\ \citenamefont {Polosukhin}}]{vaswani2017attention}%
  \BibitemOpen
  \bibfield  {author} {\bibinfo {author} {\bibfnamefont {A.}~\bibnamefont
  {Vaswani}}, \bibinfo {author} {\bibfnamefont {N.}~\bibnamefont {Shazeer}},
  \bibinfo {author} {\bibfnamefont {N.}~\bibnamefont {Parmar}}, \bibinfo
  {author} {\bibfnamefont {J.}~\bibnamefont {Uszkoreit}}, \bibinfo {author}
  {\bibfnamefont {L.}~\bibnamefont {Jones}}, \bibinfo {author} {\bibfnamefont
  {A.~N.}\ \bibnamefont {Gomez}}, \bibinfo {author} {\bibfnamefont
  {{\L}.}~\bibnamefont {Kaiser}}, \ and\ \bibinfo {author} {\bibfnamefont
  {I.}~\bibnamefont {Polosukhin}},\ }\bibfield  {title} {\enquote {\bibinfo
  {title} {{Attention is all you need}},}\ }\href@noop {} {\bibfield  {journal}
  {\bibinfo  {journal} {Proc. Adv. Neural Inf. Process. Syst.}\ }\textbf
  {\bibinfo {volume} {30}} (\bibinfo {year} {2017})}\BibitemShut {NoStop}%
\bibitem [{\citenamefont {Keskar}\ \emph {et~al.}(2019)\citenamefont {Keskar},
  \citenamefont {McCann}, \citenamefont {Varshney}, \citenamefont {Xiong},\
  and\ \citenamefont {Socher}}]{keskar2019ctrl}%
  \BibitemOpen
  \bibfield  {author} {\bibinfo {author} {\bibfnamefont {N.~S.}\ \bibnamefont
  {Keskar}}, \bibinfo {author} {\bibfnamefont {B.}~\bibnamefont {McCann}},
  \bibinfo {author} {\bibfnamefont {L.~R.}\ \bibnamefont {Varshney}}, \bibinfo
  {author} {\bibfnamefont {C.}~\bibnamefont {Xiong}}, \ and\ \bibinfo {author}
  {\bibfnamefont {R.}~\bibnamefont {Socher}},\ }\bibfield  {title} {\enquote
  {\bibinfo {title} {{Ctrl: A conditional transformer language model for
  controllable generation}},}\ }\href@noop {} {\bibfield  {journal} {\bibinfo
  {journal} {arXiv preprint arXiv:1909.05858}\ } (\bibinfo {year}
  {2019})}\BibitemShut {NoStop}%
\bibitem [{\citenamefont {Dai}\ \emph {et~al.}(2019)\citenamefont {Dai},
  \citenamefont {Yang}, \citenamefont {Yang}, \citenamefont {Carbonell},
  \citenamefont {Le},\ and\ \citenamefont
  {Salakhutdinov}}]{dai2019transformer}%
  \BibitemOpen
  \bibfield  {author} {\bibinfo {author} {\bibfnamefont {Z.}~\bibnamefont
  {Dai}}, \bibinfo {author} {\bibfnamefont {Z.}~\bibnamefont {Yang}}, \bibinfo
  {author} {\bibfnamefont {Y.}~\bibnamefont {Yang}}, \bibinfo {author}
  {\bibfnamefont {J.}~\bibnamefont {Carbonell}}, \bibinfo {author}
  {\bibfnamefont {Q.~V.}\ \bibnamefont {Le}}, \ and\ \bibinfo {author}
  {\bibfnamefont {R.}~\bibnamefont {Salakhutdinov}},\ }\bibfield  {title}
  {\enquote {\bibinfo {title} {{Transformer-xl: Attentive language models
  beyond a fixed-length context}},}\ }\href@noop {} {\bibfield  {journal}
  {\bibinfo  {journal} {arXiv preprint arXiv:1901.02860}\ } (\bibinfo {year}
  {2019})}\BibitemShut {NoStop}%
\bibitem [{\citenamefont {Li}, \citenamefont {Meidani},\ and\ \citenamefont
  {Farimani}(2022)}]{li2022transformer}%
  \BibitemOpen
  \bibfield  {author} {\bibinfo {author} {\bibfnamefont {Z.}~\bibnamefont
  {Li}}, \bibinfo {author} {\bibfnamefont {K.}~\bibnamefont {Meidani}}, \ and\
  \bibinfo {author} {\bibfnamefont {A.~B.}\ \bibnamefont {Farimani}},\
  }\bibfield  {title} {\enquote {\bibinfo {title} {{Transformer for partial
  differential equations' operator learning}},}\ }\href@noop {} {\bibfield
  {journal} {\bibinfo  {journal} {arXiv preprint arXiv:2205.13671}\ } (\bibinfo
  {year} {2022})}\BibitemShut {NoStop}%
\bibitem [{\citenamefont {Drikakis}\ \emph {et~al.}(2024)\citenamefont
  {Drikakis}, \citenamefont {Kokkinakis}, \citenamefont {Fung},\ and\
  \citenamefont {Spottswood}}]{drikakis2024generalizability}%
  \BibitemOpen
  \bibfield  {author} {\bibinfo {author} {\bibfnamefont {D.}~\bibnamefont
  {Drikakis}}, \bibinfo {author} {\bibfnamefont {I.~W.}\ \bibnamefont
  {Kokkinakis}}, \bibinfo {author} {\bibfnamefont {D.}~\bibnamefont {Fung}}, \
  and\ \bibinfo {author} {\bibfnamefont {S.~M.}\ \bibnamefont {Spottswood}},\
  }\bibfield  {title} {\enquote {\bibinfo {title} {Generalizability of
  transformer-based deep learning for multidimensional turbulent flow data},}\
  }\href@noop {} {\bibfield  {journal} {\bibinfo  {journal} {Physics of
  Fluids}\ }\textbf {\bibinfo {volume} {36}} (\bibinfo {year}
  {2024})}\BibitemShut {NoStop}%
\bibitem [{\citenamefont {Xu}\ \emph {et~al.}(2023)\citenamefont {Xu},
  \citenamefont {Zhuang}, \citenamefont {Pan},\ and\ \citenamefont
  {Wen}}]{xu2023super}%
  \BibitemOpen
  \bibfield  {author} {\bibinfo {author} {\bibfnamefont {Q.}~\bibnamefont
  {Xu}}, \bibinfo {author} {\bibfnamefont {Z.}~\bibnamefont {Zhuang}}, \bibinfo
  {author} {\bibfnamefont {Y.}~\bibnamefont {Pan}}, \ and\ \bibinfo {author}
  {\bibfnamefont {B.}~\bibnamefont {Wen}},\ }\bibfield  {title} {\enquote
  {\bibinfo {title} {{Super-resolution reconstruction of turbulent flows with a
  transformer-based deep learning framework}},}\ }\href@noop {} {\bibfield
  {journal} {\bibinfo  {journal} {Phys. Fluids}\ }\textbf {\bibinfo {volume}
  {35}} (\bibinfo {year} {2023})}\BibitemShut {NoStop}%
\bibitem [{\citenamefont {Momenifar}\ \emph {et~al.}(2021)\citenamefont
  {Momenifar}, \citenamefont {Diao}, \citenamefont {Tarokh},\ and\
  \citenamefont {Bragg}}]{momenifar2021emulating}%
  \BibitemOpen
  \bibfield  {author} {\bibinfo {author} {\bibfnamefont {M.}~\bibnamefont
  {Momenifar}}, \bibinfo {author} {\bibfnamefont {E.}~\bibnamefont {Diao}},
  \bibinfo {author} {\bibfnamefont {V.}~\bibnamefont {Tarokh}}, \ and\ \bibinfo
  {author} {\bibfnamefont {A.~D.}\ \bibnamefont {Bragg}},\ }\bibfield  {title}
  {\enquote {\bibinfo {title} {{Emulating Spatio-Temporal Realizations of
  Three-Dimensional Isotropic Turbulence via Deep Sequence Learning Models}},}\
  }\href@noop {} {\bibfield  {journal} {\bibinfo  {journal} {arXiv preprint
  arXiv:2112.03469}\ } (\bibinfo {year} {2021})}\BibitemShut {NoStop}%
\bibitem [{\citenamefont {Bi}\ \emph {et~al.}(2022)\citenamefont {Bi},
  \citenamefont {Xie}, \citenamefont {Zhang}, \citenamefont {Chen},
  \citenamefont {Gu},\ and\ \citenamefont {Tian}}]{bi2022pangu}%
  \BibitemOpen
  \bibfield  {author} {\bibinfo {author} {\bibfnamefont {K.}~\bibnamefont
  {Bi}}, \bibinfo {author} {\bibfnamefont {L.}~\bibnamefont {Xie}}, \bibinfo
  {author} {\bibfnamefont {H.}~\bibnamefont {Zhang}}, \bibinfo {author}
  {\bibfnamefont {X.}~\bibnamefont {Chen}}, \bibinfo {author} {\bibfnamefont
  {X.}~\bibnamefont {Gu}}, \ and\ \bibinfo {author} {\bibfnamefont
  {Q.}~\bibnamefont {Tian}},\ }\bibfield  {title} {\enquote {\bibinfo {title}
  {{Pangu-weather: A 3d high-resolution model for fast and accurate global
  weather forecast}},}\ }\href@noop {} {\bibfield  {journal} {\bibinfo
  {journal} {arXiv preprint arXiv:2211.02556}\ } (\bibinfo {year}
  {2022})}\BibitemShut {NoStop}%
\bibitem [{\citenamefont {Wang}\ \emph {et~al.}(2024)\citenamefont {Wang},
  \citenamefont {Solera-Rico}, \citenamefont {Vila},\ and\ \citenamefont
  {Vinuesa}}]{wang2024towards}%
  \BibitemOpen
  \bibfield  {author} {\bibinfo {author} {\bibfnamefont {Y.}~\bibnamefont
  {Wang}}, \bibinfo {author} {\bibfnamefont {A.}~\bibnamefont {Solera-Rico}},
  \bibinfo {author} {\bibfnamefont {C.~S.}\ \bibnamefont {Vila}}, \ and\
  \bibinfo {author} {\bibfnamefont {R.}~\bibnamefont {Vinuesa}},\ }\bibfield
  {title} {\enquote {\bibinfo {title} {{Towards optimal $\beta$-variational
  autoencoders combined with transformers for reduced-order modelling of
  turbulent flows}},}\ }\href@noop {} {\bibfield  {journal} {\bibinfo
  {journal} {Int J Heat Fluid Flow}\ }\textbf {\bibinfo {volume} {105}},\
  \bibinfo {pages} {109254} (\bibinfo {year} {2024})}\BibitemShut {NoStop}%
\bibitem [{\citenamefont {Janny}\ \emph {et~al.}(2023)\citenamefont {Janny},
  \citenamefont {Beneteau}, \citenamefont {Thome}, \citenamefont {Nadri},
  \citenamefont {Digne},\ and\ \citenamefont {Wolf}}]{janny2023eagle}%
  \BibitemOpen
  \bibfield  {author} {\bibinfo {author} {\bibfnamefont {S.}~\bibnamefont
  {Janny}}, \bibinfo {author} {\bibfnamefont {A.}~\bibnamefont {Beneteau}},
  \bibinfo {author} {\bibfnamefont {N.}~\bibnamefont {Thome}}, \bibinfo
  {author} {\bibfnamefont {M.}~\bibnamefont {Nadri}}, \bibinfo {author}
  {\bibfnamefont {J.}~\bibnamefont {Digne}}, \ and\ \bibinfo {author}
  {\bibfnamefont {C.}~\bibnamefont {Wolf}},\ }\bibfield  {title} {\enquote
  {\bibinfo {title} {{Eagle: Large-scale learning of turbulent fluid dynamics
  with mesh transformers}},}\ }\href@noop {} {\bibfield  {journal} {\bibinfo
  {journal} {arXiv preprint arXiv:2302.10803}\ } (\bibinfo {year}
  {2023})}\BibitemShut {NoStop}%
\bibitem [{\citenamefont {Patil}, \citenamefont {Viquerat},\ and\ \citenamefont
  {Hachem}(2022)}]{patil2022autoregressive}%
  \BibitemOpen
  \bibfield  {author} {\bibinfo {author} {\bibfnamefont {A.}~\bibnamefont
  {Patil}}, \bibinfo {author} {\bibfnamefont {J.}~\bibnamefont {Viquerat}}, \
  and\ \bibinfo {author} {\bibfnamefont {E.}~\bibnamefont {Hachem}},\
  }\bibfield  {title} {\enquote {\bibinfo {title} {Autoregressive transformers
  for data-driven spatio-temporal learning of turbulent flows},}\ }\href@noop
  {} {\bibfield  {journal} {\bibinfo  {journal} {arXiv preprint
  arXiv:2209.08052}\ } (\bibinfo {year} {2022})}\BibitemShut {NoStop}%
\bibitem [{\citenamefont {Dang}\ \emph {et~al.}(2022)\citenamefont {Dang},
  \citenamefont {Hu}, \citenamefont {Cranmer}, \citenamefont {Eickenberg},\
  and\ \citenamefont {Ho}}]{dang2022tnt}%
  \BibitemOpen
  \bibfield  {author} {\bibinfo {author} {\bibfnamefont {Y.}~\bibnamefont
  {Dang}}, \bibinfo {author} {\bibfnamefont {Z.}~\bibnamefont {Hu}}, \bibinfo
  {author} {\bibfnamefont {M.}~\bibnamefont {Cranmer}}, \bibinfo {author}
  {\bibfnamefont {M.}~\bibnamefont {Eickenberg}}, \ and\ \bibinfo {author}
  {\bibfnamefont {S.}~\bibnamefont {Ho}},\ }\bibfield  {title} {\enquote
  {\bibinfo {title} {{TNT: Vision transformer for turbulence simulations}},}\
  }\href@noop {} {\bibfield  {journal} {\bibinfo  {journal} {arXiv preprint
  arXiv:2207.04616}\ } (\bibinfo {year} {2022})}\BibitemShut {NoStop}%
\bibitem [{\citenamefont {Li}, \citenamefont {Shu},\ and\ \citenamefont
  {Farimani}(2023)}]{li2023scalable}%
  \BibitemOpen
  \bibfield  {author} {\bibinfo {author} {\bibfnamefont {Z.}~\bibnamefont
  {Li}}, \bibinfo {author} {\bibfnamefont {D.}~\bibnamefont {Shu}}, \ and\
  \bibinfo {author} {\bibfnamefont {A.~B.}\ \bibnamefont {Farimani}},\
  }\bibfield  {title} {\enquote {\bibinfo {title} {{Scalable Transformer for
  PDE Surrogate Modeling}},}\ }\href@noop {} {\bibfield  {journal} {\bibinfo
  {journal} {arXiv preprint arXiv:2305.17560}\ } (\bibinfo {year}
  {2023})}\BibitemShut {NoStop}%
\bibitem [{\citenamefont {Jiang}\ \emph
  {et~al.}(2023{\natexlab{b}})\citenamefont {Jiang}, \citenamefont {Li},
  \citenamefont {Jiang}, \citenamefont {Zhang},\ and\ \citenamefont
  {Deng}}]{jiang2023transcfd}%
  \BibitemOpen
  \bibfield  {author} {\bibinfo {author} {\bibfnamefont {J.}~\bibnamefont
  {Jiang}}, \bibinfo {author} {\bibfnamefont {G.}~\bibnamefont {Li}}, \bibinfo
  {author} {\bibfnamefont {Y.}~\bibnamefont {Jiang}}, \bibinfo {author}
  {\bibfnamefont {L.}~\bibnamefont {Zhang}}, \ and\ \bibinfo {author}
  {\bibfnamefont {X.}~\bibnamefont {Deng}},\ }\bibfield  {title} {\enquote
  {\bibinfo {title} {{TransCFD: A transformer-based decoder for flow field
  prediction}},}\ }\href@noop {} {\bibfield  {journal} {\bibinfo  {journal}
  {Eng Appl Artif Intell}\ }\textbf {\bibinfo {volume} {123}},\ \bibinfo
  {pages} {106340} (\bibinfo {year} {2023}{\natexlab{b}})}\BibitemShut
  {NoStop}%
\bibitem [{\citenamefont {Pope}\ and\ \citenamefont {Pope}(2000)}]{pope2000}%
  \BibitemOpen
  \bibfield  {author} {\bibinfo {author} {\bibfnamefont {S.~B.}\ \bibnamefont
  {Pope}}\ and\ \bibinfo {author} {\bibfnamefont {S.~B.}\ \bibnamefont
  {Pope}},\ }\href@noop {} {\emph {\bibinfo {title} {Turbulent flows}}}\
  (\bibinfo  {publisher} {Cambridge university press},\ \bibinfo {year}
  {2000})\BibitemShut {NoStop}%
\bibitem [{\citenamefont {Ishihara}, \citenamefont {Gotoh},\ and\ \citenamefont
  {Kaneda}(2009)}]{Ishihara2009}%
  \BibitemOpen
  \bibfield  {author} {\bibinfo {author} {\bibfnamefont {T.}~\bibnamefont
  {Ishihara}}, \bibinfo {author} {\bibfnamefont {T.}~\bibnamefont {Gotoh}}, \
  and\ \bibinfo {author} {\bibfnamefont {Y.}~\bibnamefont {Kaneda}},\
  }\bibfield  {title} {\enquote {\bibinfo {title} {{Study of high--Reynolds
  number isotropic turbulence by direct numerical simulation}},}\ }\href@noop
  {} {\bibfield  {journal} {\bibinfo  {journal} {Annu. Rev. Fluid Mech.}\
  }\textbf {\bibinfo {volume} {41}},\ \bibinfo {pages} {165--180} (\bibinfo
  {year} {2009})}\BibitemShut {NoStop}%
\bibitem [{\citenamefont {Wang}\ \emph {et~al.}(2022)\citenamefont {Wang},
  \citenamefont {Yuan}, \citenamefont {Wang},\ and\ \citenamefont
  {Wang}}]{wyp2022}%
  \BibitemOpen
  \bibfield  {author} {\bibinfo {author} {\bibfnamefont {Y.}~\bibnamefont
  {Wang}}, \bibinfo {author} {\bibfnamefont {Z.}~\bibnamefont {Yuan}}, \bibinfo
  {author} {\bibfnamefont {X.}~\bibnamefont {Wang}}, \ and\ \bibinfo {author}
  {\bibfnamefont {J.}~\bibnamefont {Wang}},\ }\bibfield  {title} {\enquote
  {\bibinfo {title} {{Constant-coefficient spatial gradient models for the
  sub-grid scale closure in large-eddy simulation of turbulence}},}\
  }\href@noop {} {\bibfield  {journal} {\bibinfo  {journal} {Phys. Fluids}\
  }\textbf {\bibinfo {volume} {34}},\ \bibinfo {pages} {095108} (\bibinfo
  {year} {2022})}\BibitemShut {NoStop}%
\bibitem [{\citenamefont {Sagaut}(2006)}]{sagaut2006}%
  \BibitemOpen
  \bibfield  {author} {\bibinfo {author} {\bibfnamefont {P.}~\bibnamefont
  {Sagaut}},\ }\href@noop {} {\emph {\bibinfo {title} {{Large eddy simulation
  for incompressible flows: an introduction}}}}\ (\bibinfo  {publisher}
  {Springer Science \& Business Media},\ \bibinfo {year} {2006})\BibitemShut
  {NoStop}%
\bibitem [{\citenamefont {Moser}, \citenamefont {Haering},\ and\ \citenamefont
  {Yalla}(2021)}]{moser2021}%
  \BibitemOpen
  \bibfield  {author} {\bibinfo {author} {\bibfnamefont {R.~D.}\ \bibnamefont
  {Moser}}, \bibinfo {author} {\bibfnamefont {S.~W.}\ \bibnamefont {Haering}},
  \ and\ \bibinfo {author} {\bibfnamefont {G.~R.}\ \bibnamefont {Yalla}},\
  }\bibfield  {title} {\enquote {\bibinfo {title} {{Statistical properties of
  subgrid-scale turbulence models}},}\ }\href@noop {} {\bibfield  {journal}
  {\bibinfo  {journal} {Annu. Rev. Fluid Mech.}\ }\textbf {\bibinfo {volume}
  {53}},\ \bibinfo {pages} {255--286} (\bibinfo {year} {2021})}\BibitemShut
  {NoStop}%
\bibitem [{\citenamefont {Johnson}(2022)}]{johnson2022}%
  \BibitemOpen
  \bibfield  {author} {\bibinfo {author} {\bibfnamefont {P.~L.}\ \bibnamefont
  {Johnson}},\ }\bibfield  {title} {\enquote {\bibinfo {title} {{A
  physics-inspired alternative to spatial filtering for large-eddy simulations
  of turbulent flows}},}\ }\href@noop {} {\bibfield  {journal} {\bibinfo
  {journal} {J. Fluid Mech.}\ }\textbf {\bibinfo {volume} {934}},\ \bibinfo
  {pages} {A30} (\bibinfo {year} {2022})}\BibitemShut {NoStop}%
\bibitem [{\citenamefont {Lilly}(1992)}]{lilly1992}%
  \BibitemOpen
  \bibfield  {author} {\bibinfo {author} {\bibfnamefont {D.~K.}\ \bibnamefont
  {Lilly}},\ }\bibfield  {title} {\enquote {\bibinfo {title} {{A proposed
  modification of the Germano subgrid-scale closure method}},}\ }\href@noop {}
  {\bibfield  {journal} {\bibinfo  {journal} {Phys. Fluids A}\ }\textbf
  {\bibinfo {volume} {4}},\ \bibinfo {pages} {633--635} (\bibinfo {year}
  {1992})}\BibitemShut {NoStop}%
\bibitem [{\citenamefont {Liu}, \citenamefont {Meneveau},\ and\ \citenamefont
  {Katz}(1994)}]{liu1994properties}%
  \BibitemOpen
  \bibfield  {author} {\bibinfo {author} {\bibfnamefont {S.}~\bibnamefont
  {Liu}}, \bibinfo {author} {\bibfnamefont {C.}~\bibnamefont {Meneveau}}, \
  and\ \bibinfo {author} {\bibfnamefont {J.}~\bibnamefont {Katz}},\ }\bibfield
  {title} {\enquote {\bibinfo {title} {{On the properties of similarity
  subgrid-scale models as deduced from measurements in a turbulent jet}},}\
  }\href@noop {} {\bibfield  {journal} {\bibinfo  {journal} {J. Fluid Mech.}\
  }\textbf {\bibinfo {volume} {275}},\ \bibinfo {pages} {83--119} (\bibinfo
  {year} {1994})}\BibitemShut {NoStop}%
\bibitem [{\citenamefont {Yuan}, \citenamefont {Xie},\ and\ \citenamefont
  {Wang}(2020)}]{yuan2020deconvolutional}%
  \BibitemOpen
  \bibfield  {author} {\bibinfo {author} {\bibfnamefont {Z.}~\bibnamefont
  {Yuan}}, \bibinfo {author} {\bibfnamefont {C.}~\bibnamefont {Xie}}, \ and\
  \bibinfo {author} {\bibfnamefont {J.}~\bibnamefont {Wang}},\ }\bibfield
  {title} {\enquote {\bibinfo {title} {{Deconvolutional artificial neural
  network models for large eddy simulation of turbulence}},}\ }\href@noop {}
  {\bibfield  {journal} {\bibinfo  {journal} {Phys. Fluids}\ }\textbf {\bibinfo
  {volume} {32}},\ \bibinfo {pages} {115106} (\bibinfo {year}
  {2020})}\BibitemShut {NoStop}%
\bibitem [{\citenamefont {Beauzamy}(2011)}]{beauzamy2011}%
  \BibitemOpen
  \bibfield  {author} {\bibinfo {author} {\bibfnamefont {B.}~\bibnamefont
  {Beauzamy}},\ }\href@noop {} {\emph {\bibinfo {title} {{Introduction to
  Banach spaces and their geometry}}}}\ (\bibinfo  {publisher} {Elsevier},\
  \bibinfo {year} {2011})\BibitemShut {NoStop}%
\bibitem [{\citenamefont {Vapnik}(1999)}]{vapnik1999}%
  \BibitemOpen
  \bibfield  {author} {\bibinfo {author} {\bibfnamefont {V.~N.}\ \bibnamefont
  {Vapnik}},\ }\bibfield  {title} {\enquote {\bibinfo {title} {{An overview of
  statistical learning theory}},}\ }\href@noop {} {\bibfield  {journal}
  {\bibinfo  {journal} {IEEE Trans Neural Netw}\ }\textbf {\bibinfo {volume}
  {10}},\ \bibinfo {pages} {988--999} (\bibinfo {year} {1999})}\BibitemShut
  {NoStop}%
\bibitem [{\citenamefont {Li}\ \emph {et~al.}(2020{\natexlab{b}})\citenamefont
  {Li}, \citenamefont {Kovachki}, \citenamefont {Azizzadenesheli},
  \citenamefont {Liu}, \citenamefont {Bhattacharya}, \citenamefont {Stuart},\
  and\ \citenamefont {Anandkumar}}]{li2020b}%
  \BibitemOpen
  \bibfield  {author} {\bibinfo {author} {\bibfnamefont {Z.}~\bibnamefont
  {Li}}, \bibinfo {author} {\bibfnamefont {N.}~\bibnamefont {Kovachki}},
  \bibinfo {author} {\bibfnamefont {K.}~\bibnamefont {Azizzadenesheli}},
  \bibinfo {author} {\bibfnamefont {B.}~\bibnamefont {Liu}}, \bibinfo {author}
  {\bibfnamefont {K.}~\bibnamefont {Bhattacharya}}, \bibinfo {author}
  {\bibfnamefont {A.}~\bibnamefont {Stuart}}, \ and\ \bibinfo {author}
  {\bibfnamefont {A.}~\bibnamefont {Anandkumar}},\ }\bibfield  {title}
  {\enquote {\bibinfo {title} {{Neural operator: Graph kernel network for
  partial differential equations}},}\ }\href@noop {} {\bibfield  {journal}
  {\bibinfo  {journal} {arXiv preprint arXiv:2003.03485}\ } (\bibinfo {year}
  {2020}{\natexlab{b}})}\BibitemShut {NoStop}%
\bibitem [{\citenamefont {Rashid}\ \emph {et~al.}(2022)\citenamefont {Rashid},
  \citenamefont {Pittie}, \citenamefont {Chakraborty},\ and\ \citenamefont
  {Krishnan}}]{rashid2022learning}%
  \BibitemOpen
  \bibfield  {author} {\bibinfo {author} {\bibfnamefont {M.~M.}\ \bibnamefont
  {Rashid}}, \bibinfo {author} {\bibfnamefont {T.}~\bibnamefont {Pittie}},
  \bibinfo {author} {\bibfnamefont {S.}~\bibnamefont {Chakraborty}}, \ and\
  \bibinfo {author} {\bibfnamefont {N.~A.}\ \bibnamefont {Krishnan}},\
  }\bibfield  {title} {\enquote {\bibinfo {title} {{Learning the stress-strain
  fields in digital composites using Fourier neural operator}},}\ }\href@noop
  {} {\bibfield  {journal} {\bibinfo  {journal} {Iscience}\ }\textbf {\bibinfo
  {volume} {25}},\ \bibinfo {pages} {105452} (\bibinfo {year}
  {2022})}\BibitemShut {NoStop}%
\bibitem [{\citenamefont {Pathak}\ \emph {et~al.}(2022)\citenamefont {Pathak},
  \citenamefont {Subramanian}, \citenamefont {Harrington}, \citenamefont
  {Raja}, \citenamefont {Chattopadhyay}, \citenamefont {Mardani}, \citenamefont
  {Kurth}, \citenamefont {Hall}, \citenamefont {Li}, \citenamefont
  {Azizzadenesheli} \emph {et~al.}}]{pathak2022fourcastnet}%
  \BibitemOpen
  \bibfield  {author} {\bibinfo {author} {\bibfnamefont {J.}~\bibnamefont
  {Pathak}}, \bibinfo {author} {\bibfnamefont {S.}~\bibnamefont {Subramanian}},
  \bibinfo {author} {\bibfnamefont {P.}~\bibnamefont {Harrington}}, \bibinfo
  {author} {\bibfnamefont {S.}~\bibnamefont {Raja}}, \bibinfo {author}
  {\bibfnamefont {A.}~\bibnamefont {Chattopadhyay}}, \bibinfo {author}
  {\bibfnamefont {M.}~\bibnamefont {Mardani}}, \bibinfo {author} {\bibfnamefont
  {T.}~\bibnamefont {Kurth}}, \bibinfo {author} {\bibfnamefont
  {D.}~\bibnamefont {Hall}}, \bibinfo {author} {\bibfnamefont {Z.}~\bibnamefont
  {Li}}, \bibinfo {author} {\bibfnamefont {K.}~\bibnamefont {Azizzadenesheli}},
   \emph {et~al.},\ }\bibfield  {title} {\enquote {\bibinfo {title}
  {{Fourcastnet: A global data-driven high-resolution weather model using
  adaptive Fourier neural operators}},}\ }\href@noop {} {\bibfield  {journal}
  {\bibinfo  {journal} {arXiv preprint arXiv:2202.11214}\ } (\bibinfo {year}
  {2022})}\BibitemShut {NoStop}%
\bibitem [{\citenamefont {Su}\ \emph {et~al.}(2024)\citenamefont {Su},
  \citenamefont {Ahmed}, \citenamefont {Lu}, \citenamefont {Pan}, \citenamefont
  {Bo},\ and\ \citenamefont {Liu}}]{su2024roformer}%
  \BibitemOpen
  \bibfield  {author} {\bibinfo {author} {\bibfnamefont {J.}~\bibnamefont
  {Su}}, \bibinfo {author} {\bibfnamefont {M.}~\bibnamefont {Ahmed}}, \bibinfo
  {author} {\bibfnamefont {Y.}~\bibnamefont {Lu}}, \bibinfo {author}
  {\bibfnamefont {S.}~\bibnamefont {Pan}}, \bibinfo {author} {\bibfnamefont
  {W.}~\bibnamefont {Bo}}, \ and\ \bibinfo {author} {\bibfnamefont
  {Y.}~\bibnamefont {Liu}},\ }\bibfield  {title} {\enquote {\bibinfo {title}
  {Roformer: Enhanced transformer with rotary position embedding},}\
  }\href@noop {} {\bibfield  {journal} {\bibinfo  {journal} {Neurocomputing}\
  }\textbf {\bibinfo {volume} {568}},\ \bibinfo {pages} {127063} (\bibinfo
  {year} {2024})}\BibitemShut {NoStop}%
\bibitem [{\citenamefont {Hussaini}\ and\ \citenamefont
  {Zang}(1987)}]{hussaini1987spectral}%
  \BibitemOpen
  \bibfield  {author} {\bibinfo {author} {\bibfnamefont {M.~Y.}\ \bibnamefont
  {Hussaini}}\ and\ \bibinfo {author} {\bibfnamefont {T.~A.}\ \bibnamefont
  {Zang}},\ }\bibfield  {title} {\enquote {\bibinfo {title} {{Spectral methods
  in fluid dynamics}},}\ }\href@noop {} {\bibfield  {journal} {\bibinfo
  {journal} {Annu. Rev. Fluid Mech.}\ }\textbf {\bibinfo {volume} {19}},\
  \bibinfo {pages} {339--367} (\bibinfo {year} {1987})}\BibitemShut {NoStop}%
\bibitem [{\citenamefont {Chang}, \citenamefont {Yuan},\ and\ \citenamefont
  {Wang}(2022)}]{chang2022}%
  \BibitemOpen
  \bibfield  {author} {\bibinfo {author} {\bibfnamefont {N.}~\bibnamefont
  {Chang}}, \bibinfo {author} {\bibfnamefont {Z.}~\bibnamefont {Yuan}}, \ and\
  \bibinfo {author} {\bibfnamefont {J.}~\bibnamefont {Wang}},\ }\bibfield
  {title} {\enquote {\bibinfo {title} {{The effect of sub-filter scale dynamics
  in large eddy simulation of turbulence}},}\ }\href@noop {} {\bibfield
  {journal} {\bibinfo  {journal} {Phys. Fluids}\ }\textbf {\bibinfo {volume}
  {34}},\ \bibinfo {pages} {095104} (\bibinfo {year} {2022})}\BibitemShut
  {NoStop}%
\bibitem [{\citenamefont {Zhuang}\ \emph {et~al.}(2022)\citenamefont {Zhuang},
  \citenamefont {Liu}, \citenamefont {Cutkosky},\ and\ \citenamefont
  {Orabona}}]{zhuang2022understanding}%
  \BibitemOpen
  \bibfield  {author} {\bibinfo {author} {\bibfnamefont {Z.}~\bibnamefont
  {Zhuang}}, \bibinfo {author} {\bibfnamefont {M.}~\bibnamefont {Liu}},
  \bibinfo {author} {\bibfnamefont {A.}~\bibnamefont {Cutkosky}}, \ and\
  \bibinfo {author} {\bibfnamefont {F.}~\bibnamefont {Orabona}},\ }\bibfield
  {title} {\enquote {\bibinfo {title} {Understanding adamw through proximal
  methods and scale-freeness},}\ }\href@noop {} {\bibfield  {journal} {\bibinfo
   {journal} {Transactions on Machine Learning Research}\ } (\bibinfo {year}
  {2022})}\BibitemShut {NoStop}%
\bibitem [{\citenamefont {Loshchilov}\ and\ \citenamefont
  {Hutter}(2018)}]{loshchilov2018fixing}%
  \BibitemOpen
  \bibfield  {author} {\bibinfo {author} {\bibfnamefont {I.}~\bibnamefont
  {Loshchilov}}\ and\ \bibinfo {author} {\bibfnamefont {F.}~\bibnamefont
  {Hutter}},\ }\bibfield  {title} {\enquote {\bibinfo {title} {Fixing weight
  decay regularization in adam},}\ }\href@noop {} {\  (\bibinfo {year}
  {2018})}\BibitemShut {NoStop}%
\bibitem [{\citenamefont {Hendrycks}\ and\ \citenamefont
  {Gimpel}(2016)}]{hendrycks2016gaussian}%
  \BibitemOpen
  \bibfield  {author} {\bibinfo {author} {\bibfnamefont {D.}~\bibnamefont
  {Hendrycks}}\ and\ \bibinfo {author} {\bibfnamefont {K.}~\bibnamefont
  {Gimpel}},\ }\bibfield  {title} {\enquote {\bibinfo {title} {{Gaussian error
  linear units (gelus)}},}\ }\href@noop {} {\bibfield  {journal} {\bibinfo
  {journal} {arXiv preprint arXiv:1606.08415}\ } (\bibinfo {year}
  {2016})}\BibitemShut {NoStop}%
\bibitem [{\citenamefont {Xie}\ \emph {et~al.}(2018)\citenamefont {Xie},
  \citenamefont {Wang}, \citenamefont {Li}, \citenamefont {Wan},\ and\
  \citenamefont {Chen}}]{xie2018modified}%
  \BibitemOpen
  \bibfield  {author} {\bibinfo {author} {\bibfnamefont {C.}~\bibnamefont
  {Xie}}, \bibinfo {author} {\bibfnamefont {J.}~\bibnamefont {Wang}}, \bibinfo
  {author} {\bibfnamefont {H.}~\bibnamefont {Li}}, \bibinfo {author}
  {\bibfnamefont {M.}~\bibnamefont {Wan}}, \ and\ \bibinfo {author}
  {\bibfnamefont {S.}~\bibnamefont {Chen}},\ }\bibfield  {title} {\enquote
  {\bibinfo {title} {{A modified optimal LES model for highly compressible
  isotropic turbulence}},}\ }\href@noop {} {\bibfield  {journal} {\bibinfo
  {journal} {Phys. Fluids}\ }\textbf {\bibinfo {volume} {30}},\ \bibinfo
  {pages} {065108} (\bibinfo {year} {2018})}\BibitemShut {NoStop}%
\bibitem [{\citenamefont {Xie}\ \emph {et~al.}(2020)\citenamefont {Xie},
  \citenamefont {Wang}, \citenamefont {Li}, \citenamefont {Wan},\ and\
  \citenamefont {Chen}}]{xie2020approximate}%
  \BibitemOpen
  \bibfield  {author} {\bibinfo {author} {\bibfnamefont {C.}~\bibnamefont
  {Xie}}, \bibinfo {author} {\bibfnamefont {J.}~\bibnamefont {Wang}}, \bibinfo
  {author} {\bibfnamefont {H.}~\bibnamefont {Li}}, \bibinfo {author}
  {\bibfnamefont {M.}~\bibnamefont {Wan}}, \ and\ \bibinfo {author}
  {\bibfnamefont {S.}~\bibnamefont {Chen}},\ }\bibfield  {title} {\enquote
  {\bibinfo {title} {{An approximate second-order closure model for large-eddy
  simulation of compressible isotropic turbulence}},}\ }\href@noop {}
  {\bibfield  {journal} {\bibinfo  {journal} {Commun Comput Phys}\ }\textbf
  {\bibinfo {volume} {27}},\ \bibinfo {pages} {775--808} (\bibinfo {year}
  {2020})}\BibitemShut {NoStop}%
\bibitem [{\citenamefont {Kleissl}\ \emph {et~al.}(2006)\citenamefont
  {Kleissl}, \citenamefont {Kumar}, \citenamefont {Meneveau},\ and\
  \citenamefont {Parlange}}]{kleissl2006numerical}%
  \BibitemOpen
  \bibfield  {author} {\bibinfo {author} {\bibfnamefont {J.}~\bibnamefont
  {Kleissl}}, \bibinfo {author} {\bibfnamefont {V.}~\bibnamefont {Kumar}},
  \bibinfo {author} {\bibfnamefont {C.}~\bibnamefont {Meneveau}}, \ and\
  \bibinfo {author} {\bibfnamefont {M.~B.}\ \bibnamefont {Parlange}},\
  }\bibfield  {title} {\enquote {\bibinfo {title} {{Numerical study of dynamic
  Smagorinsky models in large-eddy simulation of the atmospheric boundary
  layer: Validation in stable and unstable conditions}},}\ }\href@noop {}
  {\bibfield  {journal} {\bibinfo  {journal} {Water Resour. Res.}\ }\textbf
  {\bibinfo {volume} {42}} (\bibinfo {year} {2006})}\BibitemShut {NoStop}%
\bibitem [{\citenamefont {Wang}, \citenamefont {Wang},\ and\ \citenamefont
  {Chen}(2022)}]{wang2022compressibility}%
  \BibitemOpen
  \bibfield  {author} {\bibinfo {author} {\bibfnamefont {X.}~\bibnamefont
  {Wang}}, \bibinfo {author} {\bibfnamefont {J.}~\bibnamefont {Wang}}, \ and\
  \bibinfo {author} {\bibfnamefont {S.}~\bibnamefont {Chen}},\ }\bibfield
  {title} {\enquote {\bibinfo {title} {{Compressibility effects on statistics
  and coherent structures of compressible turbulent mixing layers}},}\
  }\href@noop {} {\bibfield  {journal} {\bibinfo  {journal} {J. Fluid Mech.}\
  }\textbf {\bibinfo {volume} {947}},\ \bibinfo {pages} {A38} (\bibinfo {year}
  {2022})}\BibitemShut {NoStop}%
\bibitem [{\citenamefont {Sharan}, \citenamefont {Matheou},\ and\ \citenamefont
  {Dimotakis}(2019)}]{sharan2019turbulent}%
  \BibitemOpen
  \bibfield  {author} {\bibinfo {author} {\bibfnamefont {N.}~\bibnamefont
  {Sharan}}, \bibinfo {author} {\bibfnamefont {G.}~\bibnamefont {Matheou}}, \
  and\ \bibinfo {author} {\bibfnamefont {P.~E.}\ \bibnamefont {Dimotakis}},\
  }\bibfield  {title} {\enquote {\bibinfo {title} {{Turbulent shear-layer
  mixing: initial conditions, and direct-numerical and large-eddy
  simulations}},}\ }\href@noop {} {\bibfield  {journal} {\bibinfo  {journal}
  {J. Fluid Mech.}\ }\textbf {\bibinfo {volume} {877}},\ \bibinfo {pages}
  {35--81} (\bibinfo {year} {2019})}\BibitemShut {NoStop}%
\bibitem [{\citenamefont {Yuan}\ \emph {et~al.}(2023)\citenamefont {Yuan},
  \citenamefont {Wang}, \citenamefont {Wang},\ and\ \citenamefont
  {Wang}}]{yuan2023adjoint}%
  \BibitemOpen
  \bibfield  {author} {\bibinfo {author} {\bibfnamefont {Z.}~\bibnamefont
  {Yuan}}, \bibinfo {author} {\bibfnamefont {Y.}~\bibnamefont {Wang}}, \bibinfo
  {author} {\bibfnamefont {X.}~\bibnamefont {Wang}}, \ and\ \bibinfo {author}
  {\bibfnamefont {J.}~\bibnamefont {Wang}},\ }\bibfield  {title} {\enquote
  {\bibinfo {title} {{Adjoint-based variational optimal mixed models for
  large-eddy simulation of turbulence}},}\ }\href@noop {} {\bibfield  {journal}
  {\bibinfo  {journal} {Phys. Fluids}\ }\textbf {\bibinfo {volume} {35}},\
  \bibinfo {pages} {075105} (\bibinfo {year} {2023})}\BibitemShut {NoStop}%
\bibitem [{\citenamefont {Rogers}\ and\ \citenamefont
  {Moser}(1994)}]{rogers1994direct}%
  \BibitemOpen
  \bibfield  {author} {\bibinfo {author} {\bibfnamefont {M.~M.}\ \bibnamefont
  {Rogers}}\ and\ \bibinfo {author} {\bibfnamefont {R.~D.}\ \bibnamefont
  {Moser}},\ }\bibfield  {title} {\enquote {\bibinfo {title} {{Direct
  simulation of a self-similar turbulent mixing layer}},}\ }\href@noop {}
  {\bibfield  {journal} {\bibinfo  {journal} {Phys. Fluids}\ }\textbf {\bibinfo
  {volume} {6}},\ \bibinfo {pages} {903--923} (\bibinfo {year}
  {1994})}\BibitemShut {NoStop}%
\bibitem [{\citenamefont {Dubief}\ and\ \citenamefont
  {Delcayre}(2000)}]{dubief2000coherent}%
  \BibitemOpen
  \bibfield  {author} {\bibinfo {author} {\bibfnamefont {Y.}~\bibnamefont
  {Dubief}}\ and\ \bibinfo {author} {\bibfnamefont {F.}~\bibnamefont
  {Delcayre}},\ }\bibfield  {title} {\enquote {\bibinfo {title} {{On
  coherent-vortex identification in turbulence}},}\ }\href@noop {} {\bibfield
  {journal} {\bibinfo  {journal} {J. Turbul.}\ }\textbf {\bibinfo {volume}
  {1}},\ \bibinfo {pages} {011} (\bibinfo {year} {2000})}\BibitemShut {NoStop}%
\bibitem [{\citenamefont {Zhan}\ \emph {et~al.}(2019)\citenamefont {Zhan},
  \citenamefont {Li}, \citenamefont {Wai},\ and\ \citenamefont
  {Hu}}]{zhan2019comparison}%
  \BibitemOpen
  \bibfield  {author} {\bibinfo {author} {\bibfnamefont {J.}~\bibnamefont
  {Zhan}}, \bibinfo {author} {\bibfnamefont {Y.}~\bibnamefont {Li}}, \bibinfo
  {author} {\bibfnamefont {W.~O.}\ \bibnamefont {Wai}}, \ and\ \bibinfo
  {author} {\bibfnamefont {W.}~\bibnamefont {Hu}},\ }\bibfield  {title}
  {\enquote {\bibinfo {title} {{Comparison between the Q criterion and Rortex
  in the application of an in-stream structure}},}\ }\href@noop {} {\bibfield
  {journal} {\bibinfo  {journal} {Phys. Fluids}\ }\textbf {\bibinfo {volume}
  {31}},\ \bibinfo {pages} {121701} (\bibinfo {year} {2019})}\BibitemShut
  {NoStop}%
\bibitem [{\citenamefont {Kurth}\ \emph {et~al.}(2022)\citenamefont {Kurth},
  \citenamefont {Subramanian}, \citenamefont {Harrington}, \citenamefont
  {Pathak}, \citenamefont {Mardani}, \citenamefont {Hall}, \citenamefont
  {Miele}, \citenamefont {Kashinath},\ and\ \citenamefont
  {Anandkumar}}]{kurth2022fourcastnet}%
  \BibitemOpen
  \bibfield  {author} {\bibinfo {author} {\bibfnamefont {T.}~\bibnamefont
  {Kurth}}, \bibinfo {author} {\bibfnamefont {S.}~\bibnamefont {Subramanian}},
  \bibinfo {author} {\bibfnamefont {P.}~\bibnamefont {Harrington}}, \bibinfo
  {author} {\bibfnamefont {J.}~\bibnamefont {Pathak}}, \bibinfo {author}
  {\bibfnamefont {M.}~\bibnamefont {Mardani}}, \bibinfo {author} {\bibfnamefont
  {D.}~\bibnamefont {Hall}}, \bibinfo {author} {\bibfnamefont {A.}~\bibnamefont
  {Miele}}, \bibinfo {author} {\bibfnamefont {K.}~\bibnamefont {Kashinath}}, \
  and\ \bibinfo {author} {\bibfnamefont {A.}~\bibnamefont {Anandkumar}},\
  }\bibfield  {title} {\enquote {\bibinfo {title} {{Fourcastnet: Accelerating
  global high-resolution weather forecasting using adaptive Fourier neural
  operators}},}\ }\href@noop {} {\bibfield  {journal} {\bibinfo  {journal}
  {arXiv preprint arXiv:2208.05419}\ } (\bibinfo {year} {2022})}\BibitemShut
  {NoStop}%
\bibitem [{\citenamefont {Alkin}\ \emph {et~al.}(2024)\citenamefont {Alkin},
  \citenamefont {F{\"u}rst}, \citenamefont {Schmid}, \citenamefont {Gruber},
  \citenamefont {Holzleitner},\ and\ \citenamefont
  {Brandstetter}}]{alkin2024universal}%
  \BibitemOpen
  \bibfield  {author} {\bibinfo {author} {\bibfnamefont {B.}~\bibnamefont
  {Alkin}}, \bibinfo {author} {\bibfnamefont {A.}~\bibnamefont {F{\"u}rst}},
  \bibinfo {author} {\bibfnamefont {S.}~\bibnamefont {Schmid}}, \bibinfo
  {author} {\bibfnamefont {L.}~\bibnamefont {Gruber}}, \bibinfo {author}
  {\bibfnamefont {M.}~\bibnamefont {Holzleitner}}, \ and\ \bibinfo {author}
  {\bibfnamefont {J.}~\bibnamefont {Brandstetter}},\ }\bibfield  {title}
  {\enquote {\bibinfo {title} {Universal physics transformers},}\ }\href@noop
  {} {\bibfield  {journal} {\bibinfo  {journal} {arXiv preprint
  arXiv:2402.12365}\ } (\bibinfo {year} {2024})}\BibitemShut {NoStop}%
\bibitem [{\citenamefont {Zhao}, \citenamefont {Ding},\ and\ \citenamefont
  {Prakash}(2023)}]{zhao2023pinnsformer}%
  \BibitemOpen
  \bibfield  {author} {\bibinfo {author} {\bibfnamefont {L.~Z.}\ \bibnamefont
  {Zhao}}, \bibinfo {author} {\bibfnamefont {X.}~\bibnamefont {Ding}}, \ and\
  \bibinfo {author} {\bibfnamefont {B.~A.}\ \bibnamefont {Prakash}},\
  }\bibfield  {title} {\enquote {\bibinfo {title} {Pinnsformer: A
  transformer-based framework for physics-informed neural networks},}\
  }\href@noop {} {\bibfield  {journal} {\bibinfo  {journal} {arXiv preprint
  arXiv:2307.11833}\ } (\bibinfo {year} {2023})}\BibitemShut {NoStop}%
\end{thebibliography}%

\end{document}